\PassOptionsToPackage{table}{xcolor}
\documentclass[10pt,a4paper]{article}
\usepackage[margin=1in]{geometry}

\usepackage{filecontents}
\usepackage[pdfencoding=auto]{hyperref}
\usepackage{etoolbox}
\newtoggle{electronic}
\toggletrue{electronic}
\iftoggle{electronic}
{
        \hypersetup{
        colorlinks,
        linkcolor=blue,
        citecolor=blue
        }
}{
        \hypersetup{
        colorlinks=false,
        pdfborder=0 0 0
        }
}

\usepackage[document]{ragged2e}
\usepackage{amsmath}
\usepackage{amssymb}
\usepackage{mathtools}
\usepackage{graphicx}
\usepackage{dsfont}
\usepackage{booktabs}
\usepackage{tabularx}

\usepackage{multirow}
\usepackage[normalem]{ulem} \usepackage{siunitx}
\DeclareSIUnit\Molar{M}
\usepackage{tikz} 
\usetikzlibrary{matrix}
\hypersetup{pdfstartview=}

\usepackage{xr}
\makeatletter
\newcommand*{\addFileDependency}[1]{  \typeout{(#1)}  \@addtofilelist{#1}  \IfFileExists{#1}{}{\typeout{No file #1.}}}
\makeatother

\newcommand*{\myexternaldocument}[1]{    \externaldocument[supp-]{#1}    \addFileDependency{#1.tex}    \addFileDependency{#1.aux}}
\myexternaldocument{supplement}

\newcommand{\ikr}{$I_\text{Kr}$}
\DeclareMathOperator*{\argmax}{arg max} \DeclareMathOperator*{\argmin}{arg min} 
\newcommand{\uCoU}{u_{P}}
\newcommand{\YCoU}{Y_{P}}

\newcommand{\thetaest}{\hat{\boldsymbol{\theta}}}

\DeclareMathOperator*{\KL}{KL}
\DeclareMathOperator*{\GP}{GP}
\DeclareMathOperator*{\ARMA}{ARMA}

\providecommand{\keywords}[1]{\textbf{Keywords:} #1}

\begin{document}

\title{
Considering discrepancy when calibrating a mechanistic electrophysiology model
\vspace{0.5cm}
}

\author{Chon Lok Lei$^{1}$,
Sanmitra Ghosh$^{2}$,
Dominic G.\ Whittaker$^{3}$,
Yasser Aboelkassem$^{4}$,\\
Kylie A.\ Beattie$^{5}$,
Chris D.\ Cantwell$^{6}$,
Tammo Delhaas$^{7}$,
Charles Houston$^{6}$,\\
Gustavo Montes Novaes$^{8}$,
Alexander V.\ Panfilov$^{9,10}$,
Pras Pathmanathan$^{11}$,\\
Marina Riabiz$^{12}$,
Rodrigo Weber dos Santos$^{8}$,
John Walmsley$^{13}$,\\
Keith Worden$^{14}$,
Gary R.\ Mirams$^{3}$ and
Richard D.\ Wilkinson$^{15}$
}

\date{\FlushLeft
\scriptsize{$^{1}$ Computational Biology \& Health Informatics, Dept.\ of Computer Science, University of Oxford, UK.\\ $^{2}$ MRC Biostatistics Unit, University of Cambridge, UK\\ $^{3}$ Centre for Mathematical Medicine \& Biology, School of Mathematical Sciences, University of Nottingham, UK.\\ $^{4}$ Department of Bioengineering, University of California San Diego, USA.\\ $^{5}$ Systems Modeling and Translational Biology, GlaxoSmithKline R\&D, Stevenage, UK.\\ 
$^{6}$ ElectroCardioMaths Programme, Centre for Cardiac Engineering, Imperial College London, UK.\\
$^{7}$ CARIM School for Cardiovascular Diseases, Maastricht University, the Netherlands.\\ 
$^{8}$ Graduate Program in Computational Modeling, Universidade Federal de Juiz de Fora, Brazil.\\ 
$^{9}$ Department of Physics and Astronomy, Ghent University, Belgium.\\ 
$^{10}$ Laboratory of Computational Biology and Medicine, Ural Federal University, Ekaterinburg, Russia.\\  
$^{11}$ U.S. Food and Drug Administration, Center for Devices and Radiological Health, Office of Science and Engineering Laboratories, USA.\\ $^{12}$ Department of Biomedical Engineering King's College London and Alan Turing Institute, UK.\\ $^{13}$ James T.\ Willerson Center for Cardiovascular Modeling and Simulation, Oden Institute for Computational Engineering and Sciences, The University of Texas at Austin, USA.\\
$^{14}$ Dynamics Research Group, Department of Mechanical Engineering, University of Sheffield, UK.\\ $^{15}$ School of Mathematics and Statistics, University of Sheffield, UK.\\
\vspace{0.2cm}
(April 2020)}}
\maketitle

\begin{abstract}
Uncertainty quantification (UQ) is a vital step in using mathematical models and simulations to take decisions. 
The field of cardiac simulation has begun to explore and adopt UQ methods to characterise uncertainty in model inputs and how that propagates through to outputs or predictions.
In this perspective piece we draw attention to an important and under-addressed source of uncertainty in our predictions --- that of uncertainty in the model structure or the equations themselves.
The difference between imperfect models and reality is termed \emph{model discrepancy}, and we are often uncertain as to the size and consequences of this discrepancy.
Here we provide two examples of the consequences of discrepancy when calibrating models at the ion channel and action potential scales.
Furthermore, we attempt to account for this discrepancy when calibrating and validating an ion channel model using different methods, based on modelling the discrepancy using Gaussian processes (GPs) and autoregressive-moving-average (ARMA) models, then highlight the advantages and shortcomings of each approach.
Finally, suggestions and lines of enquiry for future work are provided.
\end{abstract}


\section{Introduction}
This perspective paper discusses the issue of model discrepancy --- the difference between a model's predictions and reality.
The concepts and issues we highlight are  applicable to any modelling situation where governing equations are approximations or assumptions;
thus our perspective paper is intended for computational, mathematical and statistical modellers within many other fields as well as within and outside biological modelling.
The focus of our examples is  cellular cardiac electrophysiology, a well-developed area of systems biology \cite{noble_how_2012}.

\subsection{Cardiac modelling}
Cardiac models are typically a collection of mathematical functions governed by systems of ordinary and/or partial (when spatial dimensions are considered) differential equations, integrated using computational techniques, which produce responses that depend on the model inputs.
Inputs can include model parameters, initial conditions, boundary conditions, and cellular, tissue or whole organ geometrical aspects. 
Inputs which have physiological meaning can sometimes be obtained by direct measurement, whilst others may need to be estimated via an indirect calibration procedure using  experimental data. There are many examples of such cardiac models, at a variety of different scales, discussed in the papers of this special issue.

Mathematical modelling and computational simulation 
has been remarkably successful at providing  insights into cardiac physiological mechanisms at cellular, tissue and whole organ scales \cite{Noble1962, Noble1992, Gray1995, Bass2009, Fink2011,  Sager2014}. 
In the majority of these quantitative efforts, models are derived based on simplified representations of complex biophysical systems and use {\emph{in vitro}} and {\emph{in vivo}} experimental data for calibration and validation purposes. 
Quantitative cardiac models have been a crucial tool for basic research for decades, and more recently have begun to transition into safety-critical clinical and pharmaceutical development applications \cite{Relan2011, Sermesant2012, Mirams2012, Nied2018, Li2018}. 
The use of cardiac mathematical models in such applications will require high levels of credibility in the predictive model outputs, as well as an accurate  quantification of the uncertainty in these predictions.

Parameters in cardiac models are often uncertain, mainly due to measurement uncertainty and/or natural physiological variability \cite{Mirams2016}. 
Thus, uncertainty quantification (UQ) methods are required to study uncertainty propagation in these models and help to establish confidence in model predictions. 
Parametric UQ is the process of determining the uncertainty in model inputs or parameters, and then estimating the resultant uncertainty in model outputs, thus testing the robustness of model predictions given our uncertainty in their inputs, and has been applied to a variety of cardiac models \cite{Nied2011, Krish2014, Pras2013, Pras2014, Pras2018, Pras2019}.

Another major source of uncertainty in modelling is uncertainty in the model structure, i.e., the form of the governing equations. 
There is always a difference between the imperfect model used to approximate reality, and reality itself; this difference is termed model discrepancy.
Assessment of the robustness of model predictions given our uncertainty in the model structure, and methods to characterise model discrepancy, has received relatively little attention in this field (and mathematical/systems biology more generally).
We have found only two published explicit treatments of discrepancy in cardiac electrophysiology models, in papers by Plumlee \emph{et al.} \cite{Plumlee2016,Plumlee2017}. 
In these studies, the assumption that ion channel rate equations follow an explicit form (such as that given, as we will see later, by Eq.~(\ref{eq:rate_equation})) was relaxed, and rates were allowed to be Gaussian processes (GPs) in voltage. 
A two-dimensional GP (in time and voltage) was then also added to the current prediction to represent discrepancy in current for a single step to any fixed voltage. 

\subsection{Notation and terminology}
Before discussing model discrepancy in detail, we introduce some notation and terminology.
As the concepts introduced here are intended to be understood not just by a cardiac modelling audience, we provide a non-exhaustive list of terminology we have encountered in different fields to describe useful concepts relating to calibration and model discrepancy (and mathematical/computational modelling in general) in Table~\ref{tab:terminology}.

\begin{table}[tbh]
        	\caption{Terminology used in different fields to refer to inverse problem concepts.}
	\label{tab:terminology}
	\centering
	\resizebox{\textwidth}{!}{	\begin{tabular}{lll}
		\toprule
		\textbf{CONCEPT} & \textbf{TERMINOLOGIES} & \\
		\midrule
		\multirow{5}{*}{Fitting parameters in a given model to data}
		& Calibration 
		& Inverse problem \\
		& Parameter inference
		& Parameter identification \\
		& Parameter estimation
		& Parameter tuning \\
		& Parameter fitting& Parameter optimisation \\
		& Model matching/fitting
		&  \\
		\midrule
				\multirow{2}{*}{\shortstack[l]{Do data from given experiment provide sufficient \\
                                       information to identify the model parameters?}}
		& Parameter identifiability
		& Practical identifiability \\
		& Structural identifiability
		& Well-posedness \\
		\midrule
		\multirow{1}{*}{Altering experiments to improve parameter identifiability}
		& Experimental design
		& Protocol design \\
		\midrule
		\multirow{2}{*}{Choosing model equations}
		& Model selection
		& Model choice \\
		& System identification
		& \\		\midrule
		\multirow{4}{*}{The difference between model and reality}
		& Model discrepancy
		& Model uncertainty \\
		& Model misspecification
		& Model mismatch \\
		& Model inadequacy
		& Model form error \\
		& Structural error
		& Model structure error\\
		\midrule
		\multirow{2}{*}{The observable measurements (data)}
		& Observables
		& Observable outputs \\
		& Quantities of Interest (QoIs) &\\
		\midrule
		\multirow{3}{*}{A  simplified version of the simulator/model}
		& Surrogate model
		& Metamodel \\
		& Proxy
		& Emulator \\ 		& Look-up table &\\
		\midrule
		\multirow{2}{*}{Checking the performance of the fitted model}
		& Validation
				& Certification \\
		& Qualification
		& Performance estimation \\
		\bottomrule
	\end{tabular}}
\end{table}

Here, we delve into some of those concepts in more detail.
Suppose a physiological system is modelled as $y = f(\boldsymbol{\theta}, u)$, where $f$ represents all governing equations used to model the system (also referred to as model form or model structure), $\boldsymbol{\theta}$ is a vector of parameters characterising the system, and $u$ are known externally applied conditions or control variables applied in the particular experimental procedure.
In a cardiac modelling context, these might represent a stimulus protocol, a drug concentration, or the applied voltage protocol in a simulated voltage clamp experiment. 
In general, $\boldsymbol{\theta} = \{\boldsymbol{\theta}_D, \boldsymbol{\theta}_C\}$, where values of $\boldsymbol{\theta}_D$ are directly measured, and where values of $\boldsymbol{\theta}_C$ are determined by calibration using the model $f$.
Here, for simplicity of exposition, we assume $\boldsymbol{\theta}_D$ is fixed (and known) and $\boldsymbol{\theta} = \boldsymbol{\theta}_C$.

We can distinguish between external conditions used for calibration, validation, and prediction (that is, the application of the model, or context of use (CoU)), $u_C$, $u_V$, $\uCoU$, say.
Suppose we have experimental data $Y_C$ for calibration and $Y_V$ for validation.
A typical workflow, without UQ, is:
\begin{itemize}
    \item \textbf{Calibration}:
        estimate $\thetaest = \textrm{argmin}_{\boldsymbol{\theta} \in \boldsymbol{\Theta}} d_C \left( f(\boldsymbol{\theta}, u_C), Y_C \right)$, using some calibration distance function $d_C(\cdot,\cdot)$ (e.g.,\ a vector norm: $d_C(\mathbf{x}, \mathbf{y}) = \|\mathbf{x} - \mathbf{y}\|$), and some subset of parameter space $\boldsymbol{\Theta}$;
                    \item \textbf{Validation}:
        compare $y_V = f(\boldsymbol{\thetaest}, u_V)$ against $Y_V$, either qualitatively or using a suitable validation distance $d_V(f(\boldsymbol{\thetaest}, u_V), Y_V)$;    \item \textbf{Context of use}:
        compute $\YCoU = f(\boldsymbol{\thetaest},\uCoU)$, or some quantity derived from this, to learn about the system or to make a model-based decision.
\end{itemize}
The calibration stage has many different names, see Table~\ref{tab:terminology}.

In practice, there a number of reasons why we cannot infer parameter values with certainty. The most commonly considered situation is when the link between the data and the model output is stochastic, e.g., because of measurement error on $Y_C$ or because of model discrepancy.
Computing the uncertainty about $\boldsymbol{\theta}$ based on noisy data $Y_C$ is referred to as `inverse UQ', and requires a statistical model of the experimental data to be specified.
For example, when considering measurement error, a common choice is to assume independent identically distributed zero-mean Gaussian errors on all data points, in which case (neglecting model discrepancy; see later) our model for the data is
\begin{equation}
    Y_{C} = f(\boldsymbol{\theta}, u_C) + \boldsymbol{\epsilon}, \label{eq:iid-noise}
\end{equation}
with $\boldsymbol{\epsilon}=(\epsilon_1, \epsilon_2, \ldots)^\top$,
where ${\epsilon}_i \sim \mathcal{N}(0, \sigma^2)$.
There are many different approaches to solving inverse UQ problems (see, e.g., \cite{tarantola2005, Dashti2017}),  most of which are based on inferring  probability distributions to describe the relative likelihood that each  different  parameter set is consistent with the available data.
Though a number of different methods to solve inverse UQ problems have been applied in cardiac electrophysiology \cite{Mirams2016}, the most common is a Bayesian approach, which combines prior information about the parameters, $\pi(\boldsymbol{\theta})$, with the probability of observing the data given each parameter $\pi(Y_C \mid \boldsymbol{\theta})$ (referred to as the likelihood of $\theta$), to find a posterior distribution over the parameters:
\begin{equation}
    \pi(\boldsymbol{\theta} \mid Y_{C}) = \frac{\pi(Y_C \mid \boldsymbol{\theta})\pi(\boldsymbol{\theta})}{\pi(Y_C)}.\label{eq:Gaussian}
\end{equation}
For an introduction to Bayesian methods, see \cite{Gelman2013,lambert2018BayesBook}.
For the  i.i.d.\ Gaussian error model (Eq.~\ref{eq:iid-noise}), the likelihood is given by
\begin{equation}
    \pi(Y_C \mid \boldsymbol{\theta}) =(2\pi \sigma^2)^{-n/2}\exp\left(-\frac{||Y_C - f(\boldsymbol{\theta}, u_C)||_2^2}{2\sigma^2}\right),\label{eq:ll}
\end{equation}
where $||x||_2^2=\sum_i x_i^2$, and $n$ is the number of data points.

Another potential source of uncertainty about $\boldsymbol{\theta}$ can occur when the parameter varies across the (or a) population.
Estimating population variability in $\boldsymbol{\theta}$ requires multiple $Y_C$ recordings, $\{Y^{(1)}_C, Y^{(2)}_C, \ldots\}$. Multilevel or hierarchical models can then be used: we assume the parameters for population $i$ are drawn from some distribution $\boldsymbol{\theta}^{(i)} \sim \pi(\boldsymbol{\theta} \mid \psi)$, and infer the population parameters $\psi$, see \cite{lei_rapid_2019-1}.

Once uncertainty in $\boldsymbol{\theta}$ (given the data) has been determined, the impact of this uncertainty on validation simulations $Y_V$ or CoU simulations $\YCoU$ can be computed by propagating the uncertainty through the model $f$ in the validation/CoU simulations, e.g.,
$$
\pi(Y_P \mid Y_C) = \int \pi(Y_P \mid \boldsymbol{\theta})\pi(\boldsymbol{\theta} \mid Y_C) {\rm d} \boldsymbol{\theta}.
$$
This is referred to as `uncertainty propagation' or the `posterior predictive distribution'.
Uncertainty in the prediction of $Y_V$ helps provide a more informed comparison to the observed validation data  (especially if experimental error in $Y_V$ is also accounted for).
Uncertainty in $\YCoU$ enables a more informed model-based decision-making process.

\subsection{Model discrepancy}
UQ as outlined above does not account for the fact that the model is always an imperfect representation of reality, due to limited understanding of the true data-generating mechanism and perhaps also any premeditated abstraction of the system.
The model discrepancy is the difference between the model and the true data-generating mechanism, and its existence has  implications for model selection, calibration and validation, and CoU simulations.

For calibration, the existence of model discrepancy can change the meaning of the estimated parameters. If we fail to account for the model discrepancy in our inference, our parameter estimates, instead of being physically meaningful quantities, will have their meaning  intimately tied to the model  used to estimate them (we end up estimating  `pseudo-true' values; see Section \ref{sect:motivation}\ref{sect:statistics}). 
The estimated  parameter values depend on the chosen model form, and the uncertainty estimates obtained during inverse parameter UQ tell us nothing about where the true value is (only how confident we are about the pseudo-true values).
In other words, there is no guarantee the obtained $\boldsymbol{\theta}$ will match true physiological values of any parameters that have a clear physiological meaning. 

We can try to restore meaning to the estimated parameters by including a term to represent the model discrepancy in our models.
Validation, in particular, provides an opportunity for us to identify possible model discrepancy. 
In many cases, validation, rather than being considered as an activity for confirming a `model is correct', is better considered as a method for estimating the model discrepancy.
To maximise the likelihood that the validation can discern model discrepancy, the validation data should ideally be `far' from the calibration data, and as close to the CoU as possible.

\section{A motivating example of discrepancy}\label{sect:motivation}
To illustrate the concept of model discrepancy and some of its potential consequences, we have created a cardiac example inspired by previous work \cite{Brynjarsdottir2014}, using  mathematical models of the action potential of human 
ventricular cells. These models have a high level of electrophysiological detail,  including most of the major ionic currents as well as basic calcium dynamics, and have been used to study reentrant arrhythmias. 
We  assume that the Ten Tusscher~\textit{et al.} ventricular myocyte electrophysiology model \cite{ten2004model} (Model~T) represents the ground truth, and use this model to generate  data traces in three different situations: for {\it calibration} data we use the action potential under \SI{1}{\hertz} pacing; to generate {\it validation} data we use \SI{2}{\hertz} pacing; and for {\it context of use} (CoU) data we use \SI{1}{\hertz} pacing with the 75\% \ikr\ block ($g_{Kr}$  multiplied by a scaling factor of $0.25$).

\begin{figure}[tbh]
	\centering
	\includegraphics[width=\textwidth]{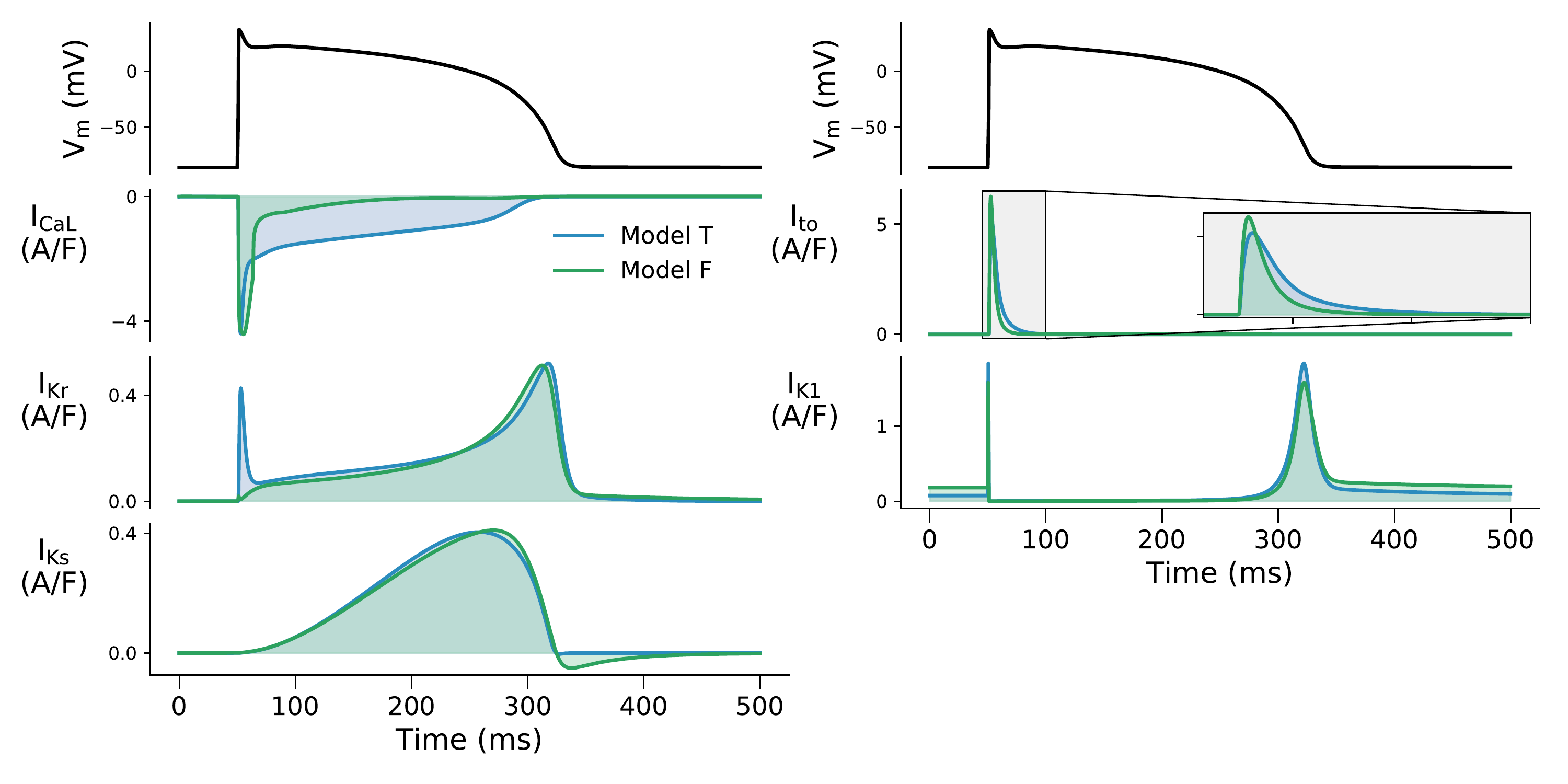}
	\caption{\label{fig:tutorial-action-potential-models}
		A comparison of the Ten Tusscher (Model~T \cite{ten2004model}, blue) and Fink (Model~F \cite{fink2008contributions}, green) kinetics.
		These currents are voltage clamp simulations under the same action potential clamp (shown in the top row panels).
				Only those currents with different kinetics are shown; the kinetics of $I_\text{Na}$, $I_\text{NaCa}$, and $I_\text{NaK}$ are identical in both models.
		Two of the gates in $I_\text{CaL}$ are identical in the two models, one gate has a different formulation, and Model~F has one extra gate compared to Model~T.
		The two models use different formulations for $I_\text{Kr}$ ($I_\text{Kr}$ activates during depolarization in Model~T but not Model~F),
		different parameterisations of the kinetics for $I_\text{Ks}$ and $I_\text{to}$,
		and different equations for $I_\text{K1}$ steady state.
		Currents are normalised in this plot by minimising the squared-difference between the two models' currents such that we emphasise the differences in kinetics rather than the conductances (which are rescaled during the calibration).
		Only $I_\text{CaL}$ shows what we would typically consider to be a large difference in repolarization kinetics, with the rest of the currents apparently being close matches between Model~T and Model~F.
	}
\end{figure}

To illustrate the problem of fitting a model under model discrepancy, 
we assume we do not know the ground truth model and
instead fit an alternative model, the Fink \textit{et al.} model \cite{fink2008contributions} (Model~F), to the synthetic data generated from Model~T.
Both models F and T were built for human ventricular cardiomyocytes, with  Model~F being a modification of Model~T that improves the descriptions of repolarising currents, especially of the hERG (or \ikr) channel (which is a major focus for Safety Pharmacology). A comparison of the differences in the current kinetics between the two models is shown in Figure~\ref{fig:tutorial-action-potential-models}, and the model equations are given in  Section~\ref{supp-sec:model-t-f-equations} of the supplementary material.
Only five currents have kinetics that vary between the two models, and importantly, no currents or compartments are missing (unlike when attempting to fit a model to  real data).

In this example the control variables are the stimulus current and \ikr\ block, the model outputs are the membrane voltage, and the parameters of interest are the maximum conductance/current density of the ionic currents.
We use Model~T to generate synthetic current clamp experiments by simulating the different protocols (control variables) then adding i.i.d.\ Gaussian noise $\sim \mathcal{N}(0,\sigma^2)$ to the resulting voltage traces (model outputs), with $\sigma$ chosen to be \SI{1}{\milli\volt}.
We use the calibration data (\SI{1}{\hertz} pacing) to estimate eight maximal conductance/current density parameters for $I_\text{Na}$, $I_\text{CaL}$, $I_\text{Kr}$, $I_\text{Ks}$, $I_\text{to}$, $I_\text{NaCa}$, $I_\text{K1}$, and $I_\text{NaK}$ using Model~F.
We will investigate whether the calibrated Model~F  makes accurate predictions in the validation and CoU situations (using the parameters estimate from the calibration data, as is commonly done in electrophysiology modelling \cite{Kaur2014,Groenendaal2015,Johnstone2016,lei_tailoring_2017,Pouranbarani2019}).
The code to reproduce all of the results in this paper are available at \url{https://github.com/CardiacModelling/fickleheart-method-tutorials}.

\subsection{Model calibration}
We calibrate the model using a train of five action potentials stimulated under a \SI{1}{\hertz} pacing protocol as the calibration data.
Before attempting to do this fitting exercise, the appropriately sceptical reader might ask whether we are attempting to do something sensible. 
Will we get back information on all the parameters we want, or will we just find one good fit to the data amongst many equally plausible ones, indicating non-identifiability of the parameters?

To address these questions, we first look at inferring the parameters of the original Model~T (as well as inferring the noise model parameter, $\sigma$).
We use Eq.~(\ref{eq:iid-noise}) with Gaussian noise giving the likelihood in Eq.~(\ref{eq:Gaussian}), together with a uniform prior distribution from $0.1\times$ to $10\times$ the original parameters of Model~T.
We take two different approaches to calibration. Firstly, 
we find a point estimate  using a global optimisation algorithm \cite{Hansen2006} to find the optimal model parameters (with no estimate of uncertainty). Secondly, we approximate the full posterior distribution using Markov chain Monte Carlo (MCMC).
All inference is done using an open source Python package, PINTS \cite{Clerx2019Pints}, and simulations are performed in Myokit \cite{clerx_myokit_2016}.

The results are shown in Supplementary Figure~\ref{supp-fig:tutorial-action-potential-no-discrepancy}.
This exercise results in a narrow plausible distribution of parameters very close to the ones that generated the data, and we conclude that  the model parameters are identifiable with the given data.
Additionally, Supplementary Figure~\ref{supp-fig:tutorial-action-potential-no-discrepancy} shows that when using samples of these distributions to make predictions, all of the forward simulations are very closely grouped around the synthetic data for the \ikr\ block CoU.

\begin{figure}[htb]
	\centering
	\begin{tikzpicture}
    \matrix (fig) [matrix of nodes]{
      \includegraphics[width=0.9\textwidth]{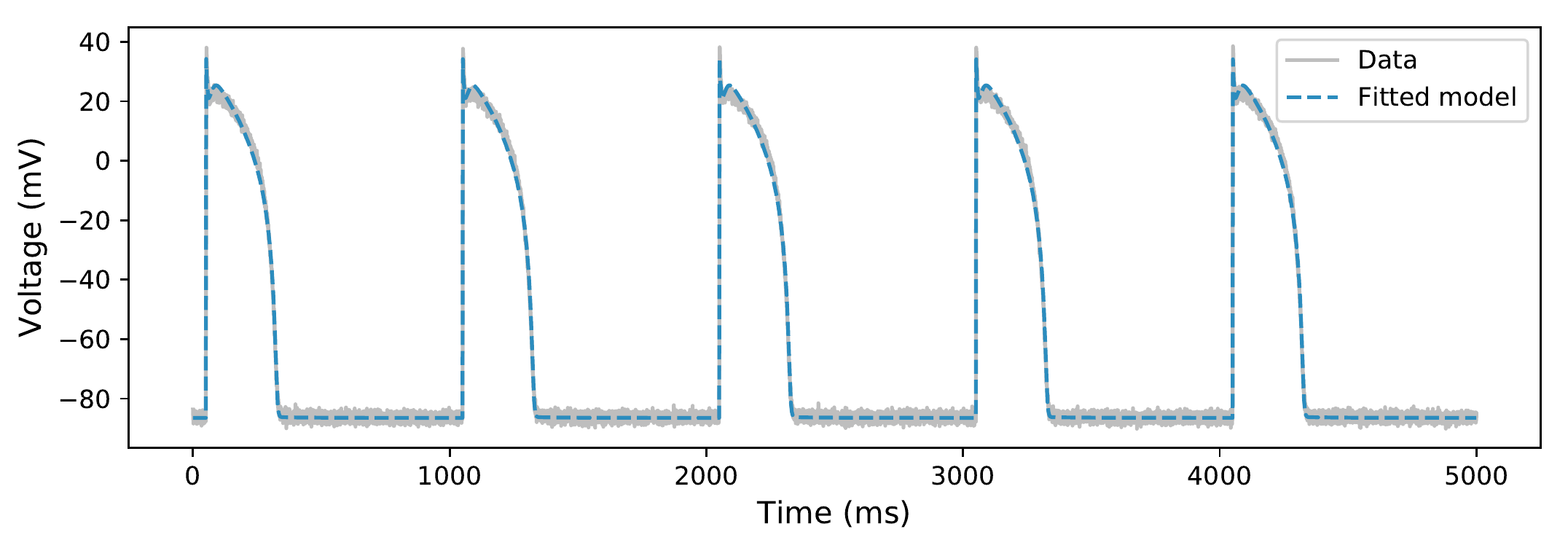} \\
  };
        \path (fig-1-1.north)  -- (fig-1-1.north) node[midway,above]{Calibration};
	\end{tikzpicture}
	\begin{tikzpicture}
    \matrix (fig) [matrix of nodes]{
    	\includegraphics[width=0.49\textwidth]{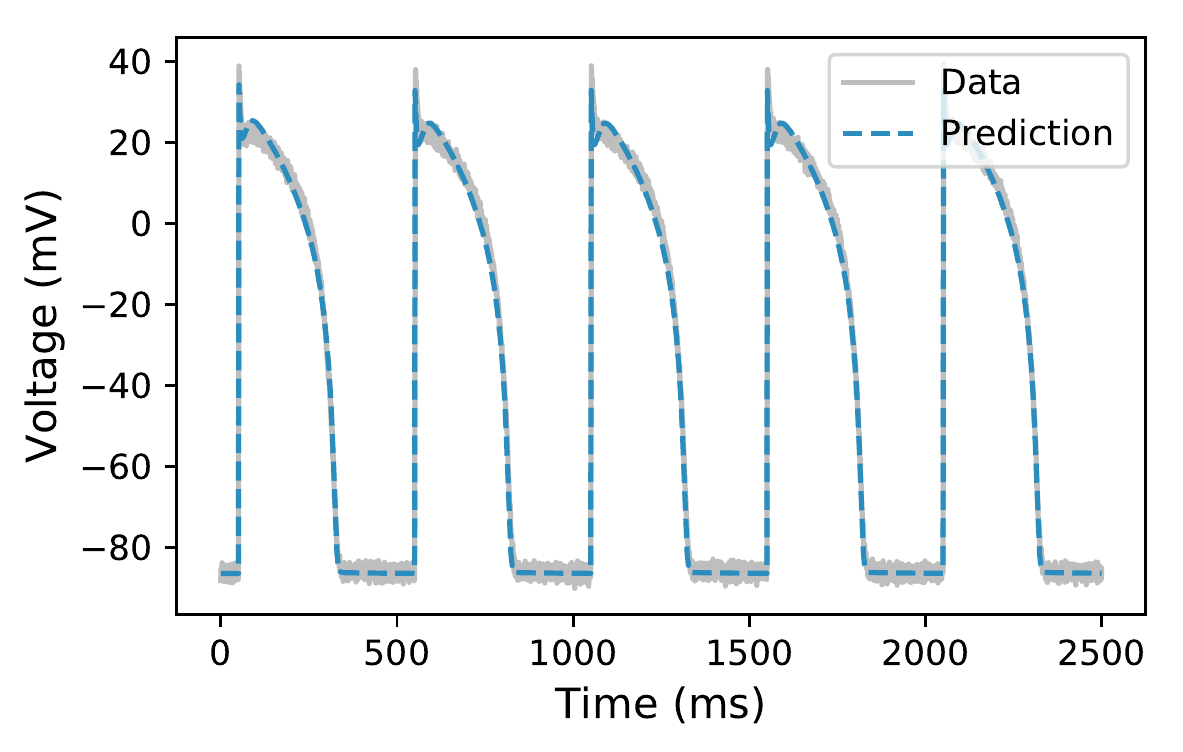} &
	    	    \includegraphics[width=0.49\textwidth]{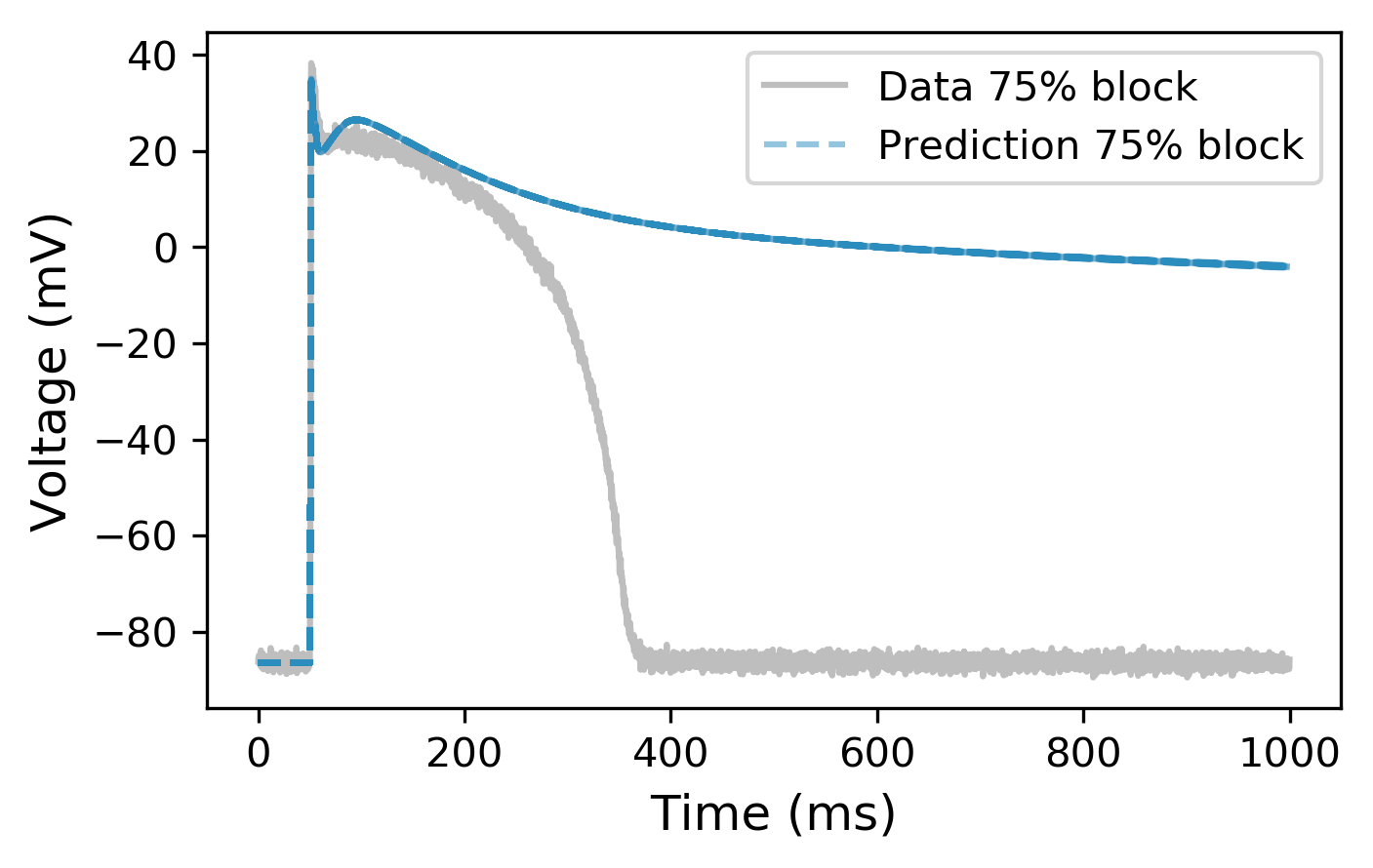} \\
	};
	    \path (fig-1-1.north)  -- (fig-1-1.north) node[midway,above]{Validation};
      \path (fig-1-2.north)  -- (fig-1-2.north) node[midway,above]{Context of Use};
	\end{tikzpicture}
	\caption{\label{fig:tutorial-action-potential-fitting-and-prediction}		
	\textbf{Model~F fitting and validation results}.
		(\textbf{Top}) Model-F is fitted to the synthetic data (generated from Model~T), using a five action potential recording under a \SI{1}{\hertz} pacing protocol.
		The calibrated Model~F (blue dashed line) shows an excellent fit to the calibration data (grey solid line).
				(\textbf{Bottom}) Model~F predctions for validation and context of use (CoU) data.
		(\emph{Left}) The calibrated Model~F predictions  closely matches the validation data (\SI{2}{\hertz} pacing), giving a (false) confidence in the model performance.
		(\emph{Right}) Notably, Model~F gives  catastrophic predictions for the $I_\text{Kr}$ block (CoU) experiments (suggesting  the validation data are not an appropriate test given the intended model use).
				The posterior predictions are model predictions made using parameter values  sampled from the posterior distribution (Figure~\ref{fig:tutorial-action-potential-posterior}); here, 200 samples/predictions are shown, but they overlay and are not distinguishable by eye.
	}
\end{figure}

We now  attempt the fitting exercise using Model~F (i.e., 
the misspecified model).
The fitted model prediction (using the maximum a-posteriori (MAP) parameter estimate), is shown in Figure~\ref{fig:tutorial-action-potential-fitting-and-prediction}\,(Top).
The agreement between the calibrated model output and the synthetic data would be considered excellent if these were real experimental data. 
Therefore, it is tempting to conclude that this calibrated model gives accurate predictions, and that the  model discrepancy is minor.
But can we trust the predictive power of the model in other scenarios based solely on the result we see in Figure~\ref{fig:tutorial-action-potential-fitting-and-prediction}\,(Top)?

\subsection{Discrepant model predictions}
Interestingly, the calibrated Model~F gives very accurate  predictions for the \SI{2}{\hertz} pacing validation protocol (data that are not used to estimate the parameters), as shown in Figure~\ref{fig:tutorial-action-potential-fitting-and-prediction}\,(bottom, left).
Such rate-adaptation predictions are used commonly as validation evidence for action potential models.
At this stage we may be increasingly tempted to conclude that we have a good model of this system's electrophysiology.

But if one now uses the model to predict the effect of drug-induced \ikr\ block, the catastrophic results are shown in the bottom right panel of Figure~\ref{fig:tutorial-action-potential-fitting-and-prediction}.
The calibrated Model~F fails to repolarise, completely missing the true $I_\text{Kr}$ block response of a modest APD prolongation.
This example highlights the need for thorough validation and the CoU-dependence of model validation, but also the difficulty in choosing appropriate validation experiments.

\begin{figure}[htb]
	\centering
	                		\includegraphics[width=\textwidth]{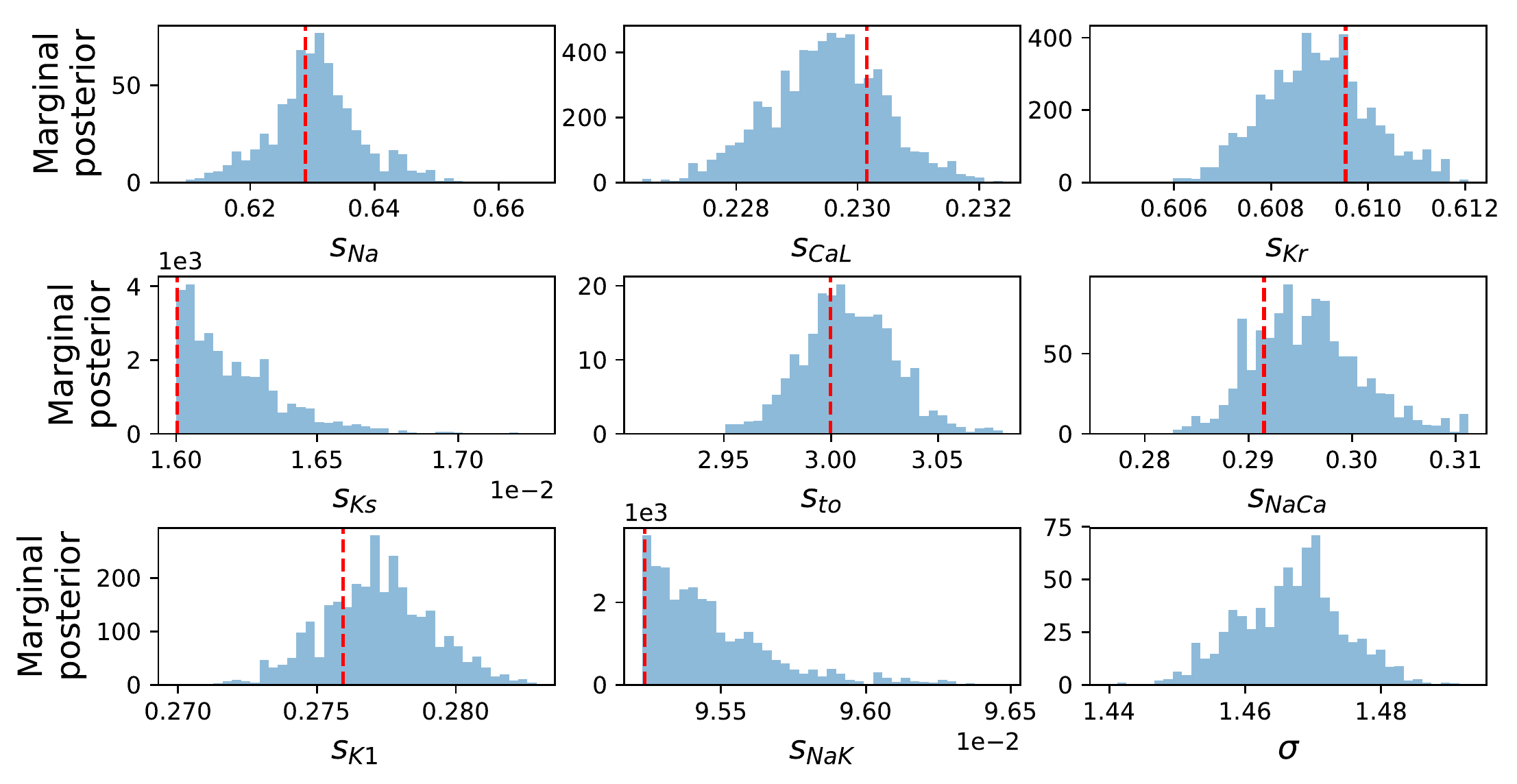}
	\caption{
												Marginals of the posterior distribution of the Model~F parameters, in terms of scaling factors for the conductances in Model~T ($s_i = g^\text{Model~F}_i/g^\text{Model~T}_i$).
		Values of \num{1} would represent the parameters of Model~T that generated the data; note that none of the inferred parameters for Model~F are close 1.
		The red dashed lines indicate the result of the global optimisation routine.
		Two of these parameters, $S_{Ks}$ and $S_{NaK}$, have distributions hitting the lower bound that was imposed by the prior, indicating that the calibration process is attempting to make them smaller than 10\% of the original Model~F parameter values.
	}
	\label{fig:tutorial-action-potential-posterior}
\end{figure}

We can also quantify the uncertainty in parameter estimates and predictions whilst continuing to ignore the discrepancy in Model~F's kinetics.
Again, we use Eq.~(\ref{eq:Gaussian}) together with a uniform prior to derive the posterior distribution of the parameters.
The marginals of the posterior distribution, estimated by MCMC, and the point estimates obtained by optimisation are shown in Figure~\ref{fig:tutorial-action-potential-posterior}.
The posterior distribution is very narrow (note the scale), which suggests that we can be confident about the parameter values.
The resulting posterior predictions, shown in Figure~\ref{fig:tutorial-action-potential-fitting-and-prediction}\,(bottom, right), give a very narrow bound.
By ignoring model discrepancy we have become highly (and wrongly) certain that the catastrophically bad predictions are correct.

It is worth noting that all of the issues above arise from the fact that the model discrepancy was ignored during calibration.
In the scenario of no model discrepancy, i.e.,\ when fitting Model~T to the data, none of  the issues above occurred, as shown in Supplementary Figure~\ref{supp-fig:tutorial-action-potential-no-discrepancy}.

To conclude our motivation of this paper, we can see that neglecting discrepancy in the model's equations is dangerous and can lead to false confidence in predictions for a new context of use. We discuss methods that have been suggested to remedy this in Section \ref{sect:tutorial}.

\subsection{A statistical explanation}\label{sect:statistics}

To understand what happens when we fit an incorrect model to data, let's first consider the well-specified situation where the data generating process (DGP) has probability density function (pdf) $g(y)$, and for which we have data $y_i \sim g(\cdot)$ for $i=1,\ldots, n$.
Then suppose we are considering the models $\mathcal{P}=\{p(y\mid {\boldsymbol{\theta}}): \boldsymbol{\theta}\in \boldsymbol{\Theta}\}$, i.e., a collection of pdfs parameterized by unknown parameter $\boldsymbol{\theta}$.
If the DGP $g$ is in $\mathcal{P}$, i.e., we have a well-specified model so that for some $\boldsymbol{\theta}_0\in \boldsymbol{\Theta}$, we have $g(\cdot) =p(\cdot \mid {\boldsymbol{\theta}_0})$, then asymptotically, as we collect more data (and under suitable conditions \cite{van2000}), the maximum likelihood estimator converges to the true value $\boldsymbol{\theta}_0$ almost surely:
$$
\hat{\boldsymbol{\theta}}_n =\argmax_{\boldsymbol{\theta}} \sum_{i=1}^n \log p(y_i\mid {\boldsymbol{\theta}}) \longrightarrow \boldsymbol{\theta}_0, \mbox{ almost surely as } n \longrightarrow \infty,
$$
or equivalently $p(\cdot \mid {\hat{\boldsymbol{\theta}}_n})$ converges to $g(\cdot)$. Similarly, for a Bayesian analysis (again under suitable conditions \cite{bernardo2009}), the posterior will converge to a Gaussian distribution centered around the true value $\boldsymbol{\theta}_0$, with variance that shrinks to zero at the asymptotically optimal rate (given by the Cram\'{e}r-Rao lower bound), i.e.,
$$\pi(\boldsymbol{\theta} \mid y_{1:n}) \approx  \mathcal{N}\left(\boldsymbol{\theta}_0, \frac{1}{n}\mathcal{I}(\boldsymbol{\theta}_0)^{-1}\right),$$
where $y_{1:n}= (y_1,\ldots,y_n)$, and $\mathcal{I}(\boldsymbol{\theta}_0)$ is the Fisher information matrix for the true parameter value $\boldsymbol{\theta}_0$.

\begin{figure}[htb]
	\centering
	\includegraphics[width=0.8\textwidth]{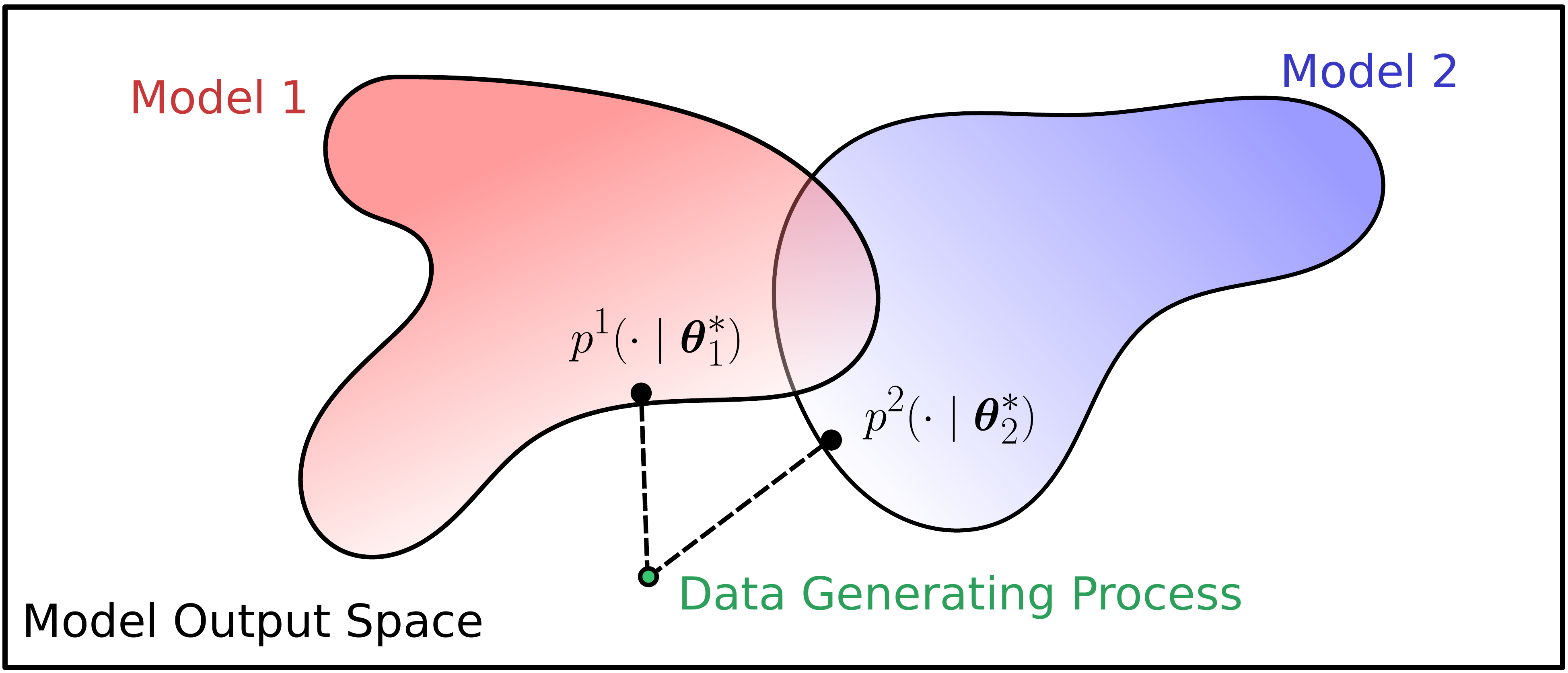}
	\caption{\label{fig:schematic-model-discrepancy}
		\textbf{
				A cartoon to illustrate the effect of model discrepancy on parameter fits in different models}.
		Each cloud represents a range of possible outputs from each model, which they can reach with different parameter values.
		The true data generating process (DGP) lies outside either of our imperfect model classes 1 and 2, and neither can fit the data perfectly due to model discrepancy.
		When we attempt to infer parameters, we will converge upon models  that generate outputs closest to the true DGP under the constraint of being in each model.
		Adding more data just increases the confidence in being constrained to model parameterisations on the boundary of the particular model, i.e., we become certain about $\boldsymbol{\theta}^*$, the  pseudo-true parameter value for each model.
		Note that different models will have different pseudo-true parameter values.
	}
\end{figure}

However, when our model is misspecified, i.e., $g \not\in \mathcal{P}$ (there is no $\boldsymbol{\theta} \in \boldsymbol{\Theta}$ for which $g(\cdot)=f(\cdot \mid {\boldsymbol{\theta}})$), if we do inference for $\boldsymbol{\theta}$ ignoring the discrepancy, then we usually still get asymptotic convergence of the maximum likelihood estimator and Bayesian posterior \cite{Kleijn2006, DeBlasi2013}. 
However, instead of converging to a true value (which does not exist), we converge to the {\it pseudo-true} value 
$$
\boldsymbol{\theta}^{*} = \argmin_{\boldsymbol{\theta}\in\boldsymbol{\Theta}} \KL( g(\cdot) \;|| \;p(\cdot\mid {\boldsymbol{\theta}}))
$$
where $\KL(g||p) = \int g(x) \log \frac{g(x)}{p(x)} \mathrm{d} x$ is the Kullback-Leibler divergence from $p$ to $g$ (a measure of the difference between two distributions). 
In other words, we converge upon the model, $p(\cdot \mid {\boldsymbol{\theta}^*})$, which is closest to the DGP as measured by the Kullback-Leibler divergence (see Figure~\ref{fig:schematic-model-discrepancy}).

Perhaps more importantly from a UQ perspective, as well as getting a point estimate that converges to the wrong value, we still usually get asymptotic concentration at rate $1/n$, i.e., the posterior variance shrinks to zero. 
That is, we have found model parameters that are wrong, and yet we are certain about this wrong value.
The way to think about this is that the Bayesian approach is not quantifying our uncertainty about a meaningful physical parameter $\boldsymbol{\theta}_0$, but instead, it gives our uncertainty about the pseudo-true value  $\boldsymbol{\theta}^*$.
Consequently, we can not expect our calibrated predictions
\begin{equation*}
\pi(y' \mid y) = \int p(y' \mid \boldsymbol{\theta}) \pi(\boldsymbol{\theta} \mid y_{1:n}) {\rm d} \boldsymbol{\theta}
\end{equation*}
to perform well, as we saw in the action potential example above. 

This leaves us with two options.
We can either extend our model class $\mathcal{P}$ in the hope that we can find a class of models that incorporates the DGP (and which is still sufficiently simple that we can hope to learn the true model from the data), or we can change  our inferential approach.

\section{Accounting for model discrepancy}\label{sect:tutorial}

Once we have acknowledged that a model is misspecified, we are then faced with the challenge of how to handle the misspecification.
The approach taken should depend upon the aim of the analysis.
Using the model to predict independent events, for example, a current time-series for some experimental protocol, will require a different approach to if our aim is inference/calibration, i.e., if interest lies in the physical value of a particular parameter.
In the first case (prediction), it can often suffice to fit the model to the data ignoring discrepancy, and then to correct the predictions in some way\footnote{Note, however, that jointly fitting model and discrepancy can make the problem easier, for example, by making the discrepancy a better behaved function more amenable to being modelled.}, although this may not work well if the prediction involves extrapolating into a regime far away from the data.
The latter case (calibration) is more challenging, as we need to jointly fit the model and the discrepancy model, which can lead to problems of non-identifiability.

The most common approach for dealing with discrepancy is to try to correct the simulator by expanding the model class. 
The simplest approach is simply to add a flexible, non-parametric term to the simulator output, i.e., instead of assuming the data arose from Eq.~(\ref{eq:iid-noise}), to assume
\begin{equation}
    y = f(\boldsymbol{\theta}, u_C)+\delta(v_C)+\epsilon.\label{eqn:KOH}
    \end{equation}
Here, $\delta(v_C)$ is the model discrepancy term, and $\epsilon$ remains an unstructured white noise term.
Note that $v_C$ is used as the input to $\delta$ as it is not necessary to have the same input as the mechanistic model: $v_C$  could include some or all of $u_C$, but may also include information from internal model variables (see Section \ref{sect:tutorial}\ref{sect:3d}). 
To train this model, one option is to first estimate $\boldsymbol{\theta}$ assuming Eq.~(\ref{eq:iid-noise}), and then to train $\delta$ to mop up any remaining structure in the residual. 
However, a  better approach is to jointly estimate $\delta$ and $\theta$ in a Bayesian approach \cite{Kennedy2001}.
Unfortunately, as demonstrated below, this often fails as it creates a non-identifiability between $\boldsymbol{\theta}$ and $\delta$ when $\delta$ is sufficiently flexible: for any $\boldsymbol{\theta}$, there exists a functional form $\delta(\cdot)$ for which Eq.~(\ref{eqn:KOH}) accurately represents the data generating process.
Brynjarsd{\'{o}}ttir et al.~\cite{Brynjarsdottir2014} suggested that the solution  is to strongly constrain the functional form of $\delta(\cdot)$ using prior knowledge.
They present a toy situation in which $\delta(0)=0$ and $\delta(x) $ is monotone increasing, and show that once armed with this knowledge, the posterior $\pi(\boldsymbol{\theta} \mid y)$ more accurately represents our uncertainty about $\boldsymbol{\theta}$.
However, knowledge of this form is not available in many realistic problems.

\subsection{Ion channel model example}

We now illustrate the difficulty of accounting for model discrepancy in a tutorial example. We demonstrate that it can be hard to determine the appropriate information to include in  $\delta$, and that different functional forms for $\delta$ can lead to different parameter estimates.

We consider three structurally different models: Models~A, B, and C.
We take Model~C as the ground truth model in this particular example, and use it to perform synthetic voltage clamp experiments and generate synthetic data.
The goal is to use Models~A and B to explain the generated synthetic data, assuming we have no knowledge about the ground truth Model~C.
This tutorial aims to demonstrate the importance of considering model discrepancy, jointly with model selection, to represent given data with unknown true DGP. 

\begin{figure}[htb]
	\centering
	\includegraphics[width=\textwidth]{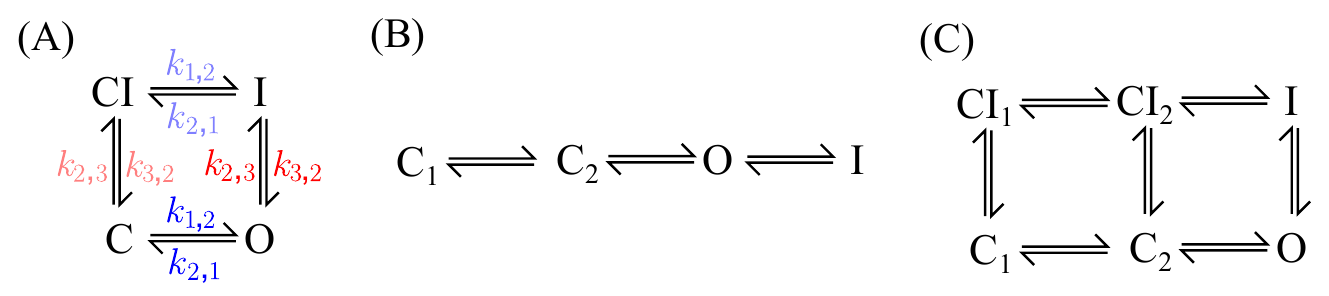}
	\caption{\label{fig:tutorial-ion-channel-models}
		Markov model representation of Models A, B, and C used in the ion channel model tutorial where Model~C is taken as ground truth and used to generate synthetic data, whilst Models~A and B are candidate models that we attempt to fit and use for predictions, demonstrating the challenge of both model discrepancy and model selection.
	}
\end{figure}

We use the hERG channel current as an example, 
and use three different model structures (shown in Figure~\ref{fig:tutorial-ion-channel-models}).
Model~A is a variant of the traditional Hodgkin-Huxley model, described in Beattie \textit{et al.}~\cite{beattie2018sinusoidal};
Model~B is used in Oehmen \textit{et al.}~\cite{oehmen2002mathematical};
and Model~C is adapted from Di Veroli \textit{et al.}~\cite{di2012high}.

All three ion channel models can be expressed using a Markov model representation.
For a model with a state vector, $\boldsymbol{x} = (x_1, x_2, \cdots)^T$, then in each case $\boldsymbol{x}$ evolves according to
\begin{equation}
    \frac{\text{d}\boldsymbol{x}}{\text{d}t} = \mathbf{M} \boldsymbol{x},
\end{equation}
where $\mathbf{M}$ is the Markov matrix describing the transition rates between states. Markov models are linear coupled ordinary differential equations (ODEs) with respect to time, $t$, and states, $\boldsymbol{x}$.
Typically the components in the Markov matrix, $\mathbf{M}$, are nonlinear functions of voltage, $V(t)$, which in these voltage-clamp experiments is an externally prescribed function of time known as the `voltage clamp protocol' (i.e.,\ $u_C$ in Eq.~\ref{eq:iid-noise}).
The observable, the macroscopic ionic current, $I$, measured under $V(t)$,  is
\begin{equation}\label{eq:ion-current}
    I(t, V) = g \cdot \mathcal{O} \cdot (V - E),
\end{equation}
where $g$ is the maximum conductance, $E$ is the reversal potential, and $\mathcal{O}$ is the sum of all `open states' in the model.

Take Model~B as an example.
Its state vector, $\boldsymbol{x}$, and Markov matrix, $\mathbf{M}$, can be written as
\begin{align}
    \boldsymbol{x} &=
        \begin{pmatrix}
            x_1 \\
            x_2 \\
            x_3 \\
            x_4
        \end{pmatrix} =
        \begin{pmatrix}
            \text{C}_2 \\
            \text{C}_1 \\
            \text{O} \\
            \text{I}
        \end{pmatrix};
    & \mathbf{M} &=
        \begin{pmatrix}
            -k_{1,2} &  k_{2,1}         &  0               &  0       \\
             k_{1,2} & -k_{2,1}-k_{2,3} &  k_{3,2}         &  0       \\
             0       &  k_{2,3}         & -k_{3,2}-k_{3,4} &  k_{4,3} \\
             0       &  0               &  k_{3,4}         & -k_{4,3} \\
        \end{pmatrix},
\end{align}
where $x_i$ is the probability a gate is in state $i$ (or equivalently, the proportion of gates  which are in  state $i$), with $\sum x_i =1$. The parameters $k_{i,j}$ represent the transition rates from state $x_i$ to state $x_j$. Note that for all our models, there is just one open state so that $\mathcal{O}=\text{O}$.
For all three models, each transition rate, $k_{i,j}$, is voltage dependent and  takes the form
\begin{equation}
    k_{i,j}(V) = A_{i,j} \exp(B_{i,j} V), \label{eq:rate_equation}
\end{equation}
with two parameters ($A_{i,j}, B_{i,j}$) to be inferred.
This yields a total of \num{12} parameters for Model~B which we denote as $\{p_1, \dots, p_{12}\}$, together with the maximum conductance, $g$, to be found.
Similarly for Model~A, it has \num{8} parameters $\{p_1, \dots, p_{8}\}$ together with $g$, to be inferred.

\subsection{Synthetic experiments}
We let  Model~C be the ground truth DGP and simulate data from it (using parameter values estimated from real room temperature data by Beattie \textit{et al.}~\cite{beattie2018sinusoidal}, where $g=204$\,\si{\nano\siemens}).
We add i.i.d.\ Gaussian noise with zero mean  and standard deviation $\sigma = 25$\,\si{\pico\ampere} to the simulated data.
We generate data under three different voltage clamp protocols, $V(t)$. 
These are a sinusoidal protocol (see top plot in Figure~\ref{fig:tutorial-ion-channel-prediction}) and an action potential (AP) series protocol from Beattie \emph{et al.}\ \cite{beattie2018sinusoidal} (see Figure~\ref{supp-fig:tutorial-ion-channel-prediction2a} in Supplementary Material), and the staircase protocol from Lei \emph{et al.}\ \cite{lei_rapid_2019-1, lei_rapid_2019-2} (see bottom plot in Figure~\ref{fig:tutorial-ion-channel-prediction}).

\subsection{Standard calibration ignoring model discrepancy}

To calibrate the model (without considering any model discrepancy), we assume a statistical model of the form of Eq.~(\ref{eq:iid-noise}), which has the same observation noise model as our synthetic data.
The likelihood of model parameter $\boldsymbol{\theta}$, having 
observed the data $\mathbf{y}=y_{1:n}$, is given by Eq.~(\ref{eq:ll}).

\begin{figure}[tbh]
	\centering
	\begin{tikzpicture}
    \matrix (fig) [matrix of nodes]{
     	\includegraphics[width=4.5in]{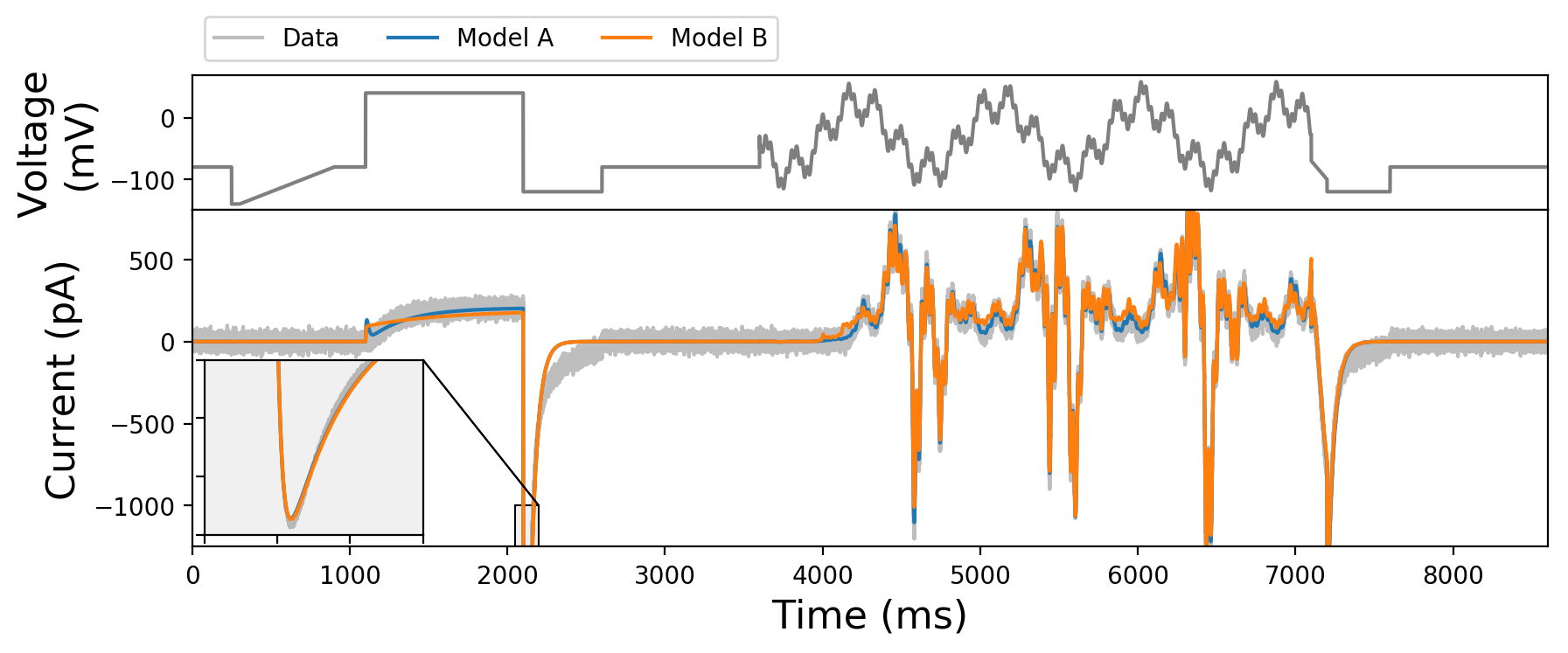} \\
    	    	\includegraphics[width=4.5in]{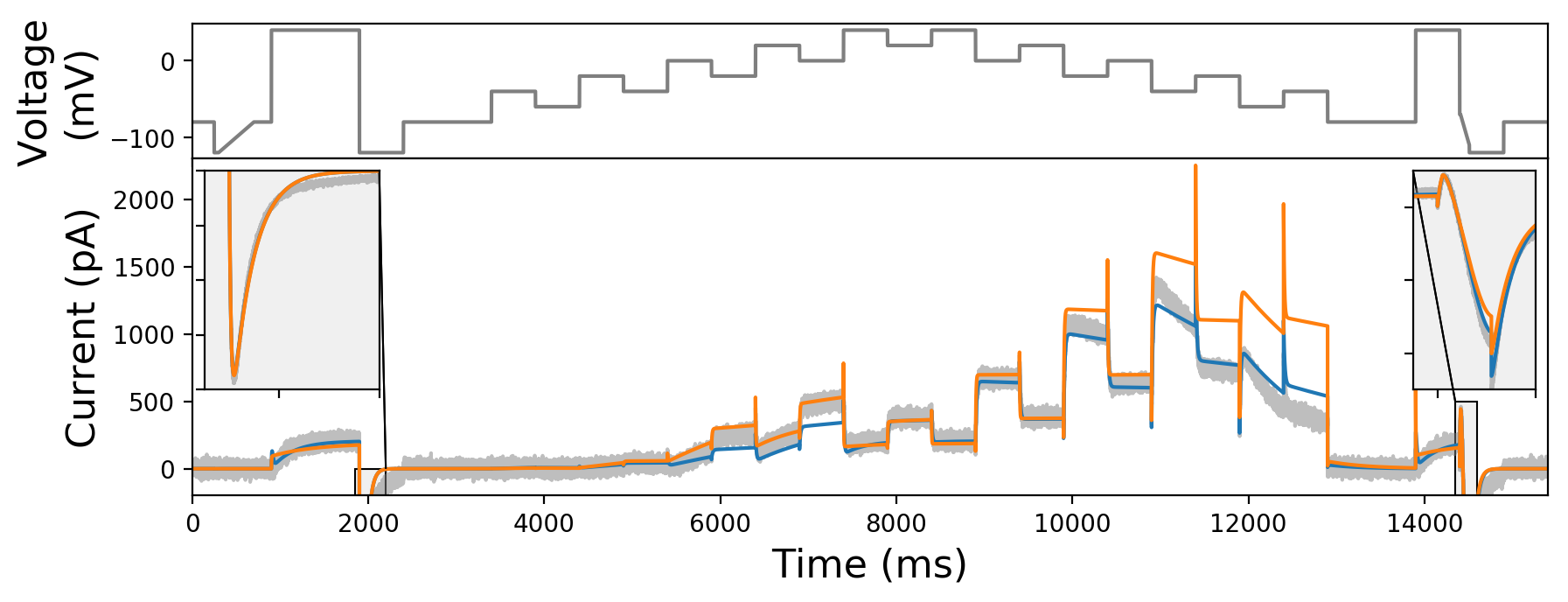} \\
	};
        \path (fig-1-1.south west)  -- (fig-1-1.north west) node[midway,above,sloped]{Calibration};
                \path (fig-2-1.south west)  -- (fig-2-1.north west) node[midway,above,sloped]{Validation};
	\end{tikzpicture}
	\caption{\label{fig:tutorial-ion-channel-prediction}
		\textbf{(Top)} The fitted model predictions for Models~A (blue) and B (orange) for the ion channel example (i.e., model predictions for the data they were trained with). Both models have been fitted to synthetic calibration data (grey) generated using Model~C using the sinusoidal voltage clamp protocol \cite{beattie2018sinusoidal}. 
		\textbf{(Bottom)} Models~A (blue) and B (orange) predictions for the  validation data (grey) generated from Model~C under the staircase protocol \cite{lei_rapid_2019-1} (not used in training).
		Note that there are significant discrepancies around \SI{12000}{\milli\second}.
	}
\end{figure}

We use the sinusoidal protocol (top, Fig.~\ref{fig:tutorial-ion-channel-prediction}) as the calibration protocol;
the action potential series protocol (top, Fig.~\ref{supp-fig:tutorial-ion-channel-prediction2a}) and the staircase protocol (bottom, Fig.~~\ref{fig:tutorial-ion-channel-prediction}) are used as validation data.
We use a global optimisation algorithm \cite{Hansen2006} to fit the model parameters using their maximum likelihood estimates. All inference is done using PINTS \cite{Clerx2019Pints}.

The fitting results of Models~A and B are shown in Figure~\ref{fig:tutorial-ion-channel-prediction}.
Using different starting points in the optimization gives almost exactly the same parameter sets each time.
Although both models fit the calibration data reasonably well, neither match perfectly, due to model discrepancy.
While the exact forms of the model discrepancy differs between the two models, both models notably fail to reproduce the correct form of the current decay following the step to \SI{-120}{\milli\volt} shortly after \SI{2000}{\milli\second}.

The validation predictions for the staircase protocol are also shown in Figure~\ref{fig:tutorial-ion-channel-prediction}.
Unlike in the sinusoidal protocol, where Model~A generally gives a better prediction than Model~B, in the staircase protocol different it is more evident that the model discrepancy traits are different for each model.
For example, Model~B appears to give slightly better predictions of the current during the first \SI{10000}{\milli\second}, whereas after this point Model~A begins to give better predictions.

\subsection{Calibration with model discrepancy}\label{sect:3d}

We now consider an approach that allows us to incorporate model discrepancy when doing parameter inference and making predictions.
We adapt the method proposed in \cite{Kennedy2001} and instead of assuming independent errors in Eq.~(\ref{eq:iid-noise}), which corresponds to assuming a diagonal covariance matrix for the vector of errors $\boldsymbol{\epsilon}$, we consider an additive discrepancy model of the form given by Eq.~(\ref{eqn:KOH}), giving a correlated (non-diagonal) error structure. 
We consider three different choices for the discrepancy $\delta(v_C)$, and jointly infer $\theta$ and $\delta$.
Note that we allow for a different choice of input $v_C$, compared to the input of model $f$, $u_C$.

First, we model $\delta$ as a sparse-Gaussian process (GP) \cite{Rasmussen2006, quinonero2005unifying}, for which we adapted the implementation in PyMC3 \cite{salvatier2016probabilistic} using Theano \cite{2016arXiv160502688short}.
The radial basis function (RBF) were used for the results presented here;
we also tried two other GP covariance functions (the exponential covariance function and the Mat\'{e}rn 3/2 covariance function) in Supplementary Section~\ref{supp-sec:ion-channel-supp}\ref{supp-sub:GPCovFuncs}, where we found the impact of the choice of covariance functions in this problem is not as sensitive as the formulation of the discrepancy models.
We explore two possibilities: choosing $v_C$ to be either (i) $t$ (time); or (ii) $O$,$V$ (the open probability, $\mathcal{O}$ in Eq.~(\ref{eq:ion-current}), and the voltage, $V$). 
In fitting the model, we estimate hyperparameters associated with the GP covariance function, and condition the model on the observed discrepancies. For the $\GP(O,V)$ model, this means that we assume that the discrepancy is a function $O$ and $V$ so that we use the observed combinations of $(O,V, \delta)$ to predict future discrepancies; in the $\GP(t)$ model, it means that we assume the discrepancy process is always similarly distributed in time (which will not be a sensible assumption in many situations). 
Full details are provided in the Supplementary Material, Section~\ref{supp-sec:gp-detail}.

As a third approach, we model discrepancy $\delta$ and the white noise error $\epsilon$, as an autoregressive-moving-average (ARMA) model of order $p, q$ \cite{durbin2012time}. If $e_t = \delta_t(v_c) + \epsilon_t$ is the residual at time $t$, then an $\ARMA(p, q)$ model for $e_t$ is
\begin{equation}\label{eq:arma-residual}
   e_t= \nu_t + \sum_{t'=1}^p \varphi_t e_{t - t'} + \sum_{t'=1}^q\zeta_{t'} \nu_{t - t'}
\end{equation}
where $\nu_t \sim \mathcal{N}(0, \tau^2)$, and $\varphi_1,\ldots, \varphi_p$ and $\zeta_1, \ldots, \zeta_q$ are, respectively, the coefficients of the autoregressive and moving-average part of the model.
We used the StatsModels \cite{seabold2010statsmodels} implementation, and assumed $p=q=2$ throughout. Note that when using the ARMA model, we do not condition on the observed discrepancy sequence (so the mean of the ARMA process remains zero, unlike in the GP approaches), but only use it to correlate the discrepancy structure in time. In general, there is an interesting connection  between GPs discretely sampled regularly in time, and autoregressive processes \cite{Rasmussen2006}, but here we treat the ARMA process differently to how we use GP discrepancies, and use the data only to estimate the ARMA parameters, not to condition the process upon the observed temporal structure, i.e., 
we use the ARMA process as a simple approach for introducing correlation into the residuals to better account for the discrepancy, not to correct the discrepancy (as is done with the GP).
The motivation is that if the mechanistic model is correct, the residuals should be uncorrelated, but for misspecified models, they will typically be correlated. For further details, please refer to the supplementary information, Section~\ref{supp-sec:arma-detail}.

\begin{figure}[htb]
	\centering
	\includegraphics[width=\textwidth]{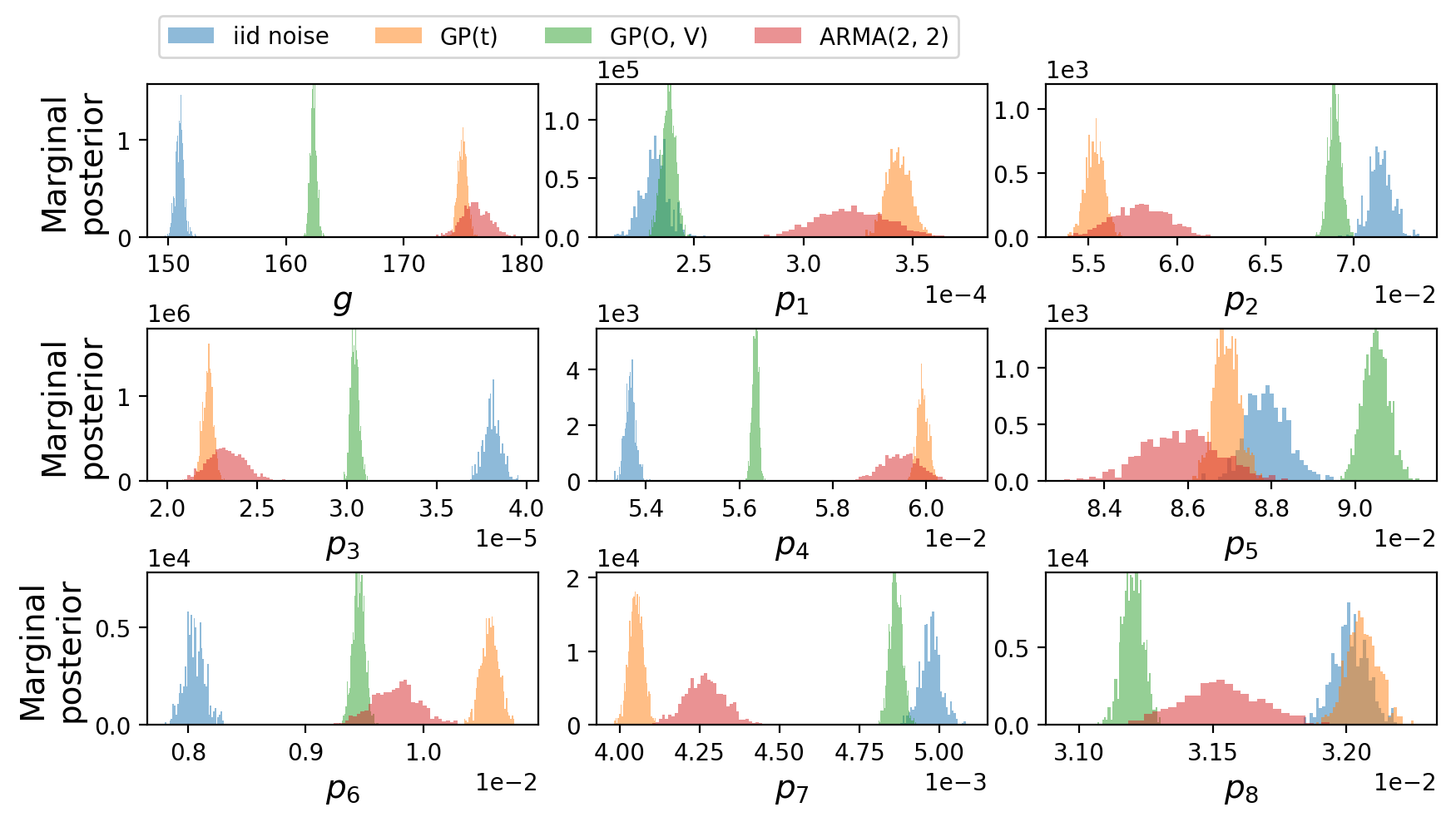}
	\caption{\label{fig:tutorial-ion-channel-posteriors}
		Model~A inferred marginal posterior distributions for the conductance, $g$ in Eq.~(\ref{eq:ion-current}), and kinetic parameters $p_1,\dots,p_8$ (a list of parameters referring to $A_{i,j}$ and $B_{i,j}$ in Eq.~\ref{eq:rate_equation}) with different discrepancy models: i.i.d.\ noise (blue), $\GP(t)$ (orange), $\GP(O, V)$ (green), and $\ARMA(2, 2)$ (red).
					}
\end{figure}

For all methods, i.i.d.\ noise, $\GP(t)$, $\GP(O, V)$, and $\ARMA(2, 2)$, we infer the posterior distribution of the parameters (Eq.~\ref{eq:Gaussian}), where the priors are specified in Supplementary Material,  Section~\ref{supp-sec:priors}.
We use an adaptive covariance MCMC method in PINTS \cite{Clerx2019Pints, Johnstone2016} to sample from the posterior distributions.
The trace plots of the samples are shown in Supplementary Section~\ref{supp-sec:ion-channel-supp}.
The inferred (marginal) posterior distributions for Model~A are shown in Figure~\ref{fig:tutorial-ion-channel-posteriors}, and they are used to generate the posterior predictive distributions shown in Figure~\ref{fig:tutorial-ion-channel-fitting2}.
Supplementary Figure~\ref{supp-fig:tutorial-ion-channel-posteriors-b} shows the same plots for Model~B.
Note that  the choice of the discrepancy model can shift the posterior distribution significantly, both in terms of its location and spread.
In particular, the $\ARMA(2, 2)$ model gives a much wider posterior than the other discrepancy models.

Figure~\ref{fig:tutorial-ion-channel-fitting2} shows the posterior predictive distributions of  Model~A with the calibration protocol using the four discrepancy models (Supplementary Figure~\ref{supp-fig:tutorial-ion-channel-fitting2-b} for Model~B), i.e., predicting the data used in training.
The top panel shows the sinusoidal voltage protocol, and the panels underneath are calibrated model predictions with i.i.d.\ noise (blue), $\GP(t)$ (orange), $\GP(O, V)$ (green), and $\ARMA(2, 2)$ (red).
The calibration data are shown in grey.
Visually, we can see that the two GP models, $\GP(t)$ (orange) and $\GP(O, V)$ (green), fit  the data with  high accuracy; later we will see one of them is overfitting, while the other is not.
The $\ARMA(2, 2)$ model (red)  increases the width of the posterior (compared to i.i.d.\ noise), but its posterior mean prediction does not follow the data as closely as the two GP models.

\begin{figure}[htb]
	\centering
	\includegraphics[width=\textwidth]{compare-model_A-sinewave-sinewave-pp.png}
	\caption{\label{fig:tutorial-ion-channel-fitting2}
		Model~A fitted to the sinusoidal calibration protocol using the different discrepancy models: i.i.d.\ noise, $\GP(t)$, $\GP(O, V)$, and $\ARMA(2, 2)$.
		The plots show the mean (solid lines) and 95\% credible intervals (shaded) of the posterior prediction for each model.
	}
\end{figure}

Table~\ref{fig:tutorial-ion-channel-rmse-table} shows the root mean square errors (RMSE) of the posterior mean predictions for all of the models, 
and is coloured so that yellow highlights the best performing model and red the worst.
The first row of the table shows the results for the calibration (sine wave ) protocol, and it is clear that the $\GP(t)$ and $\GP(O, V)$ models give the best RMSE values for the calibration data.
Note that the RMSE only assesses the accuracy of the point estimate (given by the posterior mean). Table~\ref{supp-fig:tutorial-ion-channel-likelihood-table} in the supplementary material gives the 
posterior predictive log-likelihoods; the log-likelihood is a proper scoring rule \cite{Gneiting2007} which assesses the entire predictive distribution, not just the mean prediction. The $\ARMA(2, 2)$ and $\GP(O, V)$ models achieve the highest log-likelihood scores on the calibration data (best all round predictions when accounting for uncertainty).

\begin{table}[htb]
	\centering
	\includegraphics[width=\textwidth]{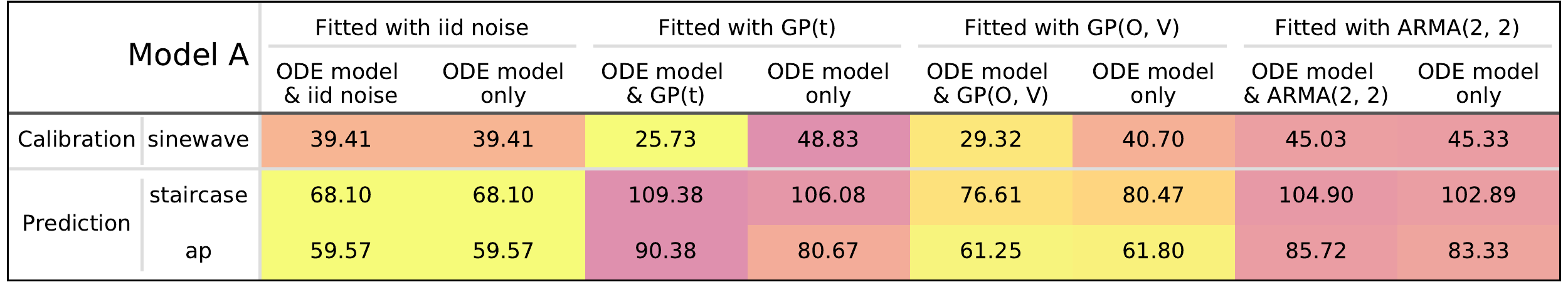}\\
	\includegraphics[width=\textwidth]{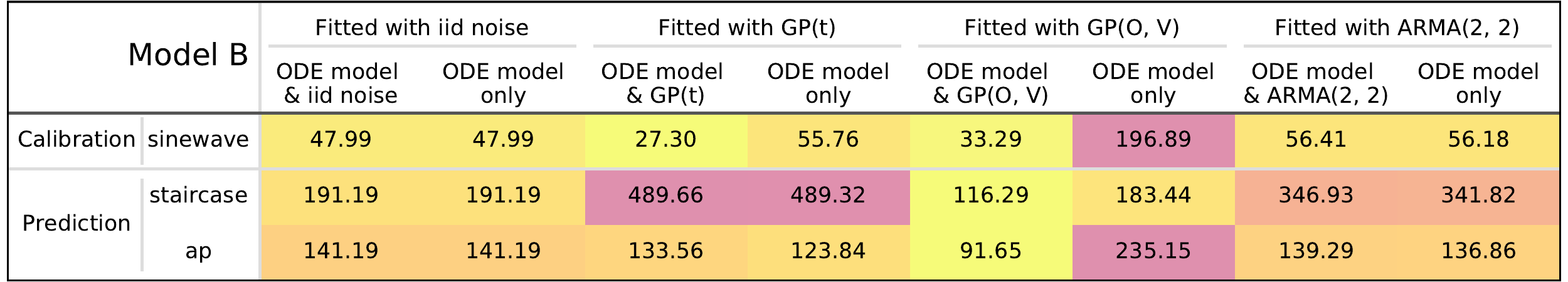}\\
	\caption{\label{fig:tutorial-ion-channel-rmse-table}
		Models~A (top) and B (bottom) RMSEs with different discrepancy models: i.i.d.\ noise, $\GP(t)$, $\GP(O, V)$, and $\ARMA(2, 2)$ for each of the three voltage protocols.
    Here `ODE model-only' refers to the predictions using only the calibrated ODE model under different discrepancy models (i.e., the model is calibrated assuming Eq.~\ref{eqn:KOH}, but prediction is done using only $f(\hat{\theta}, u_C)$). See also Supplementary Figures~\ref{supp-fig:tutorial-ion-channel-fitting2-model}--\ref{supp-fig:tutorial-ion-channel-prediction2b-model} for Model~A and Supplementary Figures~\ref{supp-fig:tutorial-ion-channel-fitting2-model-b}--\ref{supp-fig:tutorial-ion-channel-prediction2b-model-b} for Model~B.
	}
\end{table}

\begin{figure}[tbh]
  \centering
  \includegraphics[width=\textwidth]{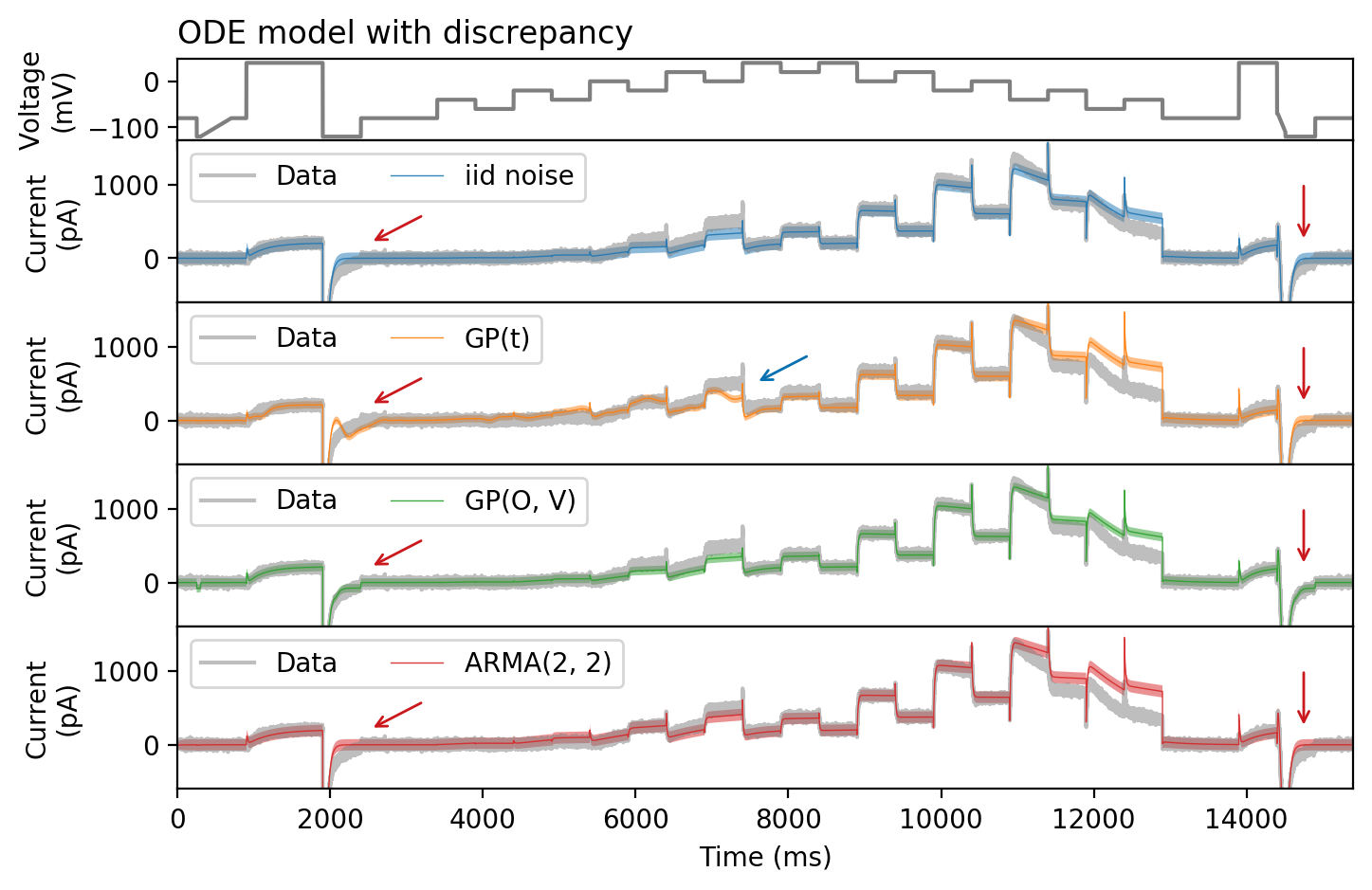}
  \caption{\label{fig:tutorial-ion-channel-prediction2}
    Model~A's prediction using the discrepancy models (i.i.d.\ noise, $\GP(t)$, $\GP(O, V)$, and $\ARMA(2, 2)$), trained using the staircase voltage clamp protocol \cite{lei_rapid_2019-1}.
    We plot the posterior predictive mean (solid lines) with 95\% credible intervals (shaded).
    The red arrows point to the tail current after the two activation steps, and mark an area of visible model mismatch: note the different performance of the four discrepancy models in this region.
    The blue arrow points to an obvious artefact at $\sim7000$\,\si{\milli\second} induced by the $\GP(t)$ prediction which was trained on the sinusoidal protocol, and which doesn't take into account that we are now predicting for the  staircase protocol.
  }
\end{figure}

Figure~\ref{fig:tutorial-ion-channel-prediction2} shows the prediction results for the staircase validation protocol for Model~A (Supplementary Figure~\ref{supp-fig:tutorial-ion-channel-prediction2-b} for Model~B) using different discrepancy models, with the same layout as Figure~\ref{fig:tutorial-ion-channel-fitting2}.
Similar figures for the action potential (AP) protocol predictions are shown in Supplementary Figures~\ref{supp-fig:tutorial-ion-channel-prediction2a} (Model~A) and~\ref{supp-fig:tutorial-ion-channel-prediction2a-b} (Model~B). The $\GP(t)$ discrepancy model is conditioned to give the same temporal discrepancy pattern as in the calibration protocal, and
is unable to change its predicted discrepancy in any way for  the validation protocol; i.e., the $GP(t)$ discrepancy predicts as if it were still under the sinusoidal protocol.
Thus, there is some residual from the calibration protocol shown in the $\GP(t)$ (orange) prediction for the staircase protocol, e.g.,\ see `wobbly' current at $\sim7000$\,\si{\milli\second} as pointed at by the blue arrow.

For Model~A, it is interesting to see that the RMSE of the point prediction (the posterior mean) in Table~\ref{fig:tutorial-ion-channel-rmse-table} is best for the  i.i.d.\ noise model with the $\GP(O, V)$ model only a little worse. Note that the $\GP(O, V)$ model is able to capture and accurately predict the tail current after the two activation steps, as indicated by the red arrows in Figure~\ref{fig:tutorial-ion-channel-prediction} --- a visible area of model mismatch in our calibration without model discrepancy. The uncertainty quantification in the predictions is poor for all of the discrepancy models, but from  Table~\ref{supp-fig:tutorial-ion-channel-likelihood-table} we can see that when we assess the uncertainty in the prediction, the i.i.d.\ noise model is the worst performing model (as for intervals where the prediction is wrong, each error is equally surprising, whereas in correlated models, the first error in any interval makes subsequent errors more probable). The unstructured $\ARMA(2, 2)$ and $\GP(O, V)$ models score highest for their uncertainty quantification.

For Model~B, the $\GP(O, V)$ discrepancy model gives the best overall predictions for both the staircase and the AP protocols, although when we examine the contributions of the mechanistic and discrepancy models, we see that an element of non-identifiability between them has arisen (Supplementary Section~\ref{supp-sec:ion-channel-supp}\ref{supp-sub:ModelBFullResults}).
In terms of the posterior predictive log-likelihood, Table~\ref{fig:tutorial-ion-channel-rmse-table} (bottom) again highlights that the $\ARMA(2, 2)$ and $\GP(O, V)$ models tend to be better than the i.i.d.\ noise and $\GP(t)$ models.

Supplementary Figures~\ref{supp-fig:tutorial-ion-channel-fitting2-residual}, \ref{supp-fig:tutorial-ion-channel-prediction2a-residual}, and \ref{supp-fig:tutorial-ion-channel-prediction2b-residual} show the model discrepancy for Model~A for the sine wave protocol, AP protocol, and staircase protocol, respectively;
Supplementary Figures~\ref{supp-fig:tutorial-ion-channel-fitting2-residual-b}, \ref{supp-fig:tutorial-ion-channel-prediction2a-residual-b}, and \ref{supp-fig:tutorial-ion-channel-prediction2b-residual-b} show the same plots for Model~B.
Supplementary Figures~\ref{supp-fig:tutorial-ion-channel-prediction2b-residual} and~\ref{supp-fig:tutorial-ion-channel-prediction2b-residual-b} in particular highlight that the $\GP(t)$ model has, by design, learnt nothing of relevance about model discrepancy for extrapolation under an independent validation protocol (in which $V(t)$, and indeed the range of $t$, differs from that of the training protocol).
Furthermore, the discrepancy model is based only on information extending to \SI{8000}{\milli\second} (the duration of the training protocol), after which the credible interval resorts to the width of the GP prior variance.
In contrast, the $\GP(O, V)$ model learns, independently of $t$,  the discrepancy under combinations of $(O,V)$ present in the training data (such as the activation step to \SI{40}{\milli\volt} followed by a step to \SI{-120}{\milli\volt}), which is why it is able to better predict the tail current after the two activation steps.
Finally, the $\ARMA(2,2)$ model has zero mean with similar 95\% credible intervals to the  i.i.d.\ noise model, but has correlated errors and so scores better in terms of the posterior predictive log-likelihood. 
The ion channel (ODE) model-only predictions for the sine wave protocol, AP protocol, and staircase protocol are shown in Supplementary Figures~\ref{supp-fig:tutorial-ion-channel-fitting2-model}, \ref{supp-fig:tutorial-ion-channel-prediction2a-model}, and \ref{supp-fig:tutorial-ion-channel-prediction2b-model} for Model~A and Supplementary Figures~\ref{supp-fig:tutorial-ion-channel-fitting2-model-b}, \ref{supp-fig:tutorial-ion-channel-prediction2a-model-b}, and \ref{supp-fig:tutorial-ion-channel-prediction2b-model-b} for Model~B.

For a given dataset, the RMSE and log-likelihood values in Tables~\ref{fig:tutorial-ion-channel-rmse-table} \&~\ref{supp-fig:tutorial-ion-channel-likelihood-table}  are comparable across models.
Note that Model~A is more accurate than Model~B on all datasets and with all discrepancy models. 
With Model~A, none of the discrepancy models are able to improve the mean predictions over the  i.i.d.\ noise model performance,
but the $\GP(O,V)$ comes close (in RMSE) whilst being able to capture some of the nonlinear dynamics that Model~A misses, as discussed above.
With Model~B, the $\GP(O,V)$ model gives the best mean predictions (as measured by the RMSE). The $\GP(t)$ model achieves a better score on the calibration data, but by over-fitting the data.
The $\ARMA(2,2)$ model consistently gives the best posterior predictive log-likelihood values for Models~A and~B, as it gives a wider posterior distribution compared to other methods (Figure~\ref{fig:tutorial-ion-channel-posteriors}). Over-confident predictions are heavily penalised by the log-likelihood, which explains the large differences observed in these values.

To conclude, we have used two different incorrect model structures (Models~A, B) to fit synthetic data generated from  a third model (Model~C).
We considered both ignoring and incorporating discrepancy when calibrating the model. Calibrating with discrepancy  improved predictions notably for  Model~B, but not for Model~A.
Although our problem was a time-dependent (ODE) system, constructing the discrepancy model as a pure time-series based function is not necessarily  useful in predicting unseen situations;
we found the $\GP(O, V)$ model performed best at correcting the point prediction from the models. .

\section{Discussion}

In this review and perspective piece we have drawn attention to an important and under-appreciated source of uncertainty in mechanistic models --- that of uncertainty in the model structure or the equations themselves (model discrepancy).
Focusing on cardiac electrophysiology models, we provided two examples of the consequences of ignoring discrepancy when calibrating models at the ion channel and action potential scales, highlighting how this could lead to over-confident parameter posterior distributions and subsequently spurious predictions.

Statistically, we can explicitly admit discrepancy exists, and include it in the model calibration process and predictions.
We attempted to do this by modelling discrepancy using two proposals from the literature --- Gaussian processes (GPs) trained on different inputs and an autoregressive-moving-average (ARMA) model.
We saw how GPs can achieve some success in describing discrepancy in the calibration experiment.
A two-dimensional GP in voltage and time was used previously by Plumlee \emph{et al.}\ \cite{Plumlee2016,Plumlee2017}, where it was used to  extrapolate to new voltages for a given single step voltage-clamp experiment.
To use a discrepancy model to make predictions for unseen situations, it needs to be
a function of something other than time, otherwise features specific to the calibration experiment are projected into new situations.
A promising  discrepancy model was our two-dimensional GP as a function of the mechanistic model's open probability and voltage, although for Model~B  this led to ambiguity between the role of the ODE system and the role of the discrepancy (see Supplementary Section~\ref{supp-sec:ion-channel-supp}\ref{supp-sub:ModelBFullResults}).

The modelling community would hope to study any discrepancy model that is shown to improve predictions, and use insights from this process to iteratively improve the mechanistic model. 
How we handle model discrepancy may depend on whether we are more interested in learning about what is missing in the model, or in making more reliable predictions:
both related topics are worthy of more investigation.

\subsection{Recommendations}
Very rarely do computational studies use more than one model to test the robustness of their predictions to the model form.
We should bear in mind that all models are approximations and so when we are comparing to real data, all models have discrepancy.
Here we have seen, using synthetic data from an assumed true data-generating  model, how fragile the calibration process can be for models with discrepancy, and how this discrepancy manifests itself in predictions of unseen situations.
Synthetic data studies, simulating data from different parameter sets and different model structures, allow the modeller to test how well the inverse problem can be solved and how robust predictions from the resulting models are.
We strongly recommend performing such studies to learn more about the chosen, and alternative, models, as well as the effects of the model choice on parameter calibration and  subsequent predictions.
To develop our field further, it will be important to document the model-fitting process, and to make datasets and infrastructure available to perform and reproduce these fits with different models \cite{Daly2018WebLab}.

\subsection{Open questions and future work}
The apparent similarity of the action potential models we looked at (summarised in Figure~\ref{fig:tutorial-action-potential-models}) is a challenge for model calibration.
A number of papers have emphasised that more information can be gained to improve parameter identifiability with careful choice of experimental measurements, in particular by using membrane resistance \cite{Kaur2014,Pouranbarani2019}, or other protocols promoting more information-rich dynamics \cite{Groenendaal2015,Johnstone2016} and some of these measurements may be more robust to discrepancy than others. 
In synthetic data, fitting the model used to generate the data will recover the same parameter set from any different protocol (where there is sufficient information to identify the parameters). 
But in the presence of discrepancy, fitting the same model to data from different protocols/experiments will result in different parameter sets, as the models make the best possible compromise (as shown schematically in Figure~\ref{fig:schematic-model-discrepancy}).
This phenomenon may be an interesting way to approach and quantify model discrepancy.

If the difference between imperfect model predictions represented the difference between models and reality then this may also provide a way to estimate discrepancy. 
For instance, the largest difference between the ion channel Model A \& B predictions in the staircase protocol was at the point in time that both of them showed largest discrepancy (Figure~\ref{fig:tutorial-ion-channel-prediction}).
Some form of Bayesian model averaging \cite{Hoeting1999}, using variance-between-models to represent discrepancy, may be instructive if the models are close enough to each other and reality, but can be misleading if the ensemble of models is not statistically exchangeable with the data generating process \cite{Chandler2013, Rougier2013} or if there is some systematic error (bias) due to experimental artefacts \cite{lei2020b}.

In time-structured problems, rather than adding a discrepancy to the final simulated trajectory, as we have done here, we can instead change the dynamics of the model directly. 
It may be easier to add a discrepancy term to the differential equations to address misspecification, than it is to correct their solution, but the downside is that this makes inference of the discrepancy computationally challenging. 
One such approach is to convert  the ODE to a stochastic differential equation \cite{Crucifix2009, Carson2018},
i.e., replace $\frac{\mathrm{d}\mathbf{x}}{\mathrm{d}t} = f_\theta(\mathbf{x},t)$ by $\mathrm{d}\mathbf{x} = f_\theta(\mathbf{x},t) \mathrm{d}t + \Sigma^{\frac{1}{2}} \mathrm{d}W_t$
where $W_t$ is a Brownian motion with covariance matrix $\Sigma$.
This turns the deterministic ODE into a stochastic model and can improve the UQ, but cannot capture any structure missing from the dynamics.
We can go further and attempt to modify the underlying model equations, by changing the ODE system to
\begin{equation}
    \frac{\mathrm{d}\mathbf{x}}{\mathrm{d}t} = f_\theta(\mathbf{x},t) + \delta(\mathbf{x})
\end{equation}
where again $\delta(\mathbf{x})$ is an empirical term to be learnt from the data. 
For example, this has been tried with a discretized version of the equations using a parametric model for $\delta$ \cite{wilkinson2011quantifying}, with GPs \cite{Frigola2013}, nonlinear autoregressive exogenous (NARX) models \cite{Worden2018}, and deep neural networks \cite{Meeds2019}.
Computation of posterior distributions for these models is generally challenging, but is being made easier by the development of automatic-differentiation software, which allows derivative information to be used in MCMC samplers, or in variational approaches to inference (e.g., \cite{Neal2011, Ryder2018}). 

Ultimately, modelling our way out of trouble, by expanding the model class, may prove impossible given the quantity of data available in many cases.
Instead, we may want to modify our inferential approach to allow the best judgements possible about the parameters given the limitation of the model and data. Approaches such as approximate Bayesian computation (ABC) \cite{Sisson2018} and history-matching \cite{Craig1997, Craig2001} change the focus from learning a statistical model within a Bayesian setting, to instead only requiring that the simulation gets within a certain distance of the data. 
This change, from a fully specified statistical model for $\delta$ to instead only giving an upper bound for  $\delta$, is a conservative inferential approach where the aim is not to find the best parameter values, but instead rule out only obviously implausible values \cite{Wilkinson2013, Holden2018}. 

For example, in the action potential model from Section \ref{sect:motivation}, instead of taking a Bayesian approach with an i.i.d.\ Gaussian noise model, we can instead merely try to find parameter values that get us within some distance of the calibration data (see Supplement and Figure~\ref{supp-fig:tradeoff1} for details). 
In the Supplement, we describe a simple approach, based on the methods presented in \cite{Novaes2019}, where we find \num{1079} candidate parameter sets that give a reasonable match to the calibration data. 
When we use these parameters to predict the \SI{2}{\hertz} validation data, and the 75\% \ikr\ block CoU data, we get a wide range of predictions that incorporate the truth (Figure~\ref{supp-fig:tradeoff2}) --- for a small subset of 70 out of 1079, we get good predictions and not the catastrophic prediction shown in Figure~\ref{fig:tutorial-action-potential-fitting-and-prediction}.
By acknowledging the existence of model discrepancy, the use of wider error bounds (or higher-temperature likelihood functions) during the fitting process may avoid fitting parameters overly-precisely.
However, we have no way of knowing which subset of remaining parameter space is more plausible (if any) without doing these further experiments;
testing the model as close as possible to the desired context of use helps us spot such spurious behaviour.

This paper has focused on the ion channel and action potential models of cardiac electrophysiology. 
There is an audit of where uncertainty appears in cardiac modelling and simulation in this issue \cite{clayton2020}. 
The audit highlights many other areas where discrepancy may occur: in assumptions homogenising the subcellular scale to the models we have here; or at the tissue and organ scales in terms of spatial heterogeneity, cell coupling or mechanical models for tissue contraction and fluid-solid interaction.
All of these areas need attention if we are to prevent model discrepancy producing misleading scientific conclusions or clinical predictions.

\section{Conclusions}
In this paper we have seen how having an imperfect representation of a system in a mathematical model (discrepancy) can lead to spuriously certain parameter inference and overly-confident and wrong predictions. 
We have examined a range of methods that attempt to account for discrepancy in the fitting process using synthetic data studies. 
In some cases we can improve predictions using these methods, but different methods work better for different models in different situations, and in some cases the best predictions were still made by ignoring discrepancy. 
A large benefit of the calibration methods which include discrepancy is that they better represent uncertainty in predictions, although all the methods we trialled still failed to allow for a wide enough range of possible outputs in certain parts of the protocols.
Methodological developments are needed to design reliable methods to deal with model discrepancy for use in safety-critical electrophysiology predictions.

\vskip6pt

\enlargethispage{20pt}

\subsection*{Data access}
Code to reproduce the results in the tutorials is available at \url{https://github.com/CardiacModelling/fickleheart-method-tutorials}.

\subsection*{Author contributions}
CLL and SG wrote the code to perform the examples in the main paper.
CLL, SG, DGW, GRM and RDW drafted the manuscript.
All authors conceived and designed the study.
All authors read and approved the manuscript.

\subsection*{Competing interests}

The authors declare that they have no competing interests.

\subsection*{Funding}

This work was supported by the Wellcome Trust [grant numbers 101222/Z/13/Z and 212203/Z/18/Z];
the Engineering \& Physical Sciences Research Council [grant numbers EP/R014604/1, EP/P010741/1, EP/L016044/1, EP/R006768/1, EP/S014985/1, and EP/R003645/1];
the British Heart Foundation [grant numbers PG/15/59/31621, RE/13/4/30184, and SP/18/6/33805];
the Russian Foundation for Basic Research [grant number 18-29-13008].
CLL acknowledges support from the Clarendon Scholarship Fund; and the EPSRC, MRC and F.~Hoffmann-La~Roche Ltd.\ for studentship support.
CDC and CH were supported by the BHF.
MR acknowledges a BHF Turing Cardiovascular Data Science Award.
AVP was partially supported by RF Government Act No.\,211 of March 16, 2013, and RFBR.
RWS was supported by the Brazilian Government via CAPES, CNPq, FAPEMIG, and UFJF, and by an Endeavour Research Leadership Award from the Australian Government Department of Education. 
KW would like to acknowledge the support of the UK EPSRC.
GMN was supported by CEFET-MG and CAPES.
GRM \& SG acknowledge support from the Wellcome Trust \& Royal Society via a Sir Henry Dale Fellowship to GRM.
GRM \& DGW acknowledge support from the Wellcome Trust via a Wellcome Trust Senior Research Fellowship to GRM.
The authors would like to thank the Isaac Newton Institute for Mathematical Sciences for support and hospitality during the `Fickle Heart' programme.

\subsection*{Acknowledgements}

We would like to thank all the participants at the Isaac Newton Institute `Fickle Heart' programme for helpful discussions which informed this manuscript.

\bibliographystyle{ieeetr}
{\small
\bibliography{refs}}

\makeatletter\@input{xx.tex}\makeatother

\end{document}


\title{
Considering discrepancy when calibrating a mechanistic electrophysiology model: Supplementary Material
\vspace{0.5cm}
}

\author{Chon Lok Lei,
Sanmitra Ghosh,
Dominic G.\ Whittaker,
Yasser Aboelkassem,\\
Kylie A.\ Beattie,
Chris Cantwell,
Tammo Delhaas,
Charles Houston,\\
Gustavo Montes Novaes,
Alexander V.\ Panfilov,
Pras Pathmanathan,\\
Marina Riabiz,
Rodrigo Weber dos Santos,
John Walmsley,\\
Keith Worden,
Gary R.\ Mirams and
Richard D.\ Wilkinson
}

\date{}

\maketitle

\setcounter{table}{0}
\renewcommand{\thetable}{S\arabic{table}}
\setcounter{figure}{0}
\renewcommand{\thefigure}{S\arabic{figure}}
\setcounter{equation}{0}
\renewcommand{\theequation}{S\arabic{section}.\arabic{equation}}
\setcounter{section}{0}
\renewcommand{\thesection}{S\arabic{section}}

\tableofcontents

\section{Differences between Model~T and Model~F}\label{sec:model-t-f-equations}
Here we show the equations of the currents in Model~T \cite{ten2004model} and Model~F \cite{fink2008contributions} that are different in kinetics.
Below $g_{X}$ is the conductance (constant), $E_{X}$ is the reversal potential, $X_{o}$ is the extracellular concentration, $X_{i}$ is the intracellular concentration of ion $X$.
$V$ is the membrane voltage, $F$ is Faraday constant, $R$ is the ideal gas constant.

\newcommand{\ud}{\mathrm{d}}
\subsection*{\texorpdfstring{$I_\text{CaL}$}{ICaL}}
\subsubsection*{Model~T\hspace{6cm}Model~F}
\begingroup\makeatletter\def\f@size{4}\check@mathfonts
\def\maketag@@@#1{\hbox{\m@th\large\normalfont#1}}
\begin{equation*}
    \begin{aligned}[t]
    I_{CaL} &= \frac{g_{CaL} \cdot d \cdot f \cdot fCa \cdot 4 \cdot V \cdot F^{2}}{R \cdot 310} \\
            & \times \frac{\left(Ca_{i} \cdot e^{\frac{2 \cdot V \cdot F}{R \cdot 310}}-0.341 \cdot Ca_{ o }\right)}{\left(e^{\frac{2 \cdot V \cdot F}{R \cdot 310}}-1\right)} \\
    d_{\infty} &= \frac{1}{\left(1+e^{\frac{\left(-5-V\right)}{7.5}}\right)} \\
    \alpha_{d} &= \left(\frac{1.4}{\left(1+e^{\frac{\left(-35-V\right)}{13}}\right)}+0.25\right) \\
    \beta_{d} &= \frac{1.4}{\left(1+e^{\frac{\left(V+5\right)}{5}}\right)} \\
    \gamma_{d} &= \frac{1}{\left(1+e^{\frac{\left(50-V\right)}{20}}\right)} \\
    \tau_{d} &= \left(\alpha_{d} \cdot \beta_{ d }+\gamma_{ d }\right) \\
    \frac{\ud d}{\ud t} &= \frac{\left(d_{\infty}-d\right)}{\tau_{ d }} \\
    \end{aligned}\qquad\qquad\qquad\quad
    \begin{aligned}[t]
    I_{CaL} &= \frac{g_{CaL} \cdot d \cdot f \cdot f_2 \cdot fCass \cdot 4 \cdot \left(V-15\right) \cdot F^{2}}{R \cdot 310} \\
            &  \times \frac{\left(0.25 \cdot Ca_{ss} \cdot e^{\frac{2 \cdot \left(V-15\right) \cdot F}{R \cdot 310}}-Ca_{ o }\right)}{\left(e^{\frac{2 \cdot \left(V-15\right) \cdot F}{R \cdot 310}}-1\right)}\\
    d_{\infty} &= \frac{1}{\left(1+e^{\frac{\left(5-V\right)}{7.5}}\right)}\\
    \alpha_{d} &= \textrm{\tiny{as per Model T,}}\\
    \beta_{d} &=\textrm{\tiny{as per Model T,}}\\
    \gamma_{d} &=\textrm{\tiny{as per Model T,}}\\
    \tau_{d} &= \textrm{\tiny{as per Model T,}}\\
    \frac{\ud d}{\ud t} &= \textrm{\tiny{as per Model T,}}\\
                                    \end{aligned}
\end{equation*}
\begin{equation*}
    \begin{aligned}[t]
    f_{\infty} &= \frac{1}{\left(1+e^{\frac{\left(V+20\right)}{7}}\right)} \\
    \tau_{f} &= \left(1125 \cdot e^{\frac{-\left(V+27\right)^{2}}{240}}+80+\frac{165}{\left(1+e^{\frac{\left(25-V\right)}{10}}\right)}\right) \\
    \frac{\ud f}{\ud t} &= \frac{\left(f_{\infty}-f\right)}{\tau_{ f }} \\
        \alpha_{fCa} &= \frac{1}{\left(1+\left(\frac{Ca_{ i }}{0.000325}\right)^{8}\right)} \\
    \beta_{fCa} &= \frac{0.1}{\left(1+e^{\frac{\left(Ca_{ i }-0.0005\right)}{0.0001}}\right)} \\
    \gamma_{fCa} &= \frac{0.2}{\left(1+e^{\frac{\left(Ca_{ i }-0.00075\right)}{0.0008}}\right)} \\
    fCa_{\infty} &= \frac{\left(\alpha_{ fCa }+\beta_{ fCa }+\gamma_{ fCa }+0.23\right)}{1.46} \\
    \tau_{fCa} &= 2 \\
    d_{fCa} &= \frac{\left(fCa_{\infty}-fCa\right)}{\tau_{ fCa }} \\
    \frac{\ud fCa}{\ud t} &= 
    \begin{cases}0;
    &  \text{\tiny{if }}(fCa_{\infty} > fCa)\text{\tiny{ and }}(V > -60),\\d_{ fCa }&  \text{\tiny{otherwise}}.
    \end{cases}
    \end{aligned}\qquad\qquad\quad
    \begin{aligned}[t]
        f_{\infty} &= \textrm{\tiny{as per Model T,}}\\
        \tau_{f} &= \frac{1}{4} \left(1102.5 \cdot e^{\frac{-(\left(\left(V+27\right)\right)^{2})}{225}}+\frac{200}{\left(1+e^{\frac{\left(13-V\right)}{10}}\right)}+\frac{180}{\left(1+e^{\frac{\left(V+30\right)}{10}}\right)}+20\right)\\
    \frac{\ud f}{\ud t} &= \textrm{\tiny{as per Model T,}}\\
        f2_{\infty} &= \left(\frac{0.75}{\left(1+e^{\frac{\left(V+35\right)}{7}}\right)}+0.25\right)\\
    \tau_{f2} &= \frac{1}{2} \left(562 \cdot e^{\frac{-\left(V+27\right)^{2}}{240}}+\frac{31}{\left(1+e^{\frac{\left(25-V\right)}{10}}\right)}+\frac{80}{\left(1+e^{\frac{\left(V+30\right)}{10}}\right)}\right)\\
    \frac{\ud f_2}{\ud t} &= \frac{\left(f2_{\infty}-f_2\right)}{\tau_{ f2 }}\\
        fCass_{\infty} &= \left(\frac{0.4}{\left(1+\left(\frac{Ca_{ ss }}{0.05}\right)^{2}\right)}+0.6\right)\\
    \tau_{fCass} &= \left(\frac{80}{\left(1+\left(\frac{Ca_{ ss }}{0.05}\right)^{2}\right)}+2\right)\\
    \frac{\ud fCass}{\ud t} &= \frac{\left(fCass_{\infty}-fCass\right)}{\tau_{ fCass }}
    \end{aligned}
\end{equation*}
\endgroup

\clearpage
\subsection*{\texorpdfstring{$I_\text{to}$}{Ito}}
\subsubsection*{Model~T\hspace{5.5cm}Model~F}
\begin{equation*}
    \begin{aligned}[t]
    I_{to} &= g_{to} \cdot r \cdot s \cdot \left(V-E_{K}\right)\\
    s_{\infty} &= \frac{1}{\left(1+e^{\frac{\left(V+28\right)}{5}}\right)}\\
    \tau_{s} &= 1000 \cdot e^{\left(\frac{-(V + 67) ^ 2 }{ 1000 } \right)} + 8\\
    \frac{\ud s}{\ud t} &= \frac{\left(s_{\infty}-s\right)}{\tau_{s}}\\
        r_{\infty} &= \frac{1}{\left(1+e^{\frac{\left(20-V\right)}{6}}\right)}\\
    \tau_{r} &= \left(9.5 \cdot e^{\frac{-\left(V+40\right)^{2}}{1800}}+0.8\right)\\
    \frac{\ud r}{\ud t} &= \frac{\left(r_{\infty}-r\right)}{\tau_{r}}
    \end{aligned}
    \qquad\qquad\qquad
    \begin{aligned}[t]
    I_{to} &= \textrm{as per Model T,}\\
    s_{\infty} &= \frac{1}{\left(1+e^{\frac{\left(V+20\right)}{5}}\right)}\\
    \tau_{s} &= \left(85 \cdot e^{\frac{-\left(V+45\right)^{2}}{320}}+\frac{5}{\left(1+e^{\frac{\left(V-20\right)}{5}}\right)}+3\right)\\
    \frac{\ud s}{\ud t} &= \textrm{as per Model T,}\\
        & \\
    & \textrm{$r$ gate as per Model T.}\\
                \end{aligned}
\end{equation*}

\clearpage
\subsection*{\texorpdfstring{$I_\text{Kr}$}{IKr}}
\subsubsection*{Model~T\hspace{5.25cm}Model~F}
\begingroup\makeatletter\def\f@size{8}\check@mathfonts
\def\maketag@@@#1{\hbox{\m@th\large\normalfont#1}}
\begin{align*}
    \begin{aligned}[t]
    I_{Kr} &= g_{Kr} \cdot \sqrt{\frac{K_{o}}{5.4}} \cdot X_{r1} \cdot X_{r2} \cdot \left(V-E_{K}\right)\\
        xr1_{\infty} &= \frac{1}{\left(1+e^{\frac{\left(-26-V\right)}{7}}\right)}\\
    \alpha_{xr1} &= \frac{450}{\left(1+e^{\frac{\left(-45-V\right)}{10}}\right)}\\
    \beta_{xr1} &= \frac{6}{\left(1+e^{\frac{\left(V+30\right)}{11.5}}\right)}\\
    \tau_{xr1} &= \alpha_{xr1} \cdot \beta_{xr1}\\
    \frac{\ud X_{r1}}{\ud t} &= \frac{\left(xr1_{\infty}-X_{r1}\right)}{\tau_{xr1}}\\
        xr2_{\infty} &= \frac{1}{\left(1+e^{\frac{\left(V+88\right)}{24}}\right)}\\
    \alpha_{xr2} &= \frac{3}{\left(1+e^{\frac{\left(-60-V\right)}{20}}\right)}\\
    \beta_{xr2} &= \frac{1.12}{\left(1+e^{\frac{\left(V-60\right)}{20}}\right)}\\
    \tau_{xr2} &= \alpha_{xr2} \cdot \beta_{xr2}\\
    \frac{\ud X_{r2}}{\ud t} &= \frac{\left(xr2_{\infty}-X_{r2}\right)}{\tau_{xr2}}
    \end{aligned}
    &&&&
    \begin{aligned}[t]
    I_{Kr} &= g_{Kr} \cdot \left(\frac{310}{35}-\frac{55}{7}\right) \cdot \sqrt{\frac{K_{o}}{5.4}} \cdot Or4 \cdot \left(V-E_{K}\right)\\
        \alpha_{xr1} &= e^{\left(24.335+\left(0.0112 \cdot V-25.914\right)\right)}\\
    \beta_{xr1} &= e^{\left(13.688+\left(-(0.0603) \cdot V-15.707\right)\right)}\\
    \alpha_{xr2} &= e^{\left(22.746+\left(0 \cdot V-25.914\right)\right)}\\
    \beta_{xr2} &= e^{\left(13.193+\left(0 \cdot V-15.707\right)\right)}\\
    \alpha_{xr3} &= e^{\left(22.098+\left(0.0365 \cdot V-25.914\right)\right)}\\
    \beta_{xr3} &= e^{\left(7.313+\left(-(0.0399) \cdot V-15.707\right)\right)}\\
    \alpha_{xr4} &= e^{\left(30.016+\left(0.0223 \cdot V-30.888\right)\right)} \cdot \left(\frac{5.4}{K_{o}}\right)^{0.4}\\
    \beta_{xr4} &= e^{\left(30.061+\left(-(0.0312) \cdot V-33.243\right)\right)}\\
    \frac{\ud Cr1}{\ud t} &= \left(\beta_{xr1} \cdot Cr2-\alpha_{xr1} \cdot Cr1\right)\\
    \frac{\ud Cr2}{\ud t} &= \left(\left(\alpha_{xr1} \cdot Cr1+\beta_{xr2} \cdot Cr3\right)-\left(\alpha_{xr2}+\beta_{xr1}\right) \cdot Cr2\right)\\
    \frac{\ud Cr3}{\ud t} &= \left(\left(\alpha_{xr2} \cdot Cr2+\beta_{xr3} \cdot Or4\right)-\left(\alpha_{xr3}+\beta_{xr2}\right) \cdot Cr3\right)\\
    \frac{\ud Or4}{\ud t} &= \left(\left(\alpha_{xr3} \cdot Cr3+\beta_{xr4} \cdot Ir5\right)-\left(\alpha_{xr4}+\beta_{xr3}\right) \cdot Or4\right)\\
    \frac{\ud Ir5}{\ud t} &= \left(\alpha_{xr4} \cdot Or4-\beta_{xr4} \cdot Ir5\right)
    \end{aligned}
\end{align*}
\endgroup

\clearpage
\subsection*{\texorpdfstring{$I_\text{K1}$}{IK1}}
\subsubsection*{Model~T\hspace{5.25cm}Model~F}
\begingroup\makeatletter\def\f@size{7}\check@mathfonts
\def\maketag@@@#1{\hbox{\m@th\large\normalfont#1}}
\begin{align*}
    \begin{aligned}[t]
    I_{K1} &= g_{K1} \cdot xK1_{\infty} \cdot \sqrt{\frac{K_{o}}{5.4}} \cdot \left(V-E_{K}\right)\\
    \alpha_{K1} &= \frac{0.1}{\left(1+e^{0.06 \cdot \left(\left(V-E_{K}\right)-200\right)}\right)}\\
    \beta_{K1}^a &= 3 \cdot e^{0.0002 \cdot \left(\left(V-E_{K}\right)+100\right)}\\
    \beta_{K1}^b &= e^{0.1 \cdot \left(\left(V-E_{K}\right)-10\right)}\\
    \beta_{K1} &= \frac{\beta_{K1}^a + \beta_{K1}^b}{\left(1+e^{-(0.5) \cdot \left(V-E_{K}\right)}\right)}\\
    xK1_{\infty} &= \frac{\alpha_{K1}}{\left(\alpha_{K1}+\beta_{K1}\right)}
    \end{aligned}
    &&&&
    \begin{aligned}[t]
    I_{K1} &= g_{K1} \cdot \left(\frac{310}{35}-\frac{55}{7}\right) \cdot \sqrt{\frac{K_{o}}{5.4}} \cdot xK1_{\infty} \cdot \left(V-E_{K}\right)\\
        Ki_{Mg} &= 2.8 \cdot e^{\frac{-\left(V-\delta \cdot E_{K}\right)}{180}}\\
    Kb_{Mg} &= 0.45 \cdot e^{\frac{-\left(V-\delta \cdot E_{K}\right)}{20}}\\
    Kd1_{SPM} &= 0.7 -3 \cdot e^{\frac{-\left(\left(V-\delta \cdot E_{K}\right)+8 \cdot Mg_{Buf}\right)}{4.8}}\\
    Kd2_{SPM} &= 40 -3 \cdot e^{\frac{-\left(V-\delta \cdot E_{K}\right)}{9.1}}\\
    X &= \left(1+\frac{Mg_{Buf}}{Kb_{Mg}}\right)\\
    rec1 &= \frac{X^2}{\left(\frac{SPM}{Kd1_{SPM}}+\frac{Mg_{Buf}}{Ki_{Mg}}+X^3\right)}\\
    rec2 &= \frac{1}{\left(1+\frac{SPM}{Kd2_{SPM}}\right)}\\
    xK1_{\infty} &= \left(\phi \cdot rec1+\left(1-\phi\right) \cdot rec2\right)\\
    Mg_{Buf} &= 0.0356\\
    SPM &= 0.0014613\\
    \phi &= 0.8838\\
    \delta &= 1.0648
    \end{aligned}
\end{align*}
\endgroup

\subsection*{\texorpdfstring{$I_\text{Ks}$}{IKs}}
\subsubsection*{Model~T\hspace{5.25cm}Model~F}
\begin{align*}
    \begin{aligned}[t]
    I_{Ks} &= g_{Ks} \cdot X_s^{2} \cdot \left(V-E_{Ks}\right)\\
        xs_{\infty} &= \frac{1}{\left(1+e^{\frac{\left(-5-V\right)}{14}}\right)}\\
    \alpha_{xs} &= \frac{1100}{\sqrt{\left(1+e^{\frac{\left(-10-V\right)}{6}}\right)}}\\
    \beta_{xs} &= \frac{1}{\left(1+e^{\frac{\left(V-60\right)}{20}}\right)}\\
    \tau_{xs} &= \alpha_{xs} \cdot \beta_{xs}\\
    \frac{\ud X_s}{\ud t} &= \frac{\left(xs_{\infty}-X_s\right)}{\tau_{xs}}
    \end{aligned}
    &&&&
    \begin{aligned}[t]
    I_{Ks} &= \textrm{as per Model T,}\\
        xs_{\infty} &= \textrm{as per Model T,}\\
        \alpha_{xs} &= \frac{1400}{\sqrt{\left(1+e^{\frac{\left(5-V\right)}{6}}\right)}}\\
    \beta_{xs} &= \frac{1}{\left(1+e^{\frac{\left(V-35\right)}{15}}\right)}\\
    \tau_{xs} &= \left(\alpha_{xs} \cdot \beta_{xs}+80\right)\\
    \frac{\ud X_s}{\ud t} &= \textrm{as per Model T.}
    \end{aligned}
\end{align*}

\clearpage
\section{Modelling discrepancy using a Gaussian process}\label{sec:gp-detail}
In order to add a discrepancy term to our basic measurement model (see main text), we model the $i$\textsuperscript{th} observation as:
\begin{equation}\label{eq:gp-discrepancy}
    (Y_{C})_i = f_i(\thb, u^i_C) + \delta_i(\phb, \vci) + \epsilon_i,
\end{equation} 
where $\delta_i(\boldsymbol{\phb}, \vci)$ is the model discrepancy term, a function with arguments $\vc$ and parameters $\phb$. Note that the inputs $\vc$ can be independent from the inputs passed to the mechanistic model. We choose $v_C$ to be (1) time $t$, and (2) the open probability $O$ (i.e.\ $\mathcal{O}$ in Eq.~(\ref{main-eq:ion-current})) and the voltage $V$.

Following \cite{Kennedy2001} we place a zero mean Gaussian process prior on the discrepancy function given by
\begin{equation}\label{eq:gp-prior}
    \delta(\phb, \vc) \sim \mathcal{GP}\big(0,\cov(\vc,\vc';\phb)\big),
\end{equation}
where $\cov(\vc,\vc';\phb)$ is the covariance function (also known as covariance \textit{kernel}) parameterised by $\phb$. One common choice for the covariance function is the squared exponential function given by
\begin{equation}\label{eq:rbf kernel}
    \cov(\vc,\vc';\phb)= \alpha^2 \exp\left( -\sum_{j=1}^q\frac{(v_{C_j}-v'_{C_j})^2}{2\rho^2_j}\right),
\end{equation}
where $q$ is the number of covariates, such as time or open probability as mentioned above, representing $\vc$. The parameter $\rho_j$ quantifies the characteristic length-scale along the $j$\textsuperscript{th} covariate and $\alpha$ denotes the marginal variance of the GP prior. Together they constitute the parameter vector $\phb=\left[\alpha, \rho_1, \ldots, \rho_q\right]$.

Since our measurement noise is Gaussian with variance $\sigma^2$ we can analytically compute the discrepancy function to obtain the marginal likelihood of $N$ observations $\yc = (Y_{C})_{i=1}^N$, conditioned on the parameters $\thb$, $\phb$ of the mechanistic and discrepancy models respectively, as well as the calibration inputs $u_C$ and $\vc$, given by
\begin{equation}\label{eq:likelihood}
    p(\yc|u_C,\vc,\thb,\phb) \sim \mathcal{N}(\fc,\Cov_{NN} + \sigma^2 \bv{I}),
\end{equation}
where $\fc= [f_1(\thb, u_C), \ldots, f_N(\thb, u_C)]$ is a vector collecting the $N$ evaluations of the mechanistic model function, $\Cov_{NN}$ represents the covariance function ( Eq.~\ref{eq:rbf kernel}) evaluated on all $N \times N$ pairs of the calibrations inputs $\vc$:
\begin{equation}
    \Cov_{NN} = \begin{pmatrix}
        \cov(\vcone,\vcone;\phb) &  \ldots & \cov(\vcone,\vcN;\phb) \\
        \vdots   & \ddots & \vdots  \\
        \cov(\vcN,\vcone;\phb) & \cdots & \cov(\vcN,\vcN;\phb) 
    \end{pmatrix},
\end{equation}
and $\bv{I}$ is a $N \times N$ identity matrix.

\subsection*{Inference of the model and GP parameters}

We proceed by first placing suitable prior distributions, $p(\thb)$ and $p(\phb)$, on the model and GP parameters and then obtain the posterior distribution using Bayes theorem as follows:
\begin{equation}\label{eq: posterior}
    p(\thb,\phb|\yc, u_C, \vc) \propto p(\yc|u_C,\vc,\thb,\phb)p(\thb)p(\phb).
\end{equation}
Since this posterior distribution is analytically intractable due to the non-linear dependence on $\thb$ and $\phb$ we resort to Markov chain Monte Carlo (MCMC) in order to obtain samples from this distribution.

\subsection*{Predictions}

Having inferred the parameters $\thb$ and $\phb$ we may want to predict the output of the model in Eq.~\ref{eq:gp-discrepancy} for a new set of model inputs $u_V$ and $\vv$.
Note that these new model inputs are considered as validation inputs, denoted with subscript $V$. For the purpose of derivation here, we consider the number of validation points $M$ to be different than the number of measurement points $N$, although these numbers can be the same for specific choices of calibrations.

We denote the column vector for the corresponding $M$ predicted outputs as $\yv=(Y_V)_{i=1}^M$, and the model evaluations with the new inputs as $\fv= [f_1(\thb, u_V), \ldots, f_M(\thb, u_V)]^T$
Furthermore, we denote the collection of calibration inputs at the $N$ training (points corresponding to the measurements) as $\bv{I}_C = ( u_C,\vc)$, and at the prediction points as $\bv{I}_V=( u_V,\vv)$ 

Note that for a fixed value of parameters, $\thb$ and $\phb$ respectively, we can analytically obtain the predictive distribution of $\yv$ given by
\begin{equation}
    p(\yv|\bv{I}_V,\bv{I}_C,\yc,\thb,\phb) = \mathcal{N}(\bv{\mu_V}, \bv{\sigma^2_V}),
\end{equation}
where the mean and variance is given by \cite{Rasmussen2006}
\begin{equation}\label{eq:gp mean var}
    \begin{aligned}
        \bv{\mu_V} &= \fv + \Cov_{MN}[\Cov_{NN}+\sigma^2\bv{I}]^{-1}(\yc-\fc)\\
        \bv{\sigma^2_V} &= \Cov_{MM}- \Cov_{MN}[\Cov_{NN}+\sigma^2\bv{I}]^{-1}\Cov_{NM},
    \end{aligned}
\end{equation}
where $\Cov_{MN}$ and $\Cov_{NM}$ denotes the $M \times N$ and $N \times M$ matrices of covariance function evaluations between the training and prediction inputs given by
\begin{equation}\label{eq:cross-covariance-1}
    \Cov_{MN} =  \begin{pmatrix}
        \cov(\vvone,\vcone;\phb) &  \ldots & \cov(\vvone,\vcN;\phb) \\
        \vdots   & \ddots & \vdots  \\
        \cov(\vvM,\vcone;\phb) & \cdots & \cov(\vvM,\vcN;\phb) 
    \end{pmatrix},
\end{equation}
\begin{equation}\label{eq:cross-covariance-2}
    \Cov_{NM} =  \begin{pmatrix}
        \cov(\vcone,\vvone;\phb) &  \ldots & \cov(\vcone,\vvM;\phb) \\
        \vdots   & \ddots & \vdots  \\
        \cov(\vcN,\vvone;\phb) & \cdots & \cov(\vcN,\vvM;\phb) 
    \end{pmatrix},
\end{equation}
with inputs $\vc$ and $\vv$, respectively, and $\Cov_{MM}$ is the covariance evaluated at the prediction inputs $\vv$ only:
\begin{equation}
    \Cov_{MM} =  \begin{pmatrix}
        \cov(\vvone,\vvone;\phb) &  \ldots & \cov(\vvone,\vvM;\phb) \\
        \vdots   & \ddots & \vdots  \\
        \cov(\vvM,\vvone;\phb) & \cdots & \cov(\vvM,\vvM;\phb) 
    \end{pmatrix}.
\end{equation}

Finally, to obtain the marginal (i.e. integrating out the parameters) predictive distribution:
\begin{equation}\label{eq:ppc}
     p(\yv|\bv{I}_V,\bv{I}_C,\yc) = \int \mathcal{N}(\bv{\mu_V}, \bv{\sigma^2_V}) p(\thb,\phb|\yc, u_C, \vc) \operatorname{d}\thb \operatorname{d}\phb,  
\end{equation}
we use Monte Carlo integration using the samples of $\thb$ and $\phb$ obtained through MCMC.

\subsection*{Sparse Gaussian Process}

The above formulation of the discrepancy model suffers from a crucial computational bottleneck stemming from the need of inverting the covariance matrix $\Cov_{NN}$ while evaluating the marginal likelihood in Eq.~\ref{eq:likelihood},
as well as drawing posterior predictions in Eq.~\ref{eq:ppc} (in turn using Eq.~\ref{eq:gp mean var}). In all the calibration problems under consideration here, we have a large number of data points (time series measurements) where $N\geq 80000$.
Thus, it becomes infeasible to apply Gaussian processes for modelling the discrepancy without tackling this excessive computational load related to repeated inversion of a large matrix.

In order to alleviate this computational bottleneck we use a sparse approximation of the true covariance function.
Qui{\~n}onero-Candela et al. \cite{quinonero2005unifying} provides an extensive review of such sparse approximations techniques.
Following \cite{quinonero2005unifying} we use a set of $P$ or inducing inputs (or pseudo-inputs) $x_C$ with associated latent function $\delta(\phb, x_C)$ representing the discrepancy function corresponding to the inducing inputs.
This inducing function is assigned a zero mean GP prior as follows: 
\begin{equation}
    \delta(\phb, x_C) \sim \mathcal{GP}\big(0,\cov(x_C,x_C';\phb)\big).
\end{equation}

Let us denote the vector of discrepancy function evaluations at all the training points as $\delc = [\delta_1(\phb, u_C), \ldots,\delta_N(\phb, u_C)]^T$ and at inducing points as $\delx = [\delta_1(\phb, x_C), \ldots,\delta_P(\phb, x_C)]^T$.
We can then write the joint prior as a product of all the training and inducing points as $p(\delc)=\mathcal{N}(\bv{0},\Cov_{NN})$ and $p(\delx)=\mathcal{N}(\bv{0},\Cov_{PP})$ respectively, where $\Cov_{PP}$ denotes the covariance evaluated at all pairs of inducing inputs:
\begin{equation}
    \Cov_{PP} =  \begin{pmatrix}
        \cov(x_C^1,x_C^1;\phb) &  \ldots & \cov(x_C^1,x_C^P;\phb) \\
        \vdots   & \ddots & \vdots  \\
        \cov(x_C^P,x_C^1;\phb) & \cdots & \cov(x_C^P,x_C^P;\phb) 
    \end{pmatrix}.
\end{equation}

We can then approximate the prior on the true discrepancy function $\delc$ marginalising the inducing discrepancies as: 
\begin{equation}
\begin{aligned}
    p(\delc) \approx p(\delc|\delx) &= \int p(\delc|\delx)p(\delx) \operatorname{d}\delx\\
    &=\mathcal{N}(\Cov_{NP}\Cov_{PP}^{-1}\delx, \Cov_{NN}-\Cov_{NP}\Cov_{PP}^{-1}\Cov_{PN}),
\end{aligned}
\end{equation}
where $\Cov_{NP}$, $\Cov_{PN}$ denotes the covariance matrices containing the cross-covariances between the training and inducing inputs (evaluated in the same way as in Eqs.~\ref{eq:cross-covariance-1},~\ref{eq:cross-covariance-2}).
This sparse approximation was first introduced in \cite{snelson2006sparse} to scale the GP regression model.
This approximation is widely known as the \textit{fully independent training conditional} (FITC) approximation in machine learning parlance since the introduction of these inducing inputs and corresponding function values $\delx$ induces a conditional independence among all the elements of $\delc$ \cite{quinonero2005unifying}, that is we have 
\begin{equation}
    p(\delc|\delx) =\prod_{i=1}^N p(\delta_i(\phb, \vci)|\delx)=\mathcal{N}(\Cov_{NP}\Cov_{PP}^{-1}\delx, \Cov_{NN}-\Cov_{NP}\Cov_{PP}^{-1}\Cov_{PN}).
\end{equation}

Using this approximate prior $p(\delc|\delx)$ to obtain the marginal likelihood and the prediction terms we essentially approximate the true covariance $\Cov_{NN}$ as \cite{quinonero2005unifying}:
\begin{equation}
    \Cov_{NN}\approx\hat{\Cov} = \bv{Q} + \operatorname{diag}(\Cov_{NN} - \bv{Q}),
\end{equation}
where $\mathrm{diag}(A)$ is a diagonal matrix whose elements match the diagonal of $A$ and the matrix $\mathbf{Q}$ is given by
\begin{equation}
   \bv{Q}=\Cov_{NP}\Cov_{PP}^{-1}\Cov_{PN}. 
\end{equation}
$\hat{\Cov}$ has the same diagonal elements as $\Cov_{NN}$ and the off-diagonal elements are the same as for $\bv{Q}$. Thus, inversion of $\hat{\Cov}$ scales as $\mathcal{O}(NP^2)$ as opposed to $\mathcal{O}(N^3)$ for the inversion of $\Cov_{NN}$.

\clearpage
\section{Modelling residuals using an \texorpdfstring{$\ARMA(p,q)$}{ARMA(p, q)} process}\label{sec:arma-detail}
In the previous section we modelled the discrepancy as a function drawn from a GP prior. Alternatively, we can address the case of discrepancy using a correlated residual approach (see Section~$3.6.2$ in \cite{durbin2012time} for an introduction to this modelling approach). 
In this case we can model the residuals between the data $(Y_C)_i$ and the mechanistic model $f_i(\thb, u^i_C)$ as an $\ARMA(p,q)$ process as follows:
\begin{align}\label{eq:arma-ode}
    (Y_C)_i -f_i(\thb, u^i_C) &= e_i \nonumber\\
    &= \varphi_1 e_{i-1} + \ldots + \varphi_p e_{i-p} + \nu_i + \zeta_1 \nu_{i-1} + \ldots + \zeta_q \nu_{i-1-q},\\
    \shortintertext{where} \nu_i &\sim \mathcal{N}(0,\tau^2),
\end{align}
and $\bv{\varphi}=\as$, $\bv{\zeta}=\bs$ are the vectors representing the $p\geq 0$ autoregressive coefficients and $q\geq 0 $ moving-average coefficients of the $\ARMA$ process. 

The rationale behind this modelling approach comes from the fact that if the mechanistic model is able to explain the measurements adequately then the residuals are essentially uncorrelated measurement noise $\epsilon \sim \mathcal{N}(0,\sigma^2)$.
Note that we use a different symbol $\nu$, as opposed to $\epsilon$, to represent the noise term in order to highlight the difference in its interpretations.
However, the existence of discrepancy between the model output and the observations points to the fact that the residuals, for each data sample, has unexplained structure that can be modelled using a pre-determined correlation structure, as expressed through an $\ARMA(p,q)$ model.

\subsection*{Inference}

We first re-write the normally distributed error term $\nu_i$ as
\begin{equation}
    \nu_i =  (Y_C)_i - f_i(\thb, u^i_C)  - \sum_{j=1}^p \varphi_j \{(Y_C)_{i-j} - f_{i-j}(\thb, u^{i-j}_C) \} - \sum_{k=1}^q \zeta_k \nu_{i-k} ,
\end{equation}
using which we can write the conditional likelihood of the observed data for $N$ measurements as \cite{wei2006}
\begin{equation}\label{eq:arma-likelihood}
    p(\yc|\thb,\bv{\varphi},\bv{\zeta},\tau) = (2\pi \tau^2)^{N/2} \exp\bigg(-\frac{1}{2\tau^2}\sum_{i=p+1}^N \nu^2_i \bigg),
\end{equation}
where we have used $\nu_i$ for $i\geq p+1$ by assuming that $\nu_p=\nu_{p-1}=\ldots=\nu_{p+1-q}=0$, its expected value. 
Note that to calculate the likelihood for all the $N$ measurements requires us to introduce extra parameter values, as latent variables, for the past history of the data as well as the error terms before measurement commences, that is for $[(Y_C)_0,(Y_C)_{-1}, \ldots, (Y_C)_{1-p}]$ and $[\nu_0,\nu_{-1},\ldots,\nu_{1-q}]$. 
Alternatively, we can reformulate Eq.~\ref{eq:arma-ode} in a state space form and use the Kalman filter algorithm to evaluate the unconditional full likelihood for all $i$.
We refer the reader to \cite{durbin2012time} for the details of this approach. 
We point out here that the difference between these two approaches of calculating the likelihood is insignificant for long time series, which is the case for our calibration problems with $N \geq 80000$.

Having defined the likelihood we can again adopt the Bayesian framework to infer posterior distributions of the model parameters $\thb$ and the set of $\ARMA$ parameters $\phb=\left[\bv{\varphi},\bv{\zeta}\right]$, by choosing suitable prior distributions $p(\thb)$ and $p(\phb)$ respectively and using the Bayes theorem to obtain the posterior given by
\begin{equation}
    p(\thb,\phb|\yc) \propto p(\yc|\thb,\phb)p(\thb)p(\phb).
\end{equation}
Note that we have considered the noise variance known since its maximum likelihood estimate is given by $\tau^2=\frac{\sum_{i=p+1}^N \nu^2_i}{N-(2p+q+1)}$, which can be easily obtained once estimates of $\phb$ and $\thb$ are available. Similar to the GP based inference problem we again use MCMC to obtain the desired posterior distributions.

\subsection*{Predictions}

In a purely time series modelling context, where models such as the $\ARMA$ is extensively used, predictions are used to forecast ahead in time for short intervals. 
In the context of our calibration problem we are generally interested in predicting outputs for a new calibration $u_V$. 
However, considering the fact that we want to to predict $M$ output values corresponding to the new validation inputs, we can simply recast our predictions as one-step-ahead forecasts. 

We denote the $M$ predicted outputs as $\yv=(Y_V)_{m=1}^M $, while $\bv{Y}_p=(Y_C)_{N-(p-1)}^N$ and $\fp= [f_N(\thb, u_C), \ldots, f_{N-(p-1)}(\thb, u_C)]$ are column vectors representing the last $p$ observations and model evaluations with the calibration $u_C$. We denote the vector of the last $q$ errors as $\bv{\nu}_C = [\nu_{N-(q-1)},\ldots,\nu_{N}]^T$. 

Note that in our formulation here, the $m$\textsuperscript{th}, $m=1,\ldots, M$, prediction is to be considered as the $(N+1)$\textsuperscript{th} prediction from the model in Eq.~\ref{eq:arma-ode} with the following modification: we replace $f_{N+1}(\thb, u^i_C)$ with $f_{1}(\thb, u^1_V)$. Thus, for a particular value of the parameters we have 
\begin{equation}
    (Y_V)_m \sim \mathcal{N}(\mathbb{E}[ (Y_V)_m|\bv{Y}_p,\bv{\nu}_C ,\thb,\phb], \operatorname{Var}[ (Y_V)_m|\bv{Y}_p,\bv{\nu}_C ,\thb,\phb]),
\end{equation}
where the mean and the variance of the one-step ahead prediction distribution is given by
\begin{equation}
    \begin{aligned}
        \mathbb{E}[(Y_V)_m |\bv{Y}_p,\bv{\nu}_C ,\thb,\phb]&=f_m(\thb, u^m_V)+ \bv{\varphi}^T(\bv{Y}_p - \fp) + \bv{\zeta}^T\bv{\nu}_C,\\
        \operatorname{Var}[ (Y_V)_m|\bv{Y}_p,\bv{\nu}_C ,\thb,\phb] &= \operatorname{Var}\big[(Y_V)_m -\mathbb{E}[(Y_V)_m |\bv{Y}_p,\bv{\nu}_C ,\thb,\phb]\big]=\operatorname{Var}[\nu_N]=\tau^2.
    \end{aligned}
\end{equation}
In order to quantify the uncertainty in the predictions we can integrate out the model and noise parameters \cite{marriott1996bayesian}:
\begin{equation}\label{eq:pp-arma}
    p((Y_V)_m|\bv{Y}_p,\bv{\nu}_C)= \int \mathcal{N}(\mathbb{E}[ (Y_V)_m|\bv{Y}_p,\bv{\nu}_C ,\thb,\phb]p(\thb,\phb|\yc) \operatorname{d}\thb \operatorname{d}\phb, 
\end{equation}
where we use Monte Carlo integration as in Eq.~\ref{eq:ppc}.

In order to collect the full set of $M$ predictions $\yv$ we simply use the one-step-ahead forecasting distribution shown above in a recursive manner.

\clearpage
\section{Choice of priors for the ion channel example}\label{sec:priors}
Here we specify the choice of priors for the ion channel example.
\begin{itemize}
    \item For the ion channel ODE model parameters, we chose a uniform prior specified in Beattie \textit{et al.}~\cite{beattie2018sinusoidal} and Lei \textit{et al.}~\cite{lei_rapid_2019-1, lei_rapid_2019-2}.
    \item For the $\GP$ model, we have the unbiased noise parameter $\sigma$, the length-scale $\rho_i$, the marginal variance $\alpha$:
    \begin{itemize}
        \item $\sigma$: Half-Normal prior with standard deviation of \num{25};
        \item $\rho_i$: Inverse-Gamma prior with shape and scale being $(5, 5)$;
        \item $\alpha$: Inverse-Gamma prior with shape and scale being $(5, 5)$.
    \end{itemize}
    \item For the $\ARMA$ model, we have the autoregressive coefficients $\varphi_i$, and moving-average coefficients $\zeta_i$:
    \begin{itemize}
        \item $\varphi_i$: Normal prior centred on the maximum likelihood estimates with standard deviation of \num{2.5};
        \item $\zeta_i$: Normal prior centred on the maximum likelihood estimates with standard deviation of \num{2.5}.
    \end{itemize}
\end{itemize}

\clearpage
\section{Computing and representing posterior predictive}\label{sec:CI}
To compute the posterior predictive, we follow Girolami~\cite{girolami2008bayesian} and write the posterior predictive in Eqs.~\ref{eq:ppc} \&~\ref{eq:pp-arma} as
\begin{equation}
    p(Y_{V} \mid Y_{C}) = \sum_{k} p(Y_{V} \mid \boldsymbol{\theta}_k, \boldsymbol{\phi}_k, Y_{C}) p(\boldsymbol{\theta}_k, \boldsymbol{\phi}_k \mid Y_{C}),
\end{equation}
where $\boldsymbol{\theta}_k, \boldsymbol{\phi}_k$ are the $k$\textsuperscript{th} posterior sample of the parameters.
We have checked the posterior predictive of the ODE models at a given time point is symmetric and similar to a Gaussian distribution, for the sake of simplicity, we therefore use summary statistics such as the predictive mean and credible intervals computed using variance to represent the posterior predictive in this paper.
To obtain the predictive mean $\mathbb{E}[Y_{V} \mid Y_{C}]$ and variance $\operatorname{Var}[Y_{V} \mid Y_{C}]$, we use
\begin{align}
    \mathbb{E}[Y_{V} \mid Y_{C}] &= \sum_{k} \mathbb{E}[Y_{V} \mid \boldsymbol{\theta}_k, \boldsymbol{\phi}_k, Y_{C}] p(\boldsymbol{\theta}_k, \boldsymbol{\phi}_k \mid Y_{C}), \\
    \operatorname{Var}[Y_{V} \mid Y_{C}] &= \sum_{k} \left(\operatorname{Var}[Y_{V} \mid \boldsymbol{\theta}_k, \boldsymbol{\phi}_k, Y_{C}] + \mathbb{E}[Y_{V} \mid \boldsymbol{\theta}_k, \boldsymbol{\phi}_k, Y_{C}]^2\right)p(\boldsymbol{\theta}_k, \boldsymbol{\phi}_k \mid Y_{C}) - \mathbb{E}[Y_{V} \mid Y_{C}]^2.
\end{align}
Finally, to show the 95\% credible intervals of our predictions, we plot $\mathbb{E}[Y_{V} \mid Y_{C}] \pm 1.96\sigma_{Y_{V}}$ where $\sigma_{Y_{V}}^2 = \operatorname{Var}[Y_{V} \mid Y_{C}]$.

\clearpage

\section{Supplementary results for the action potential example}

\begin{figure}[htb]
	\centering
	\begin{tikzpicture}[every node/.style={anchor=north east,inner sep=0pt}, x=1mm, y=1mm]
        \node (fig1) at (0, 0)
        {\includegraphics[width=5in]{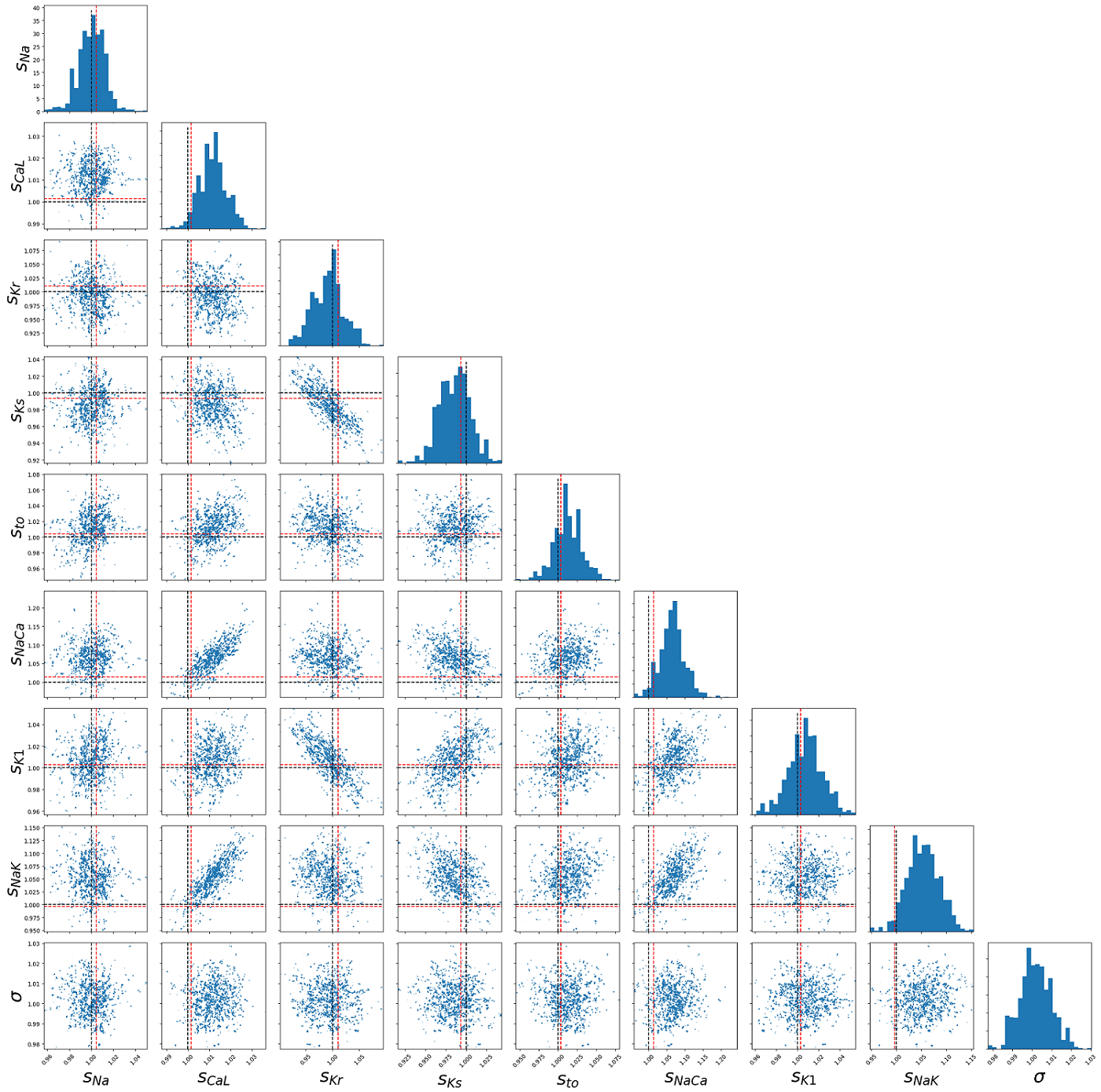}};
        \node (fig2) at (3, 3)
        {\includegraphics[width=3.25in]{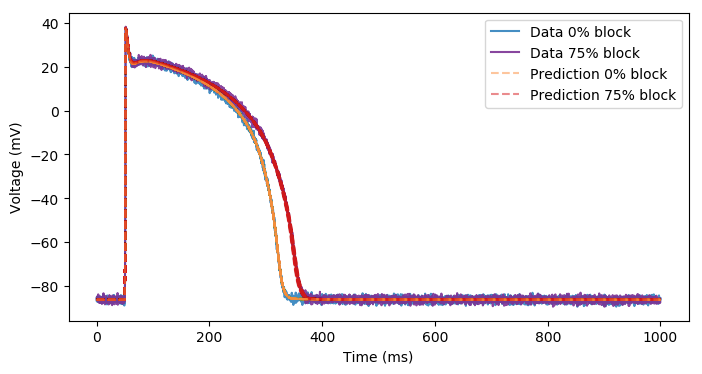}};
	\end{tikzpicture}
	\caption{
		\textbf{Matrix plot and histograms}: Posterior distribution of Model~T parameters when estimated using data from Model~T.
		The dashed black lines indicate the true (data-generating) parameters;
		the dashed red lines are the result of the global optimisation routine.
		\textbf{Inset plot}: Posterior predictions for the `context of use' (CoU) data, for the action potential model tutorial (in the scenario of no model discrepancy).
		The posterior predictions are model predictions using parameters sampled from the posterior distribution; 200 samples/predictions are shown. Model~T gives an almost-perfect prediction of the CoU data (which was not used in training).
	}
	\label{fig:tutorial-action-potential-no-discrepancy}
\end{figure}

\begin{figure}[htb]
    \centering
    \includegraphics[width=\textwidth]{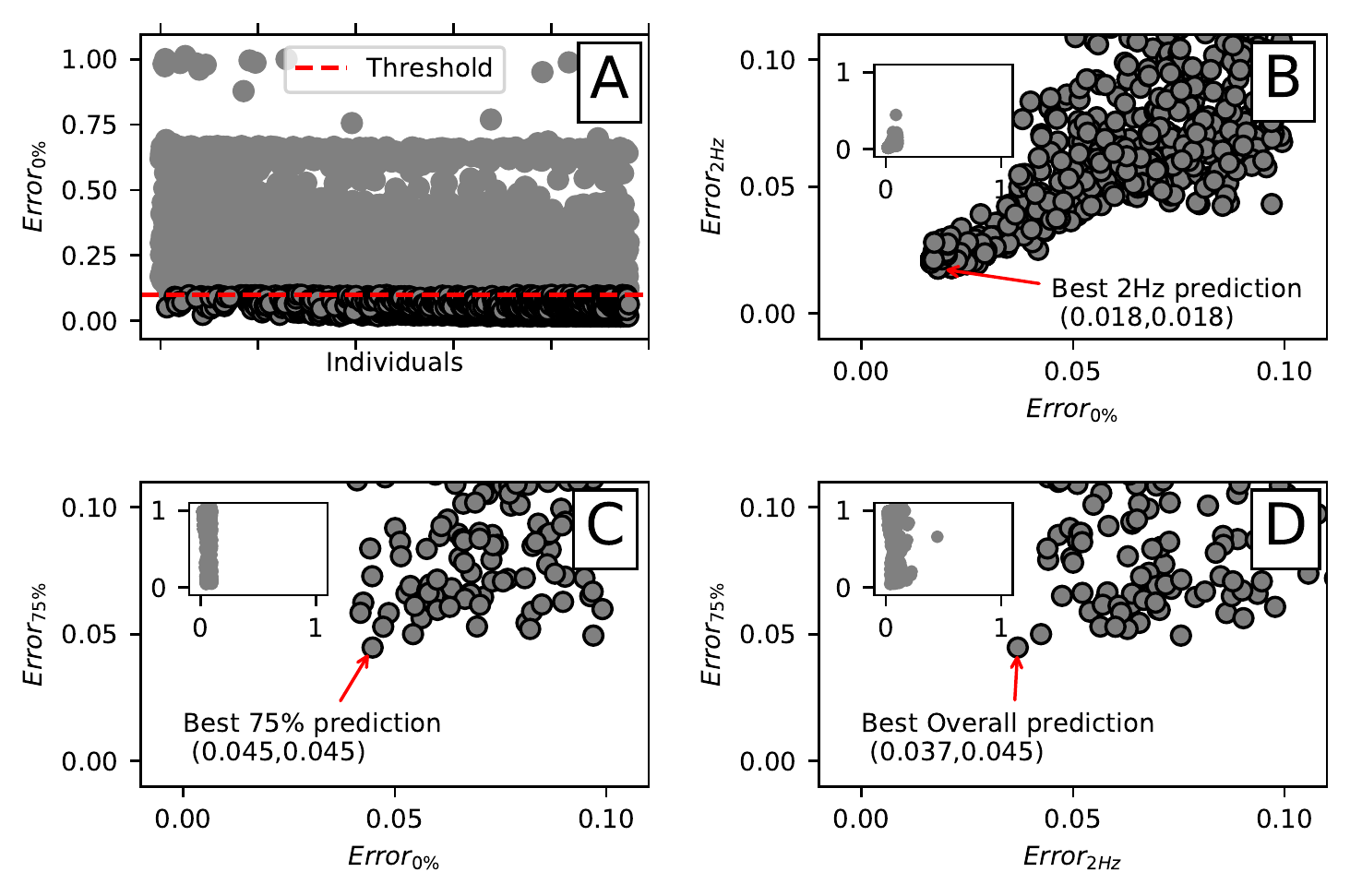}
    \caption{Results obtained via Genetic Algorithm and Differential Evolution to adjust the Fink Action Potential to ten Tusscher at control, i.e., \SI{1}{\hertz}, and 0\% IKr block configuration. The Genetic Algorithm was used with a population size of 100 individuals and 10 generations. The Differential Evolution was used with 150 individuals and 15 generations. Both algorithms were implemented using the Python library Pygmo with the standard configurations.     \textbf{A} From all the evaluations we selected the candidates (parameter sets) that satisfied $Error_{0\%} < 0.1$. Relative RMS errors are computed with $Error_{X}(i)=\frac{||T_X-{aF_X}(i)||_2}{||T_X||_2}$, where $T_X$ is the ten Tusscher model in scenario $X$, and $aF_X$ is the adjusted Fink model using the individual $i$ for scenario $X$. A total of 1079 candidates satisfied $Error_{0\%} < 0.1$, i.e., were below the displayed threshold. Using this metric, the best candidate had $Error_{0\%} = 1.6\%$.
    \textbf{B} Testing the performance of the 1079 candidates with respect to the 2Hz scenario a total of 990 candidates satisfy $Error_{0\%} < 0.1$ and $Error_{2Hz} < 0.1$. Using these two metrics, the best candidate had $Error_{0\%} = 1.8\%$ and $Error_{2Hz} = 1.8\%$.
    \textbf{C} Testing the performance of the 1079 candidates with respect to the 75\% IKr block scenario only 80 candidates satisfy $Error_{0\%} < 0.1$ and $Error_{75\%} < 0.1$. Using these two metrics, the best candidate had $Error_{0\%} = 4.5\%$ and $Error_{75\%} = 4.5\%$.
    \textbf{D} Testing the performance of the 1079 candidates with respect to both 75\% IKr block and 2Hz scenarios only 70 candidates satisfy $Error_{0\%} < 0.1$, $Error_{75\%} < 0.1$, and $Error_{2Hz} < 0.1$. Using the three metrics, the best candidate had $Error_{0\%} = 4.5\%$, $Error_{2Hz} = 3.7\%$, and $Error_{75\%} = 4.5\%$.
}
    \label{fig:tradeoff1}
\end{figure}

\begin{figure}
    \centering
    \includegraphics[width=9cm]{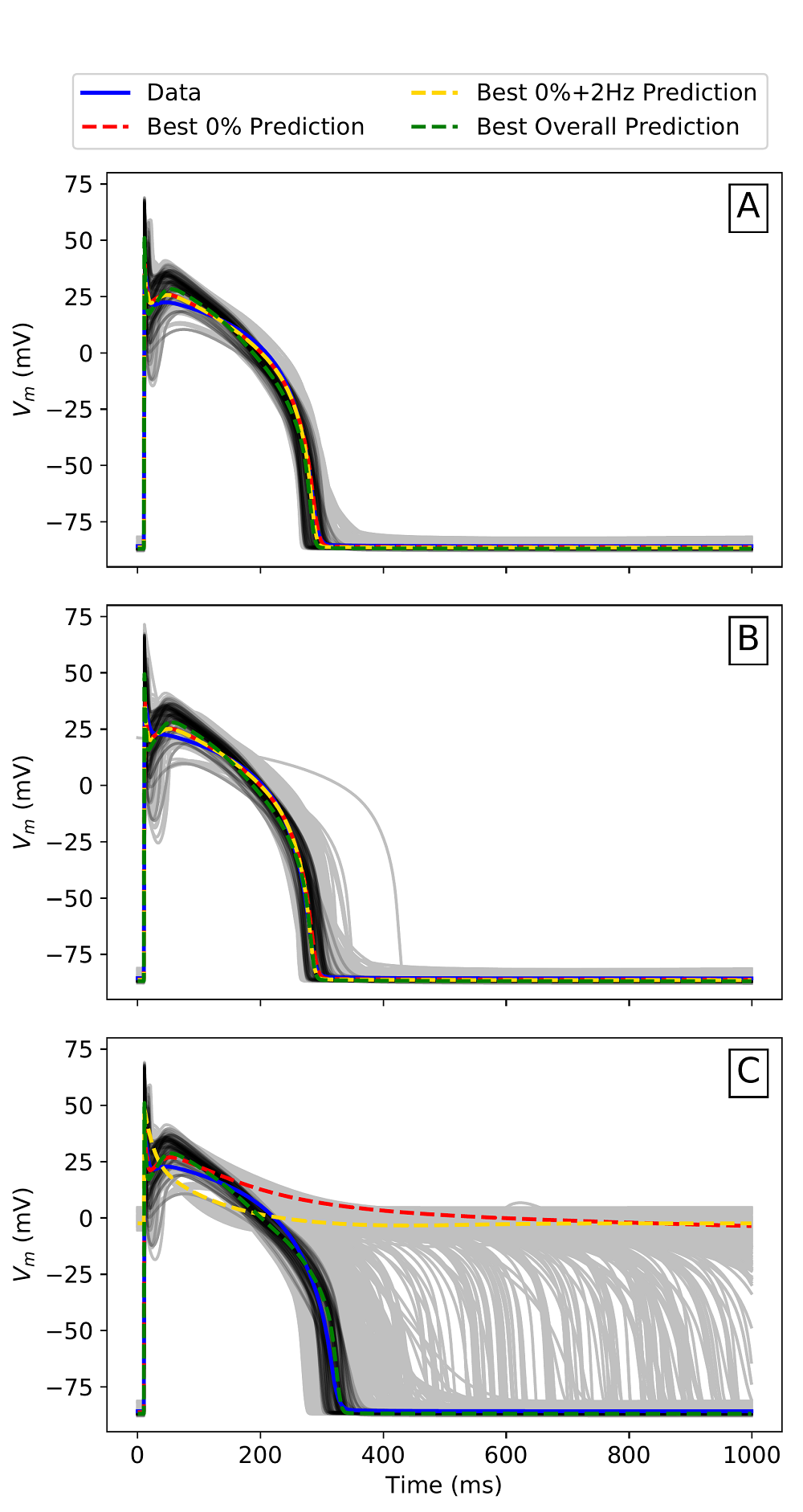}
    \caption{APs obtained by the Data (ten Tusscher model) and also by the fitting process for each scenario: \textbf{A} control, i.e., \SI{1}{\hertz}, and 0\% IKr block; \textbf{B} \SI{2}{\hertz}; and \textbf{C} 75\% IKr block. 
    ``Best $0\%$ prediction'' are the results obtained using the best candidate that satisfies $Error_{0\%} < 0.1$. 
    ``Best $0\% + 2Hz$ prediction'' are results obtained using the best candidate that satisfies $Error_{0\%} < 0.1$ and $Error_{2Hz} < 0.1$. 
    ``Best overall prediction'' are results obtained using the best candidate that satisfies $Error_{0\%} < 0.1$, $Error_{2Hz} < 0.1$, and $Error_{75\%} < 0.1$. 
    The 1079 APs that satisfy $Error_{0\%} < 0.1$ are plotted with grey lines. 
    The 70/1079 APs that satisfy $Error_{0\%} < 0.1$, $Error_{2Hz} < 0.1$, and $Error_{75\%} < 0.1$ are are plotted with black lines. 
    Note that there is no way of knowing in advance what the best candidate parameter set will be without performing the experiment, so a distribution of possibilities should generally be shown.
    }
    \label{fig:tradeoff2}
\end{figure}

\clearpage

\section{Supplementary results for the ion channel discrepancy example}\label{sec:ion-channel-supp}

\begin{figure}[htb]
	\centering
	\includegraphics[width=\textwidth]{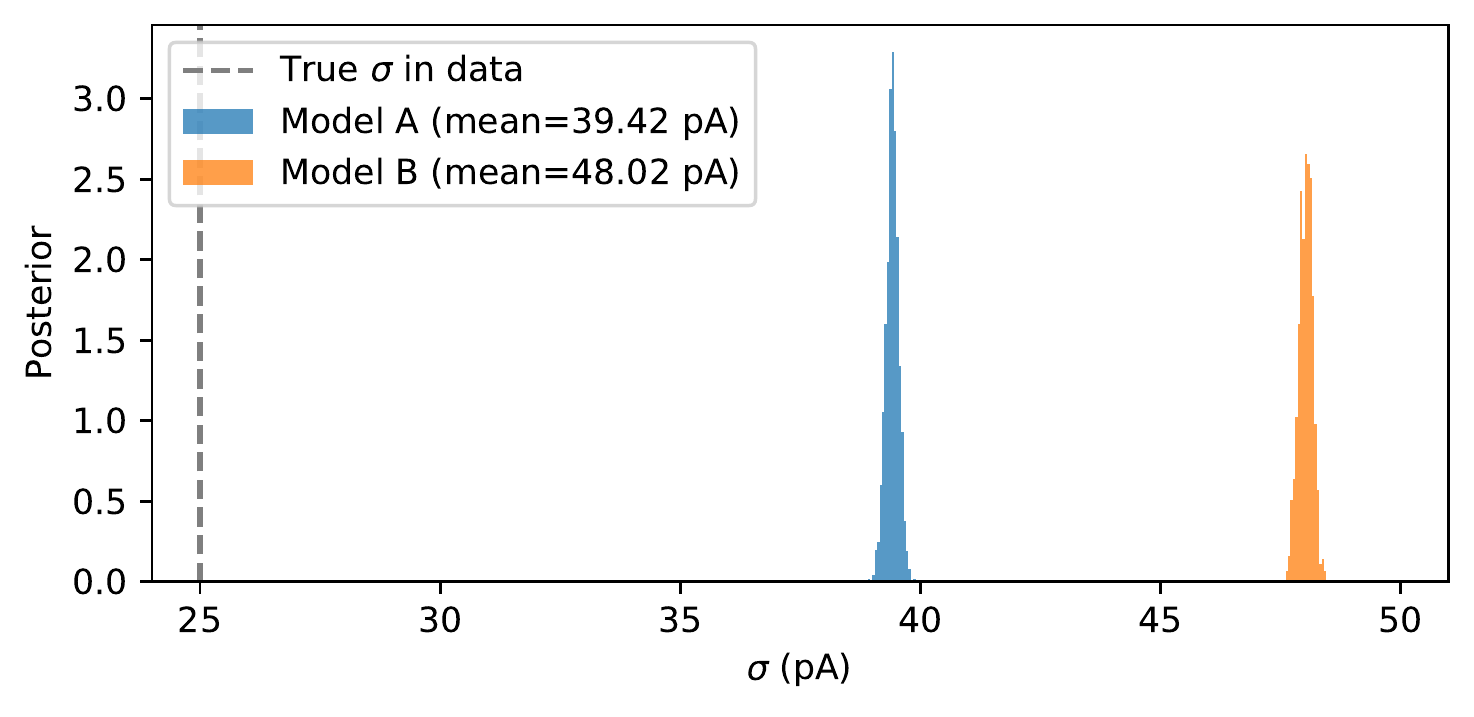}
	\caption{\label{fig:iid-sigma-comparison}
		A comparison of the $\sigma$ values in the i.i.d.\ model for Model~A and~B, and $\sigma_\text{true}$ refers to the value used in generating the data with Model~C.
		If we consider the inferred $\sigma$ value in Eq.~\ref{main-eq:ll} in the main text as $\sigma^2_\text{inferred} = \sigma^2_\text{true} + \sigma^2_\text{discrepancy}$, then we can see that both $\sigma_\text{A}, \sigma_\text{B} > \sigma_\text{true}$.
		Hence we have $\sigma^2_\text{discrepancy}$ term is non-zero for both models, which reflects the fact that there is discrepancy for both models.
		One may use the size of $\sigma_\text{inferred}$ to interpret the size of the model discrepancy here.
	}
\end{figure}

\begin{table}[htb]
	\centering
	\includegraphics[width=0.495\textwidth]{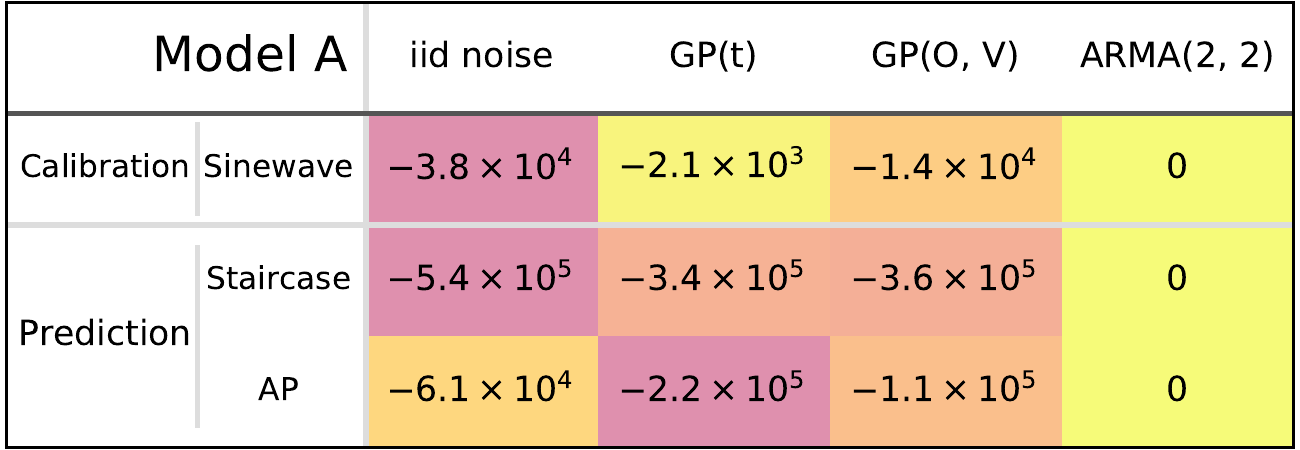}
	\includegraphics[width=0.495\textwidth]{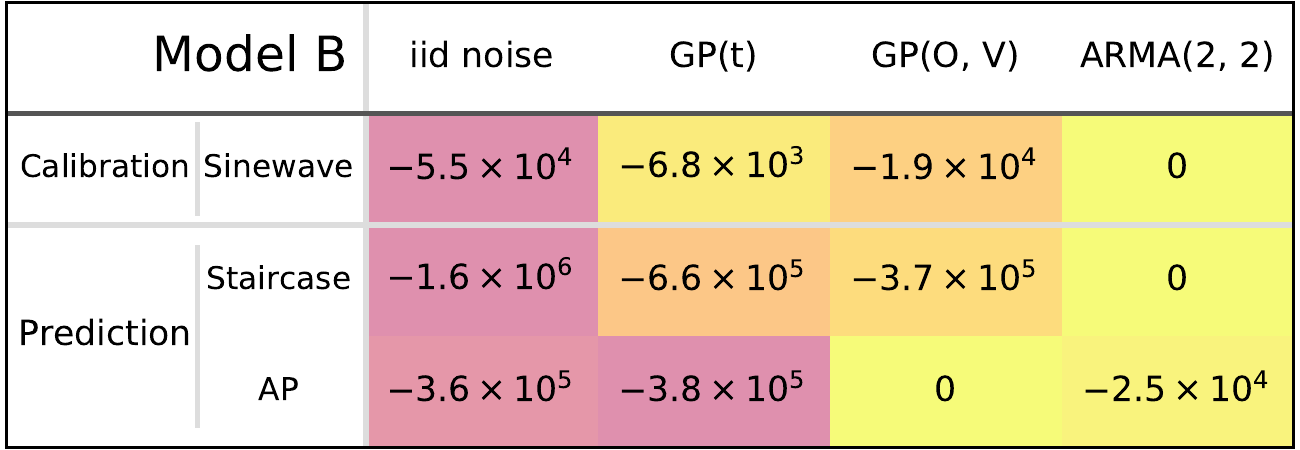}
	\caption{
		Posterior predictive log-likelihoods of fits and predictions for Models~A (left) and B (right) with different discrepancy models: i.i.d.\ noise, $\GP(t)$, $\GP(O, V)$, and $\ARMA(2, 2)$ for all three voltage protocols. The posterior predictive log-likelihood is $\pi(Y\mid Y_C) = \int \pi(Y\mid \theta) \pi(\theta \mid Y_C) {\rm d} \theta$, which we approximate by $\frac{1}{N}\sum_{n=1}^N p(Y \mid \theta_n)$ where $\theta_n$ are samples from the posterior distribution generated by MCMC.
        Only relative differences within a row are meaningful, and we therefore subtract the maximum log-likehood for each dataset from the results giving the best model in each row a score of zero. Note that care is needed when interpreting the log-likelihood values for the GP models due to  the FITC approximation used to approximate the full likelihood.
            }
		\label{fig:tutorial-ion-channel-likelihood-table}
\end{table}

\clearpage
\subsection{Model~A}

\begin{figure}[htb]
	\centering
	\includegraphics[width=\textwidth]{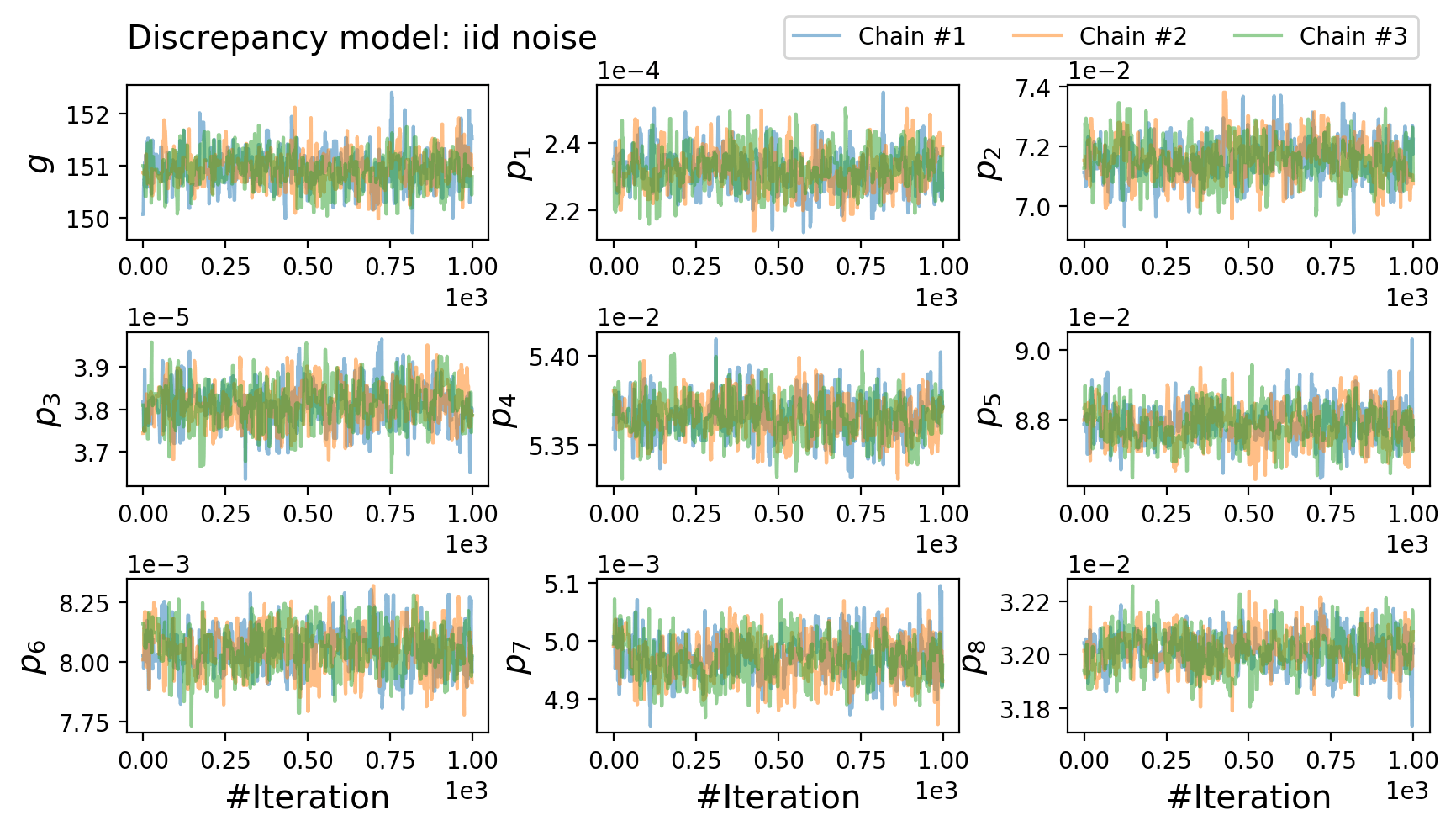}
	\caption{\label{fig:tutorial-ion-channel-posteriors-b}
		Trace plot of 3 independent MCMC runs for Model~A parameters (with the i.i.d.\ noise model): the conductance, $g$, and kinetic parameters $p_1,\dots,p_{8}$ (a list of parameters referring to $A_{i,j}$ and $B_{i,j}$ in Eq.~(\ref{main-eq:rate_equation})).
	}
\end{figure}

\begin{figure}[htb]
	\centering
	\includegraphics[width=\textwidth]{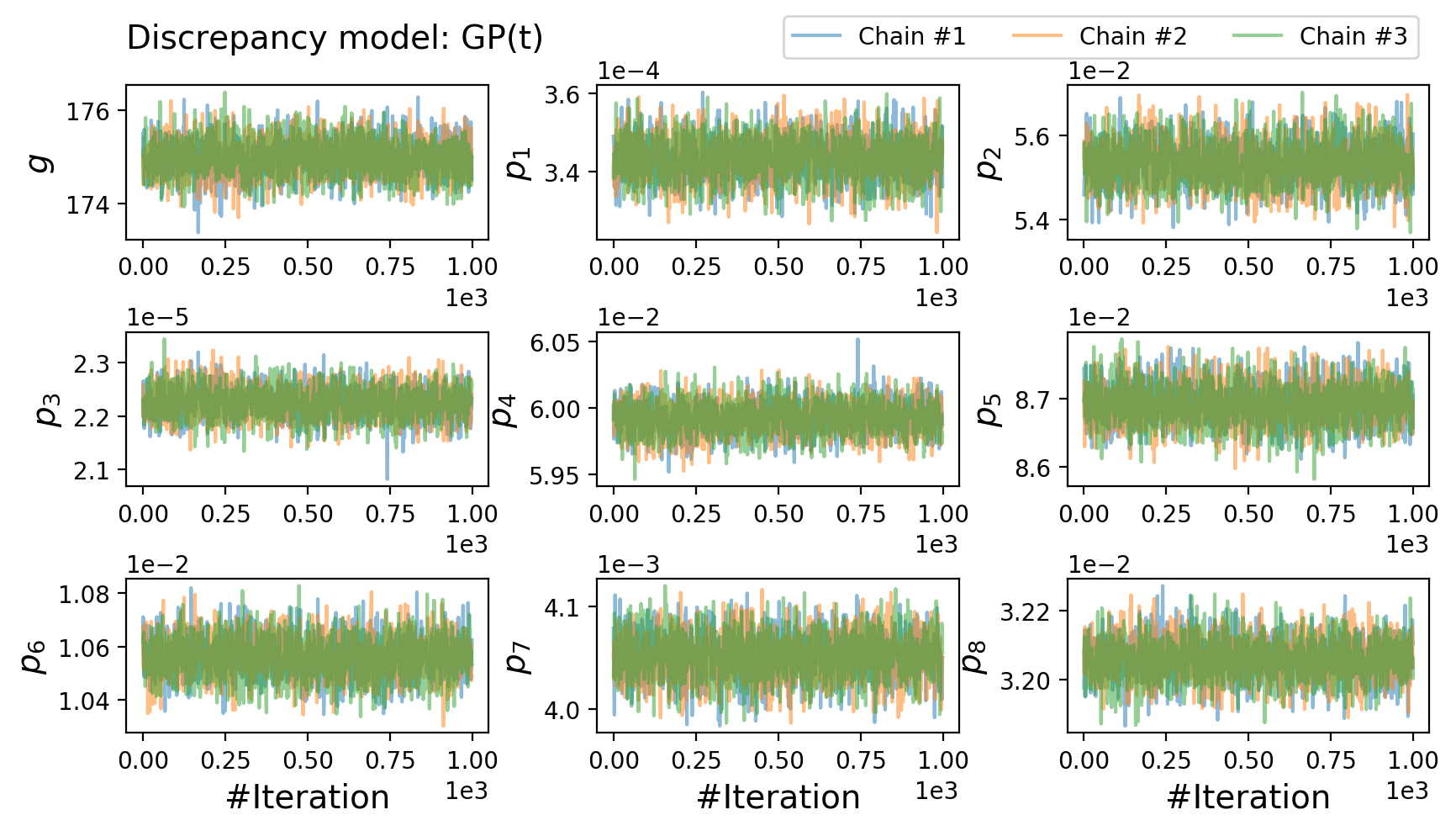}
	\caption{\label{fig:tutorial-ion-channel-posteriors-b}
		Trace plot of 3 independent MCMC runs for Model~A parameters (with the $\GP(t)$ noise discrepancy model): the conductance, $g$, and kinetic parameters $p_1,\dots,p_{8}$ (a list of parameters referring to $A_{i,j}$ and $B_{i,j}$ in Eq.~(\ref{main-eq:rate_equation})).
	}
\end{figure}

\begin{figure}[htb]
	\centering
	\includegraphics[width=\textwidth]{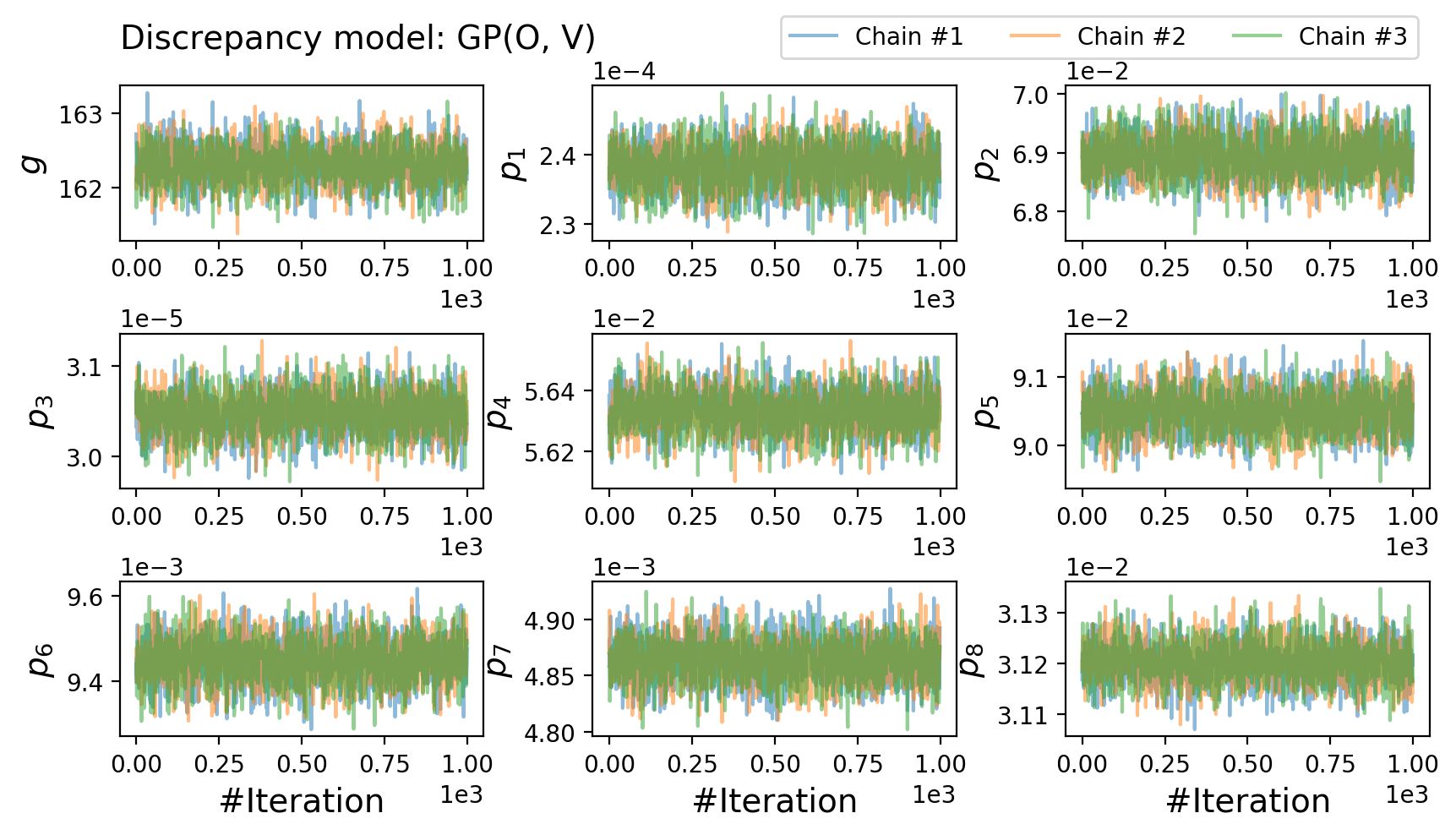}
	\caption{\label{fig:tutorial-ion-channel-posteriors-b}
		Trace plot of 3 independent MCMC runs for Model~A parameters (with the $\GP(O, V)$ noise discrepancy model): the conductance, $g$, and kinetic parameters $p_1,\dots,p_{8}$ (a list of parameters referring to $A_{i,j}$ and $B_{i,j}$ in Eq.~(\ref{main-eq:rate_equation})).
	}
\end{figure}

\begin{figure}[htb]
	\centering
	\includegraphics[width=\textwidth]{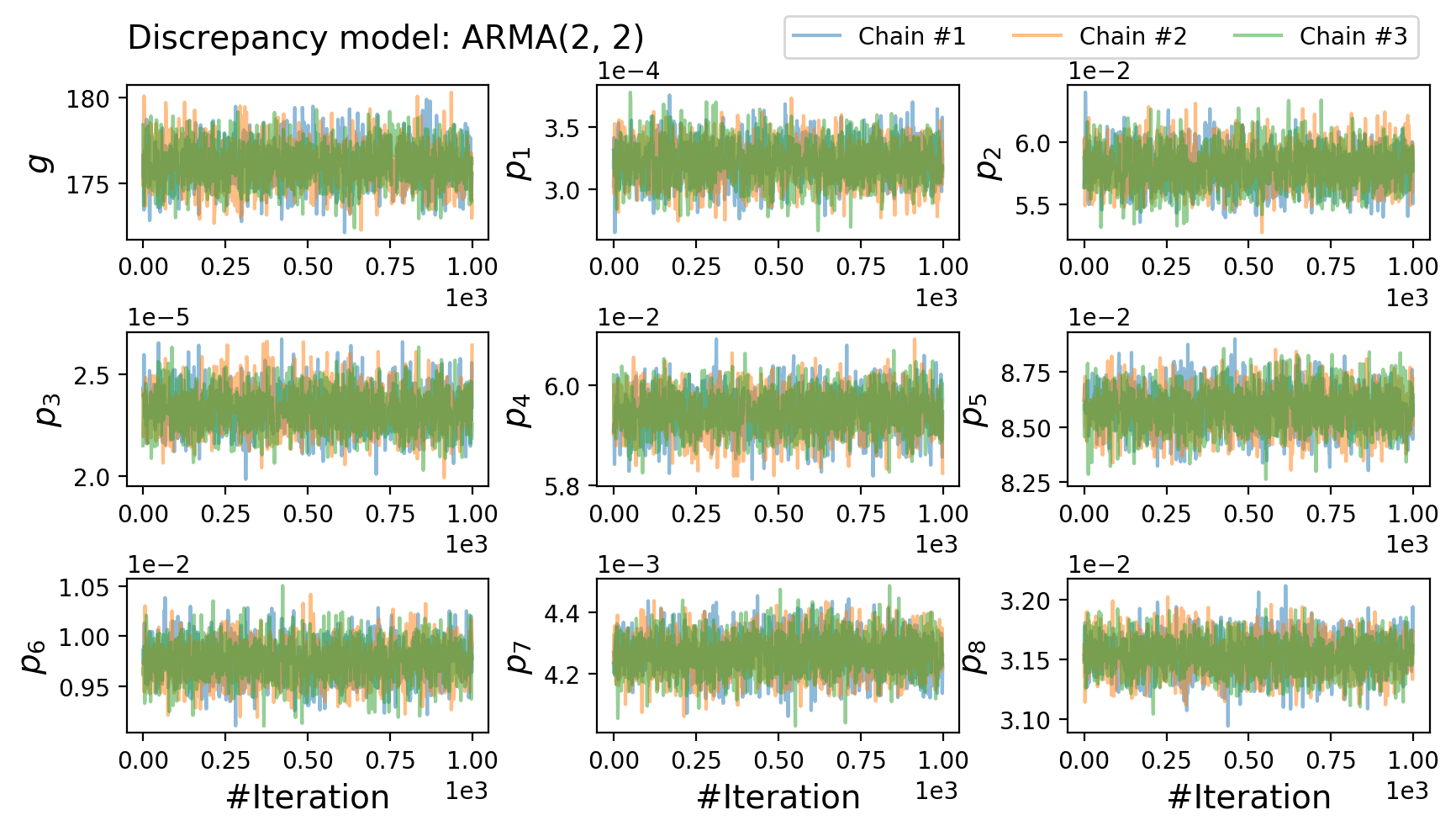}
	\caption{\label{fig:tutorial-ion-channel-posteriors-b}
		Trace plot of 3 independent MCMC runs for Model~A parameters (with the $\ARMA(2, 2)$ noise discrepancy model): the conductance, $g$, and kinetic parameters $p_1,\dots,p_{8}$ (a list of parameters referring to $A_{i,j}$ and $B_{i,j}$ in Eq.~(\ref{main-eq:rate_equation})).
	}
\end{figure}

\clearpage
\subsubsection{Model~A: Full model predictions}

\begin{figure}[htb]
	\centering
	\includegraphics[width=\textwidth]{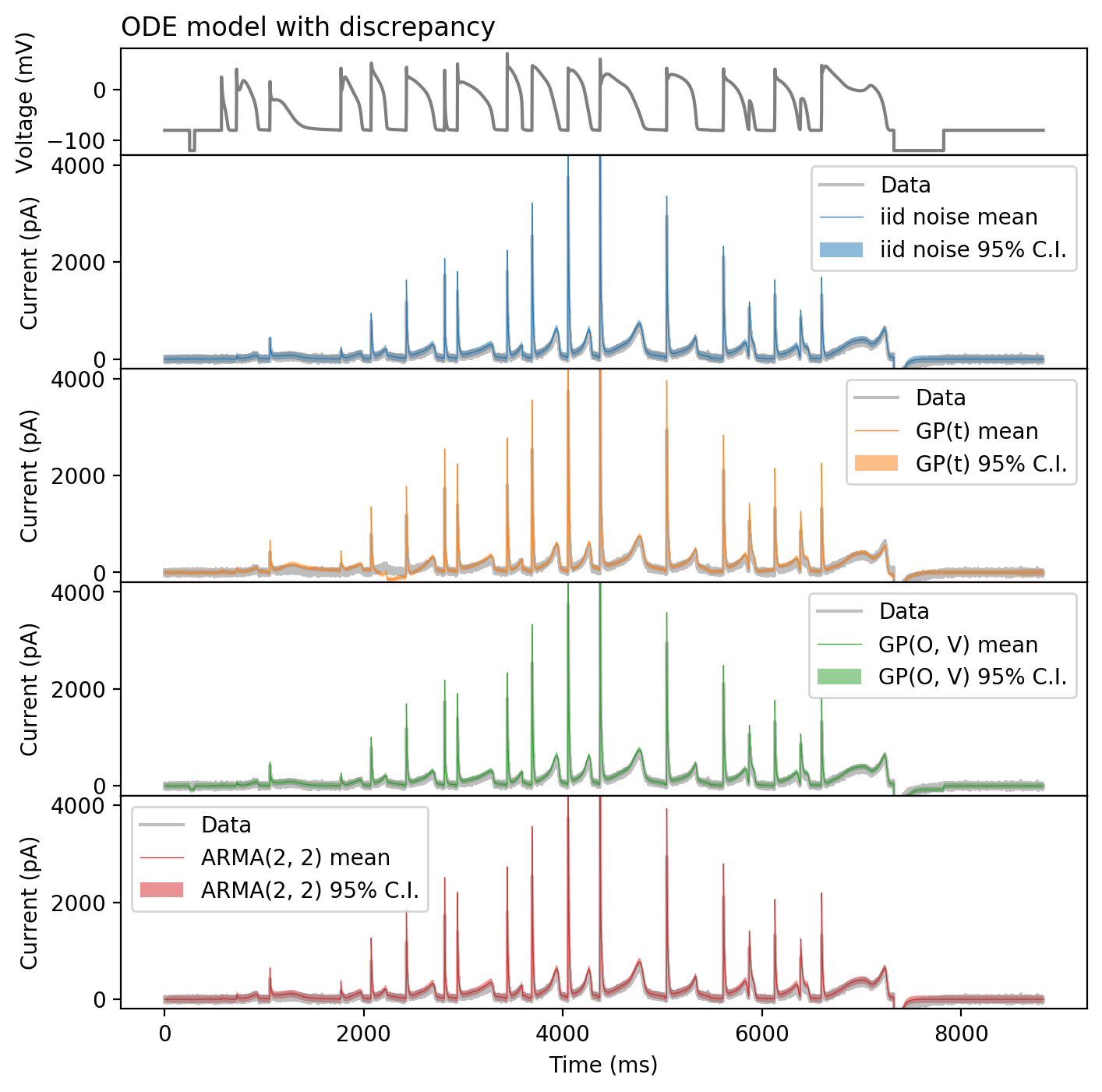}
	\caption{\label{fig:tutorial-ion-channel-prediction2a}
		Model~A prediction with different discrepancy models: no discrepancy (i.i.d.\ noise), $\GP(t)$, $\GP(O, V)$, and $\ARMA(2, 2)$.
		The voltage clamp protocol for calibration is the action potential series protocol \cite{beattie2018sinusoidal}.
	}
\end{figure}

\clearpage
\subsubsection{Model~A: Discrepancy predictions}

\begin{figure}[htb]
	\centering
	\includegraphics[width=\textwidth]{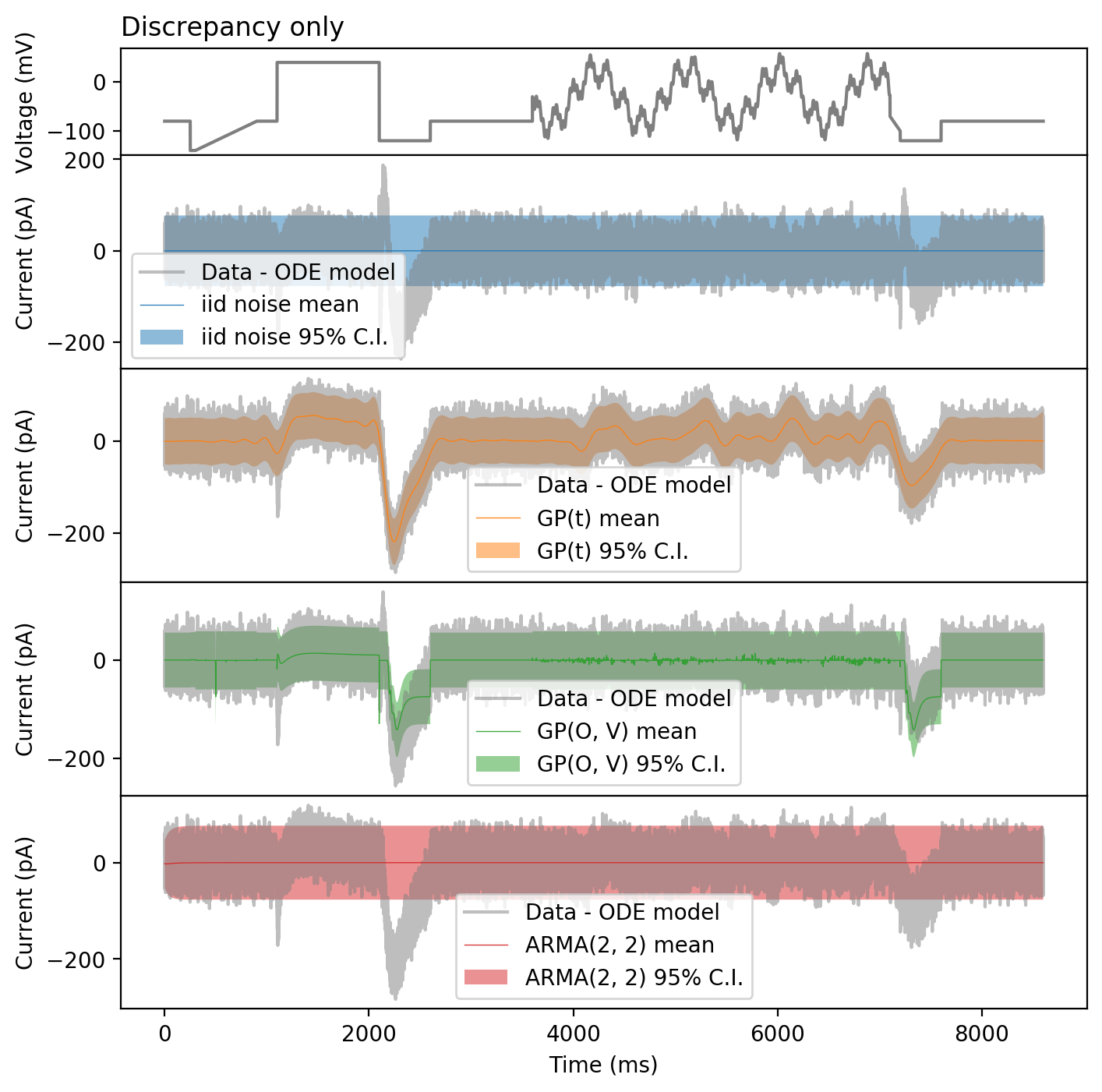}
	\caption{\label{fig:tutorial-ion-channel-fitting2-residual}
		Model~A fitting residuals of the MAP estimate accounted by different discrepancy models: no discrepancy (i.i.d.\ noise), $\GP(t)$, $\GP(O, V)$, and $\ARMA(2, 2)$.
		The voltage clamp protocol for calibration is the sinusoidal protocol \cite{beattie2018sinusoidal}.
	}
\end{figure}

\begin{figure}[htb]
	\centering
	\includegraphics[width=\textwidth]{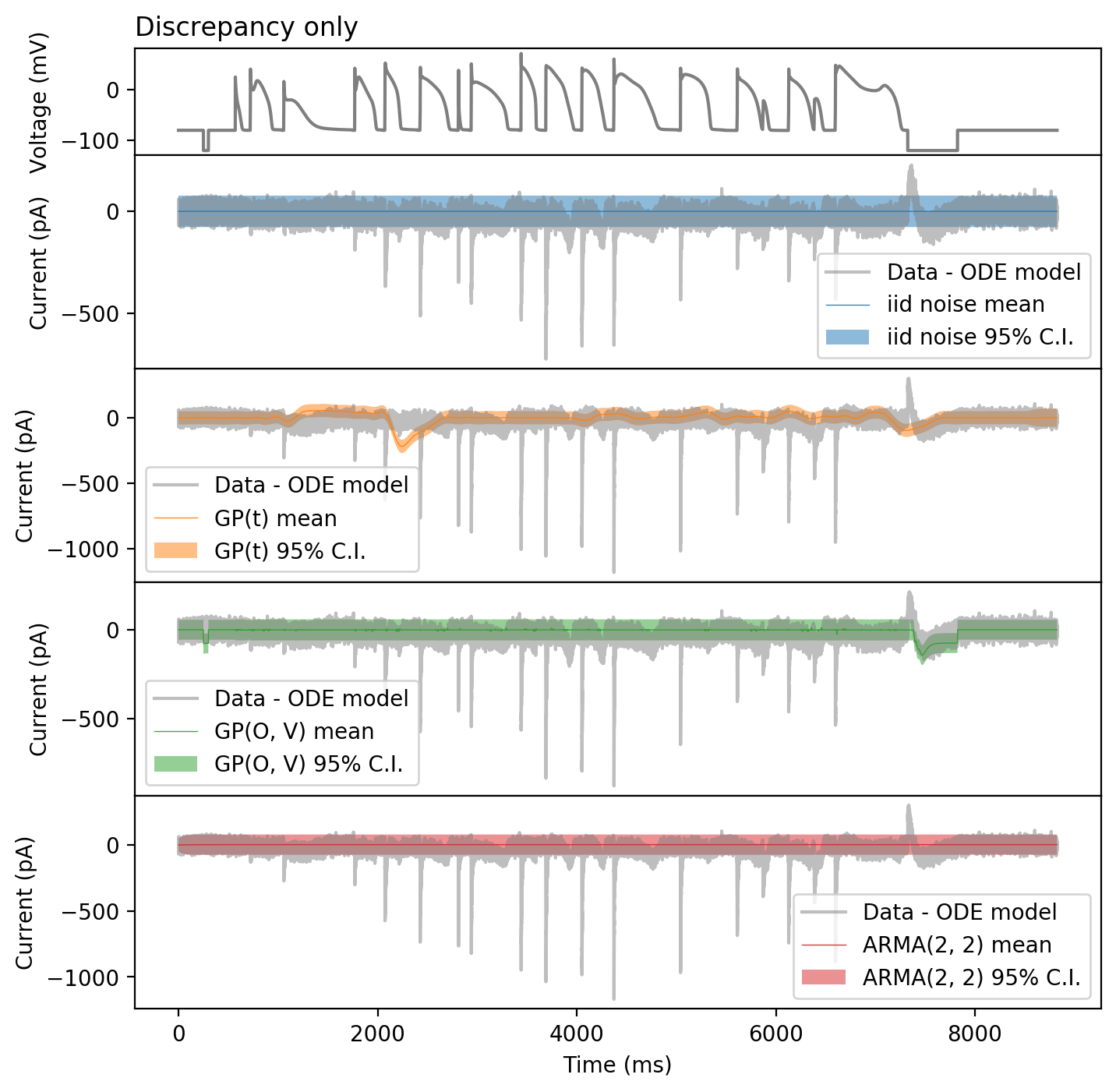}
	\caption{\label{fig:tutorial-ion-channel-prediction2a-residual}
		Model~A prediction residuals of the MAP estimate accounted by different discrepancy models: no discrepancy (i.i.d.\ noise), $\GP(t)$, $\GP(O, V)$, and $\ARMA(2, 2)$.
		The voltage clamp protocol for calibration is the action potential series protocol \cite{beattie2018sinusoidal}.
	}
\end{figure}

\begin{figure}[htb]
	\centering
	\includegraphics[width=\textwidth]{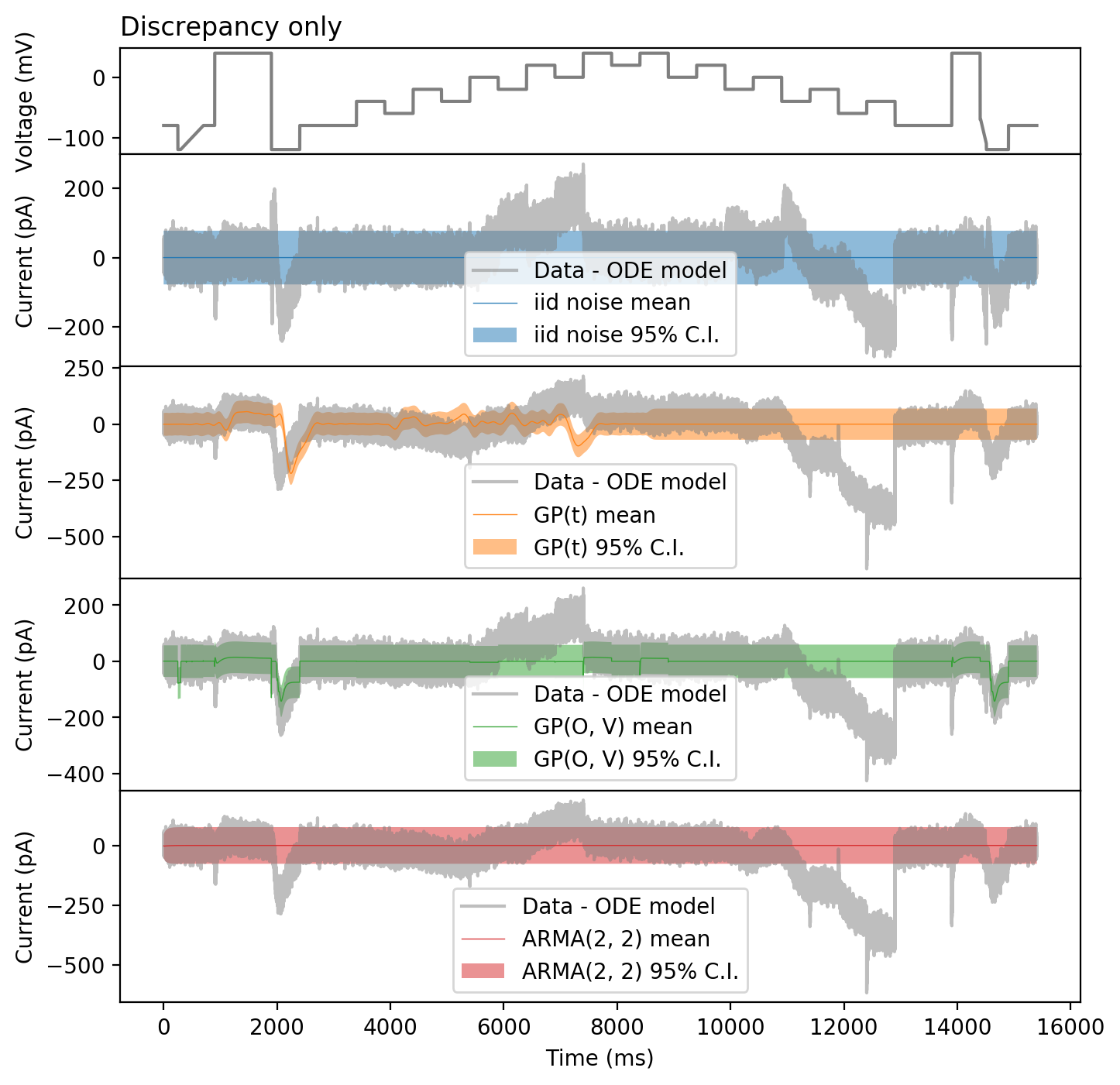}
	\caption{\label{fig:tutorial-ion-channel-prediction2b-residual}
		Model~A prediction residuals of the MAP estimate accounted by different discrepancy models: no discrepancy (i.i.d.\ noise), $\GP(t)$, $\GP(O, V)$, and $\ARMA(2, 2)$.
		The voltage clamp protocol for calibration is the staircase protocol \cite{lei_rapid_2019-1}.
	}
\end{figure}

\clearpage
\subsubsection{Model~A: ODE model predictions}

\begin{figure}[htb]
	\centering
	\includegraphics[width=\textwidth]{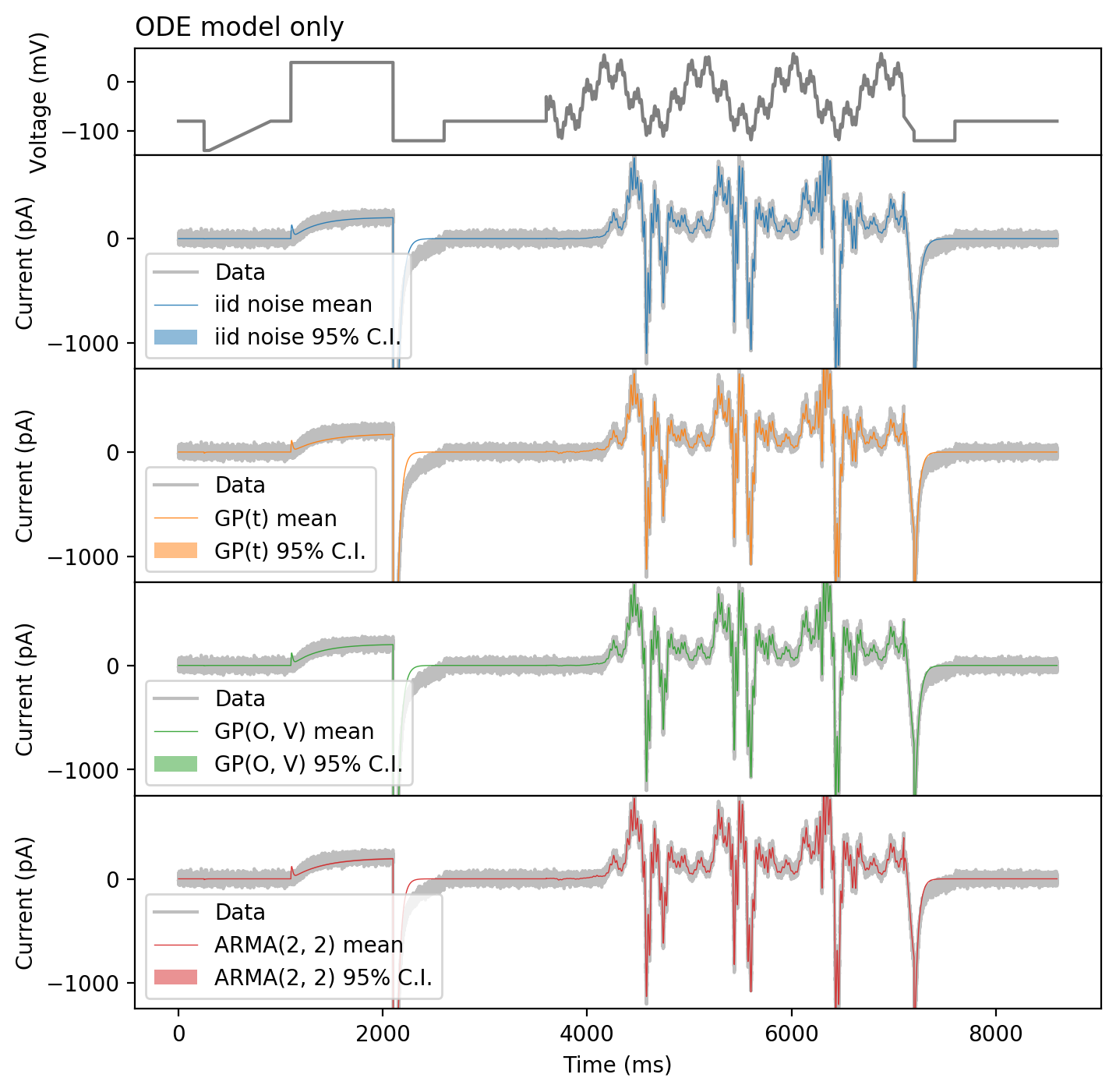}
	\caption{\label{fig:tutorial-ion-channel-fitting2-model}
		Fitting of the ODE model of Model~A, using different discrepancy models: no discrepancy (i.i.d.\ noise), $\GP(t)$, $\GP(O, V)$, and $\ARMA(2, 2)$.
		The voltage clamp protocol for calibration is the sinusoidal protocol \cite{beattie2018sinusoidal}.
	}
\end{figure}

\begin{figure}[htb]
	\centering
	\includegraphics[width=\textwidth]{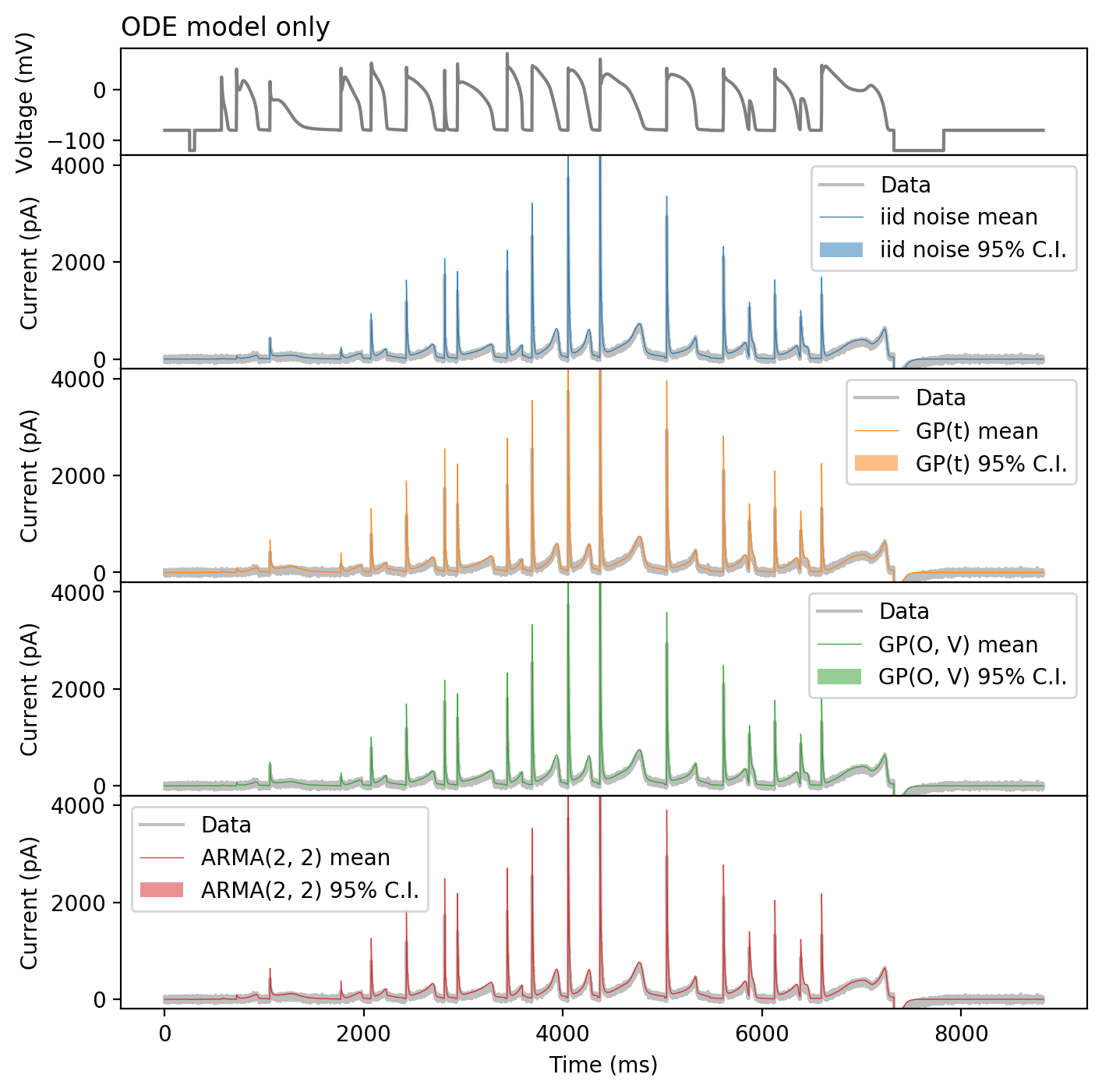}
	\caption{\label{fig:tutorial-ion-channel-prediction2a-model}
		Predictions of the ODE model of Model~A, using different discrepancy models: no discrepancy (i.i.d.\ noise), $\GP(t)$, $\GP(O, V)$, and $\ARMA(2, 2)$.
		The voltage clamp protocol for calibration is the action potential series protocol \cite{beattie2018sinusoidal}.
	}
\end{figure}

\begin{figure}[htb]
	\centering
	\includegraphics[width=\textwidth]{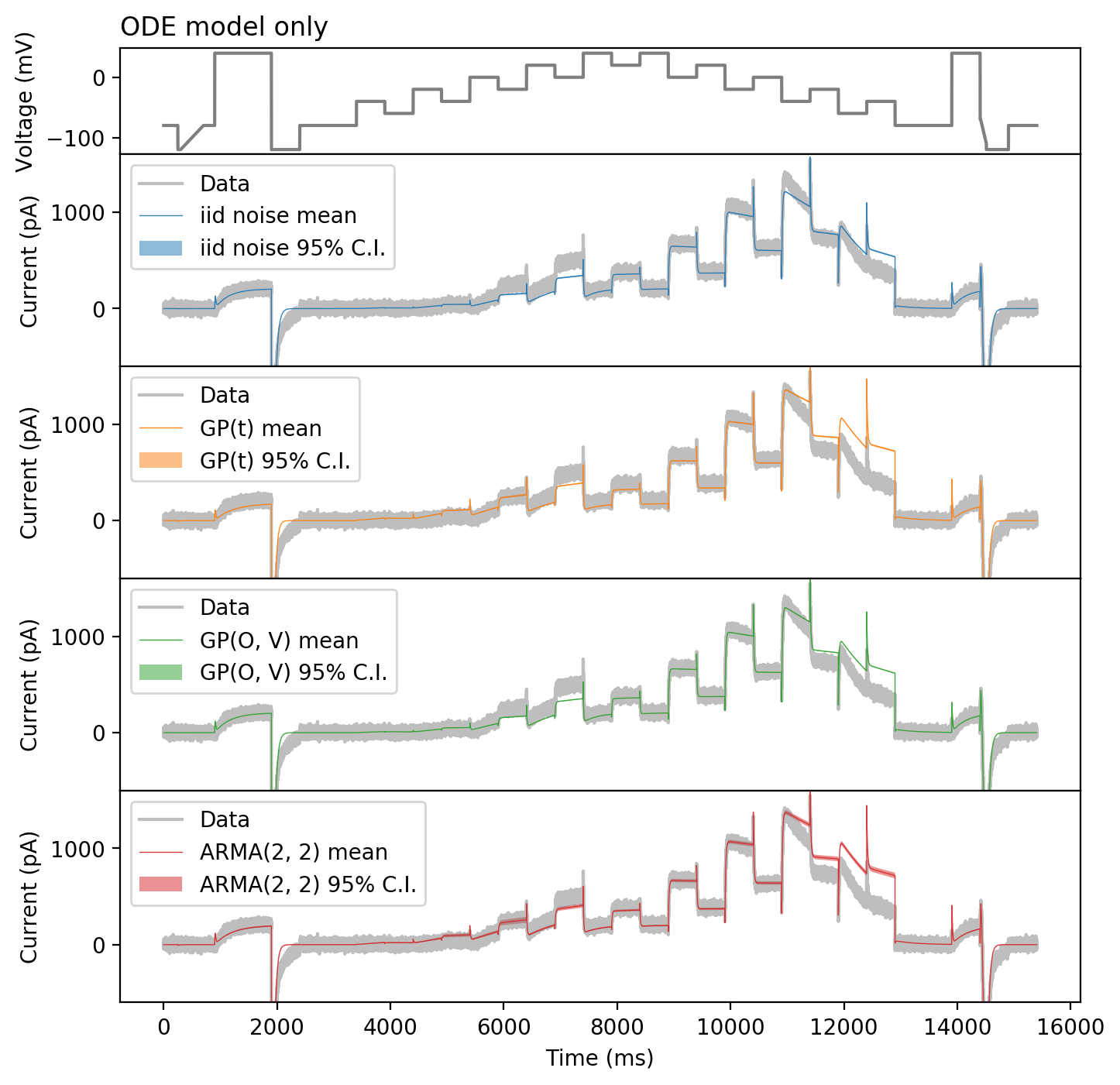}
	\caption{\label{fig:tutorial-ion-channel-prediction2b-model}
		Predictions of the ODE model of Model~A, using different discrepancy models: no discrepancy (i.i.d.\ noise), $\GP(t)$, $\GP(O, V)$, and $\ARMA(2, 2)$.
		The voltage clamp protocol for calibration is the staircase protocol \cite{lei_rapid_2019-1}.
	}
\end{figure}

\clearpage
\subsection{Model~B}
\label{sub:ModelBFullResults}

\begin{figure}[htb]
	\centering
	\includegraphics[width=\textwidth]{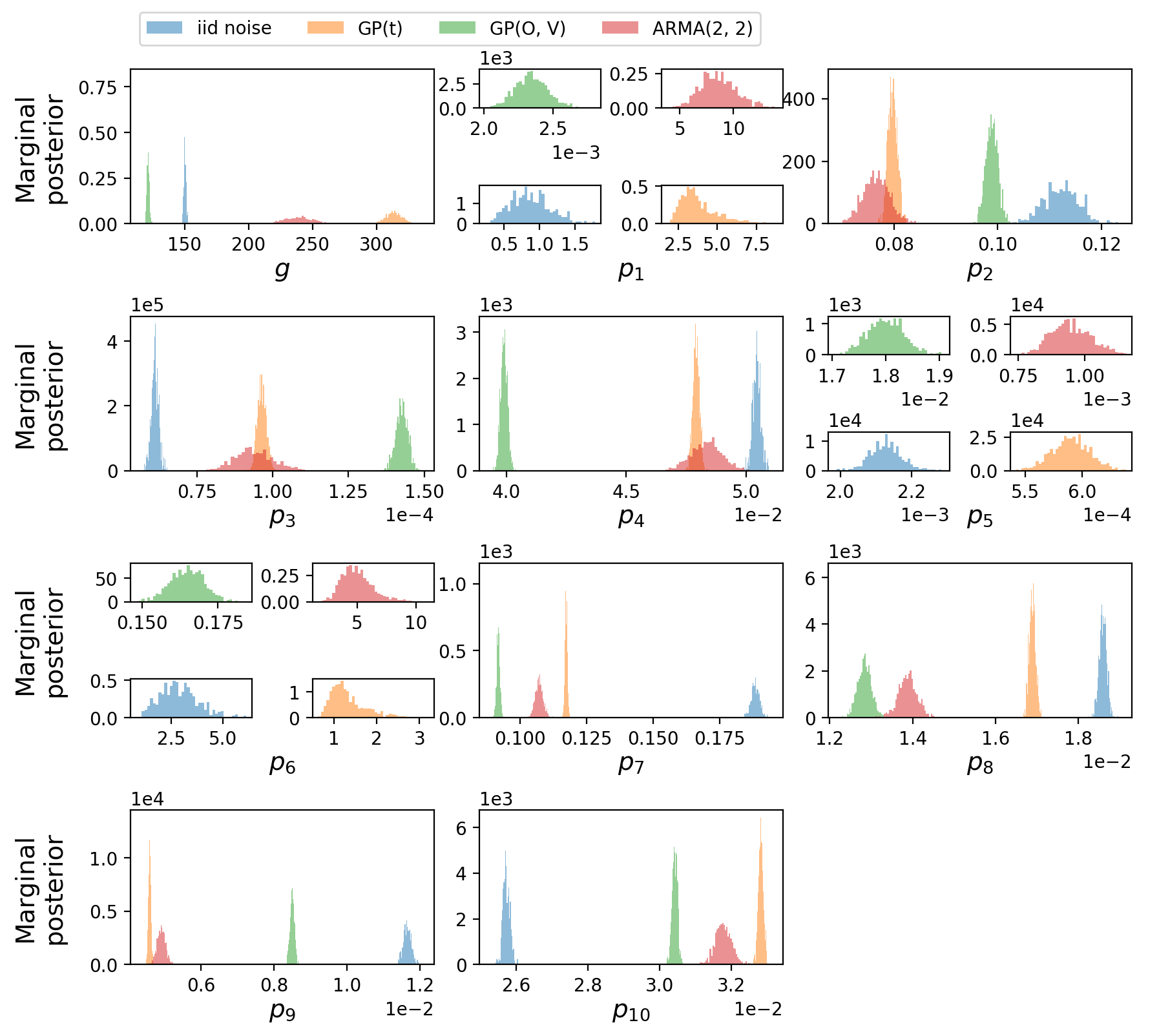}
	\caption{\label{fig:tutorial-ion-channel-posteriors-b}
		Model~B inferred marginal posterior distributions for the conductance, $g$, and kinetic parameters $p_1,\dots,p_{10}$ (a list of parameters referring to $A_{i,j}$ and $B_{i,j}$ in Eq.~(\ref{main-eq:rate_equation})) with different discrepancy models: i.i.d.\ noise (blue), $\GP(t)$ (orange), $\GP(O, V)$ (green), and $\ARMA(2, 2)$ (red).
			}
\end{figure}

\clearpage
\subsubsection{Model~B: Full model predictions}

\begin{figure}[htb]
  \centering
  \includegraphics[width=\textwidth]{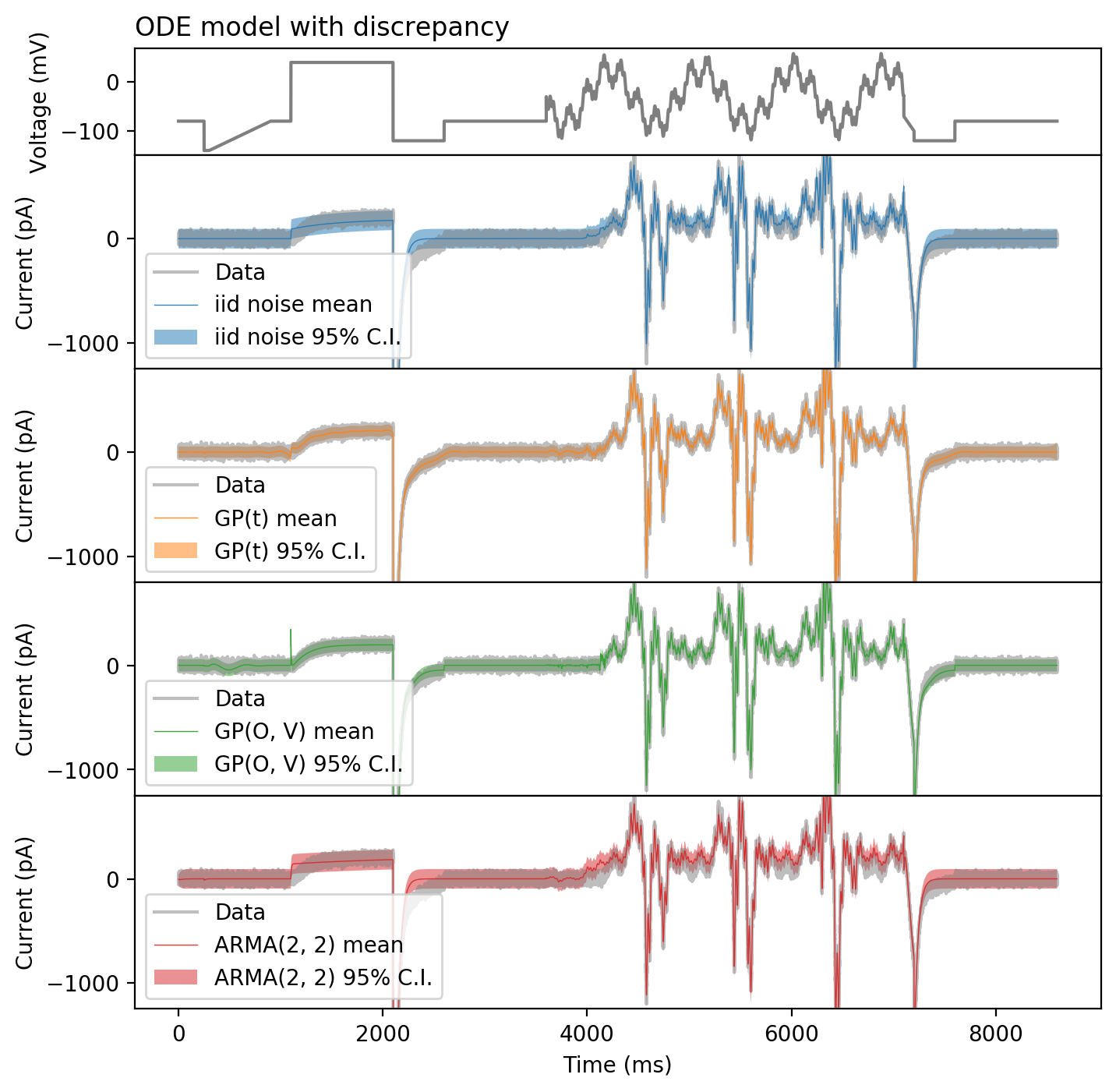}
  \caption{\label{fig:tutorial-ion-channel-fitting2-b}
    Model~B fitting results with different discrepancy models: i.i.d.\ noise, $\GP(t)$, $\GP(O, V)$, and $\ARMA(2, 2)$.
    The voltage clamp protocol for calibration is the sinusoidal protocol \cite{beattie2018sinusoidal}.
    It shows the posterior predictive with the bounds showing the 95\% credible interval.
  }
\end{figure}

\begin{figure}[tbh]
  \centering
  \includegraphics[width=\textwidth]{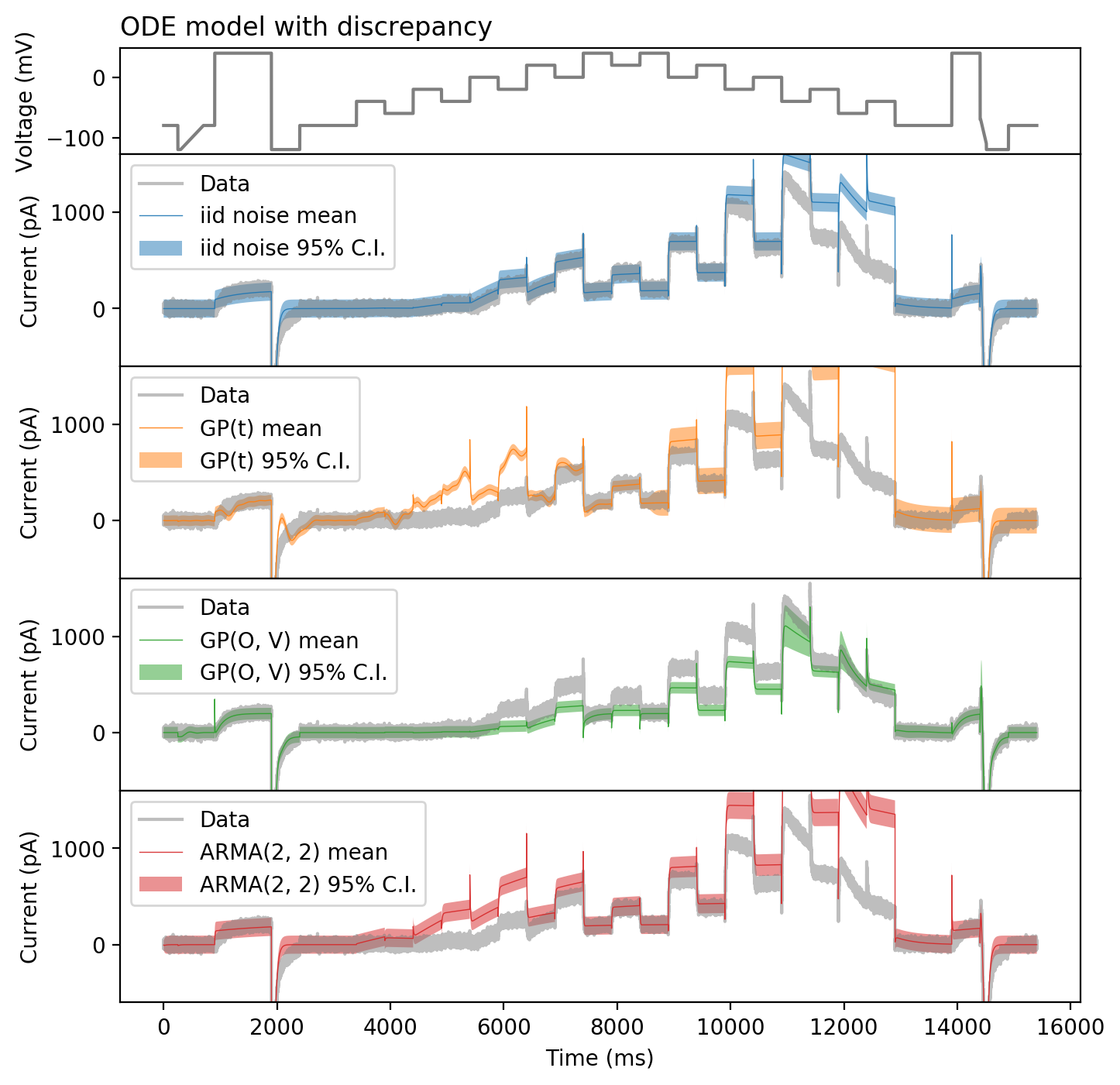}
  \caption{\label{fig:tutorial-ion-channel-prediction2-b}
    Model~B prediction with different discrepancy models: i.i.d.\ noise, $\GP(t)$, $\GP(O, V)$, and $\ARMA(2, 2)$.
    The voltage clamp protocol for calibration is the staircase protocol \cite{lei_rapid_2019-1}.
    It shows the posterior predictive with the bounds showing the 95\% credible interval.
      }
\end{figure}

\begin{figure}[htb]
  \centering
  \includegraphics[width=\textwidth]{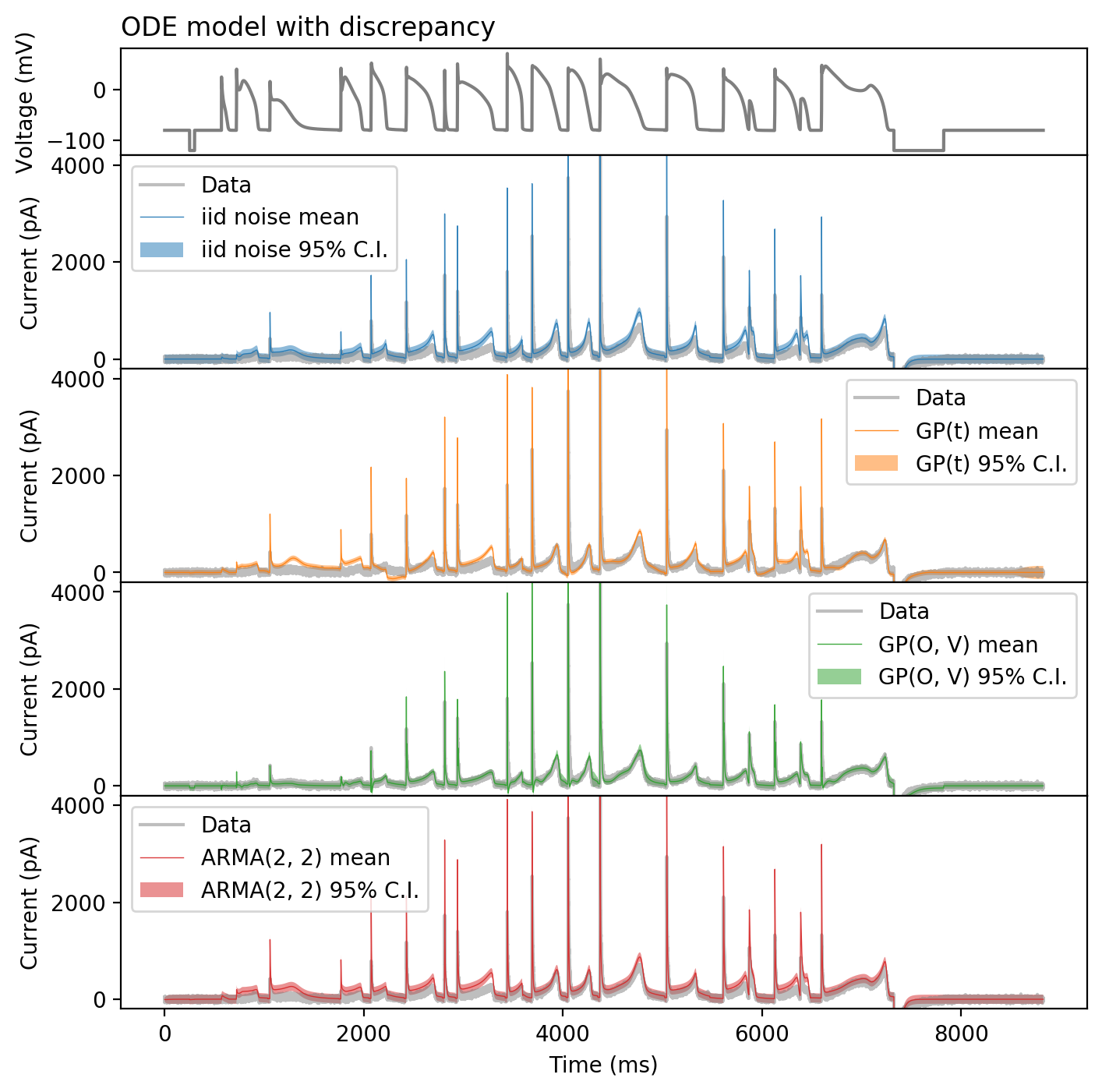}
  \caption{\label{fig:tutorial-ion-channel-prediction2a-b}
    Model~B prediction with different discrepancy models: no discrepancy (i.i.d.\ noise), $\GP(t)$, $\GP(O, V)$, and $\ARMA(2, 2)$.
    The voltage clamp protocol for calibration is the action potential series protocol \cite{beattie2018sinusoidal}.
  }
\end{figure}

\clearpage
\subsubsection{Model~B: Discrepancy predictions}\label{sec:model-b-disc-predict}

\begin{figure}[htb]
  \centering
  \includegraphics[width=\textwidth]{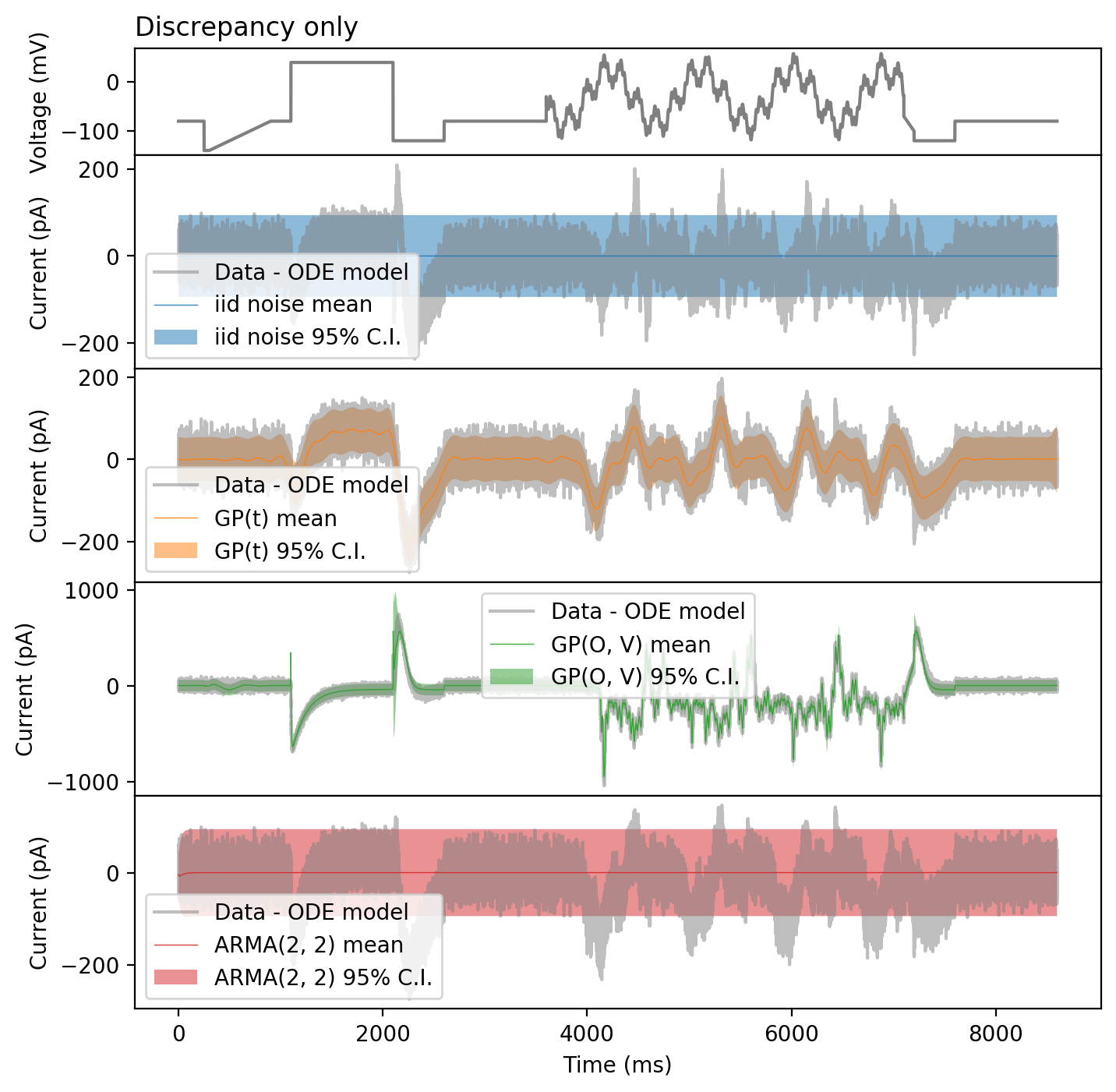}
  \caption{\label{fig:tutorial-ion-channel-fitting2-residual-b}
    Model~B fitting residuals of the MAP estimate accounted by different discrepancy models: no discrepancy (i.i.d.\ noise), $\GP(t)$, $\GP(O, V)$, and $\ARMA(2, 2)$.
    The voltage clamp protocol for calibration is the sinusoidal protocol \cite{beattie2018sinusoidal}.
  }
\end{figure}

\begin{figure}[htb]
  \centering
  \includegraphics[width=\textwidth]{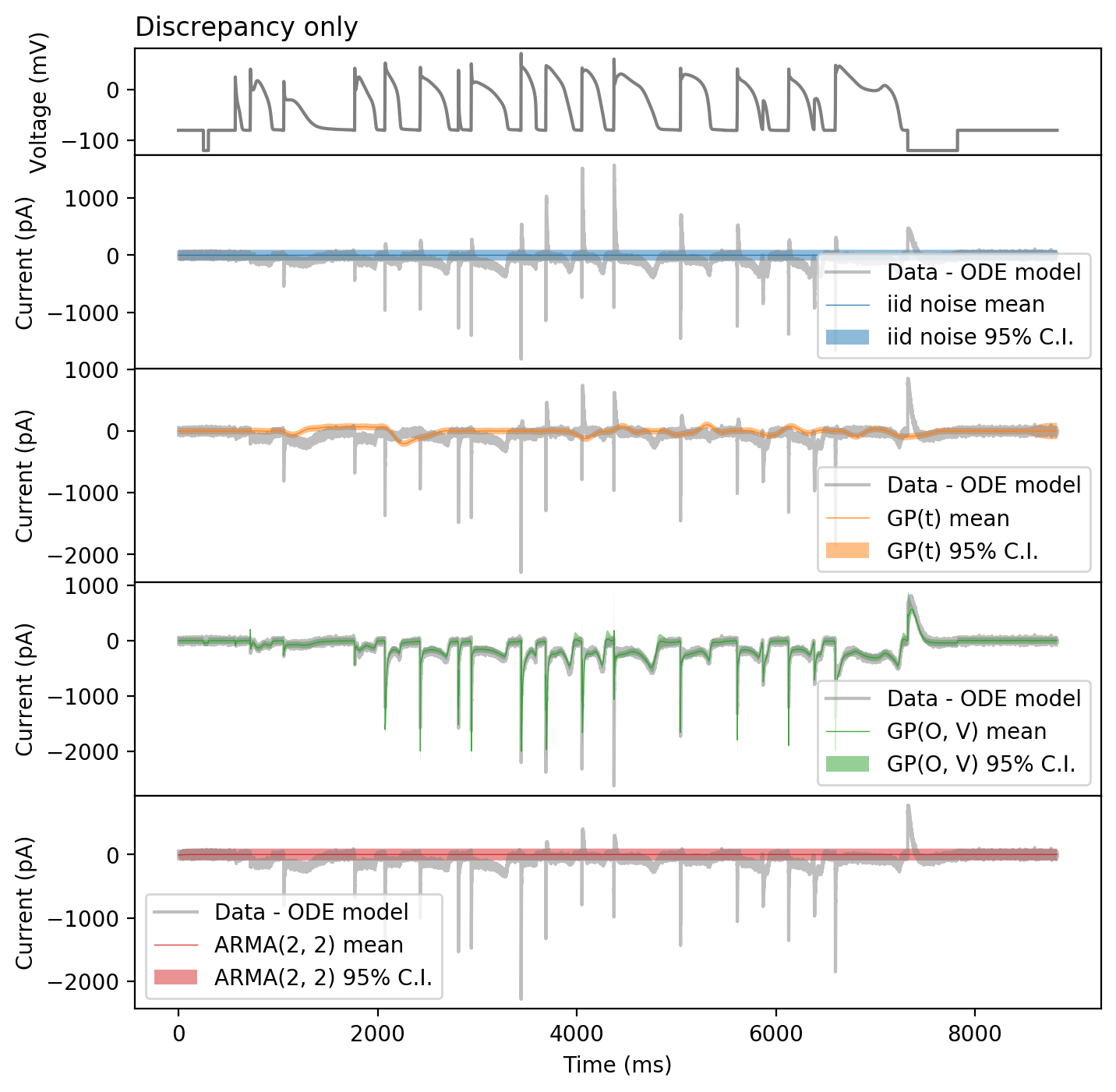}
  \caption{\label{fig:tutorial-ion-channel-prediction2a-residual-b}
    Model~B prediction residuals of the MAP estimate accounted by different discrepancy models: no discrepancy (i.i.d.\ noise), $\GP(t)$, $\GP(O, V)$, and $\ARMA(2, 2)$.
    The voltage clamp protocol for calibration is the action potential series protocol \cite{beattie2018sinusoidal}.
  }
\end{figure}

\begin{figure}[htb]
  \centering
  \includegraphics[width=\textwidth]{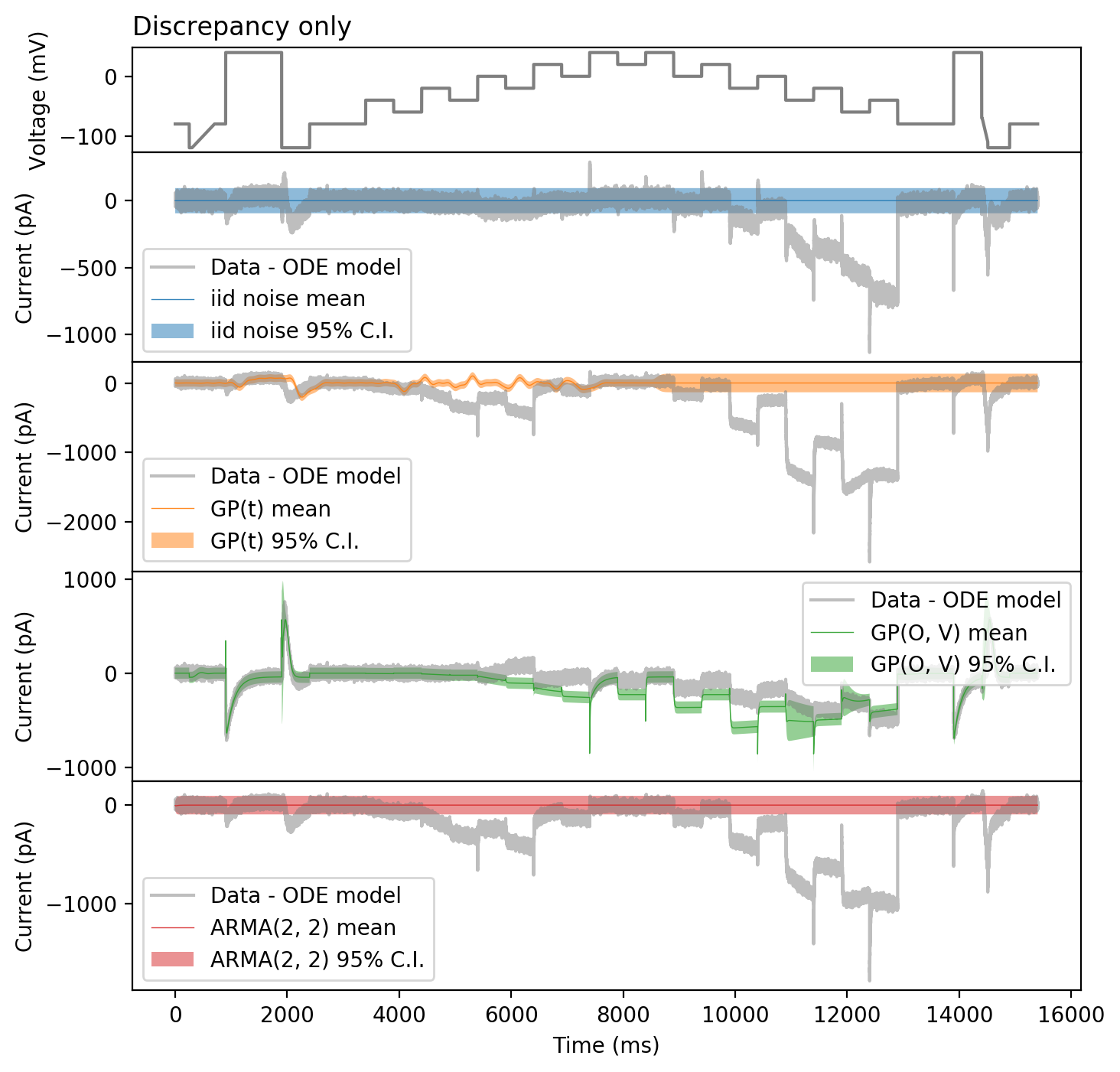}
  \caption{\label{fig:tutorial-ion-channel-prediction2b-residual-b}
    Model~B prediction residuals of the MAP estimate accounted by different discrepancy models: no discrepancy (i.i.d.\ noise), $\GP(t)$, $\GP(O, V)$, and $\ARMA(2, 2)$.
    The voltage clamp protocol for calibration is the staircase protocol \cite{lei_rapid_2019-1}.
  }
\end{figure}

\clearpage
\subsubsection{Model~B: ODE model predictions}\label{sec:model-b-model-predict}

\begin{figure}[htb]
	\centering
	\includegraphics[width=\textwidth]{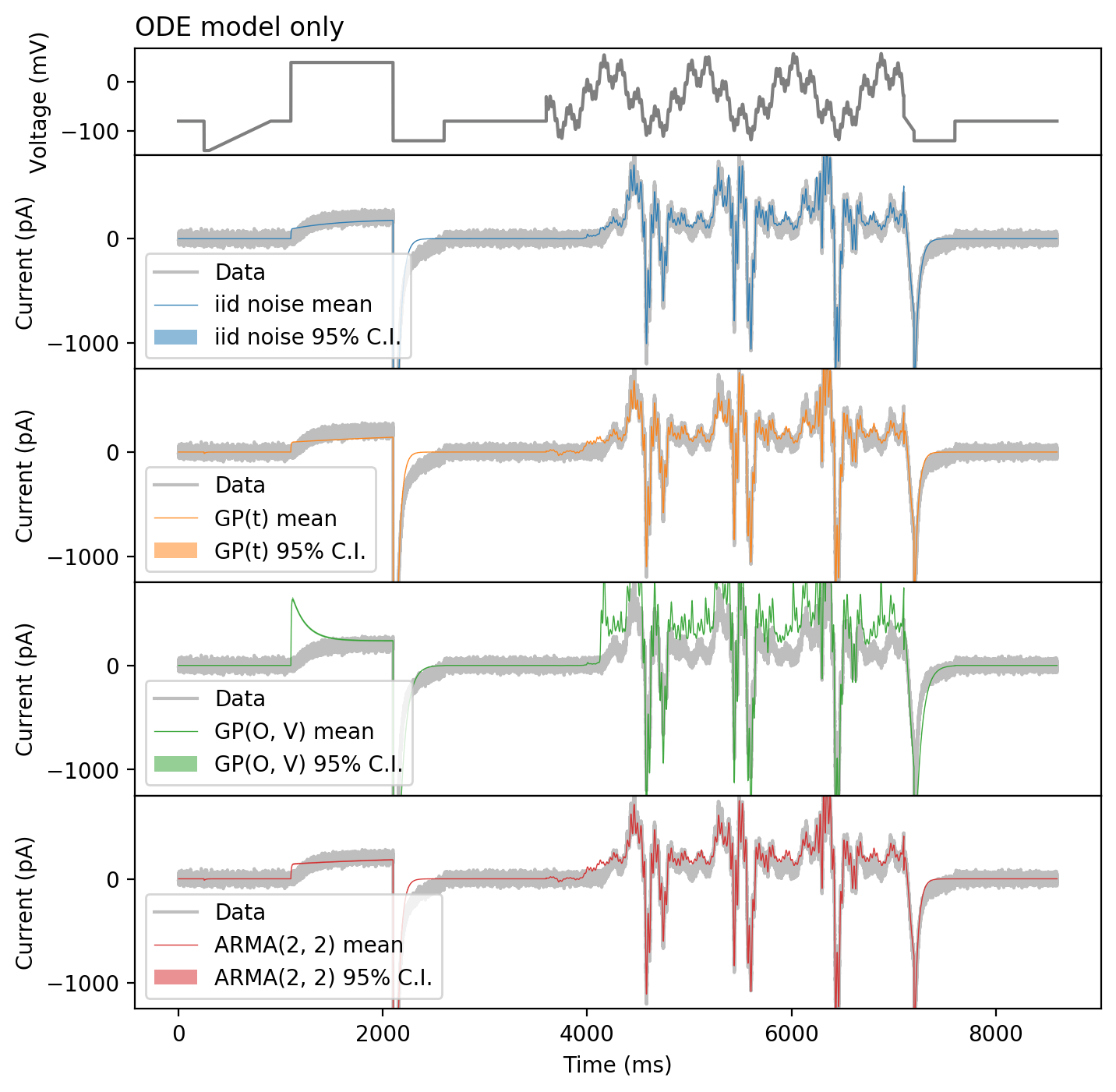}
	\caption{\label{fig:tutorial-ion-channel-fitting2-model-b}
		Fitting of the ODE model of Model~B, using different discrepancy models: no discrepancy (i.i.d.\ noise), $\GP(t)$, $\GP(O, V)$, and $\ARMA(2, 2)$.
		The voltage clamp protocol for calibration is the sinusoidal protocol \cite{beattie2018sinusoidal}.
	}
\end{figure}

\begin{figure}[htb]
	\centering
	\includegraphics[width=\textwidth]{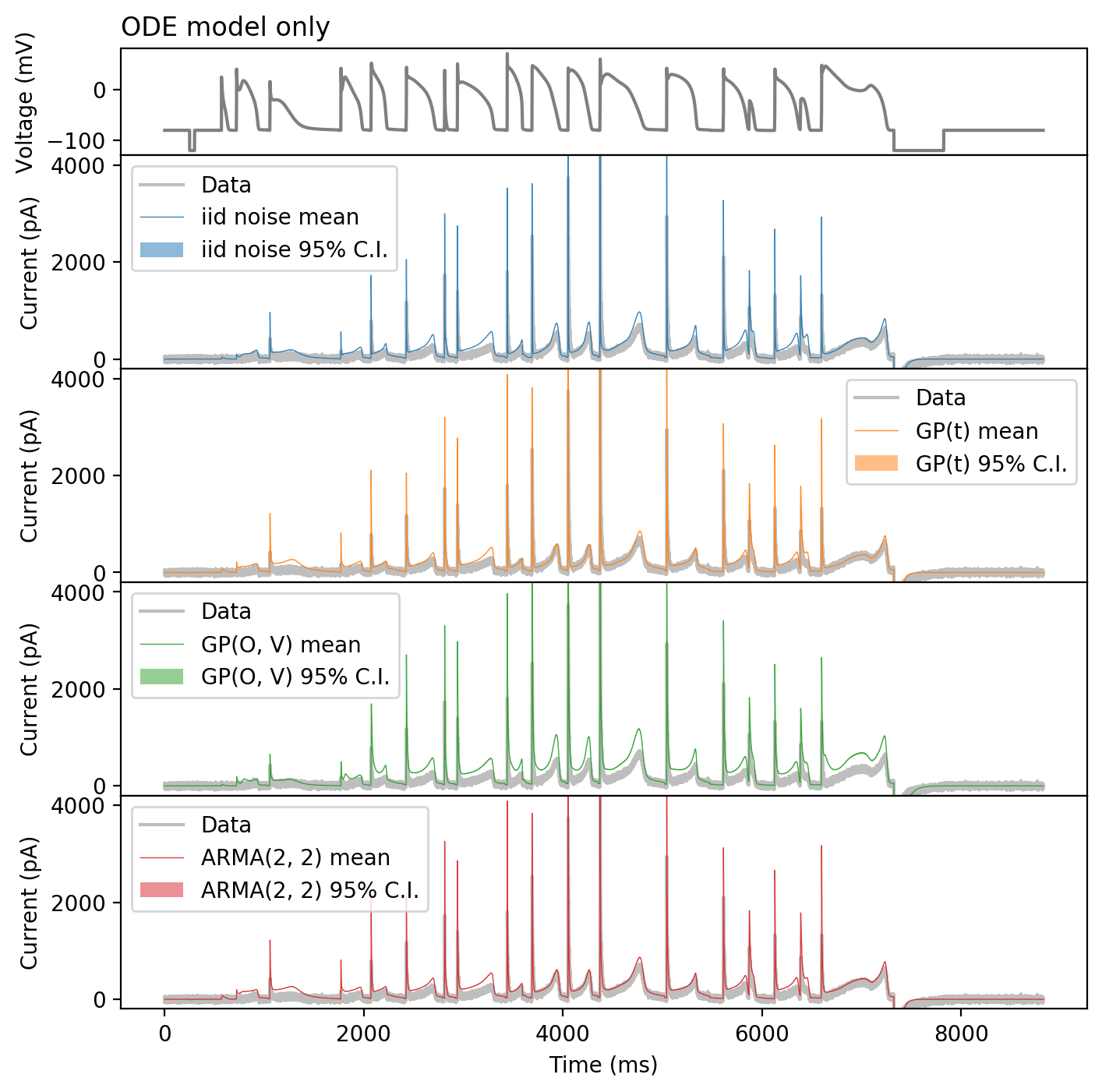}
	\caption{\label{fig:tutorial-ion-channel-prediction2a-model-b}
		Predictions of the ODE model of Model~B, using different discrepancy models: no discrepancy (i.i.d.\ noise), $\GP(t)$, $\GP(O, V)$, and $\ARMA(2, 2)$.
		The voltage clamp protocol for calibration is the action potential series protocol \cite{beattie2018sinusoidal}.
	}
\end{figure}

\begin{figure}[htb]
	\centering
	\includegraphics[width=\textwidth]{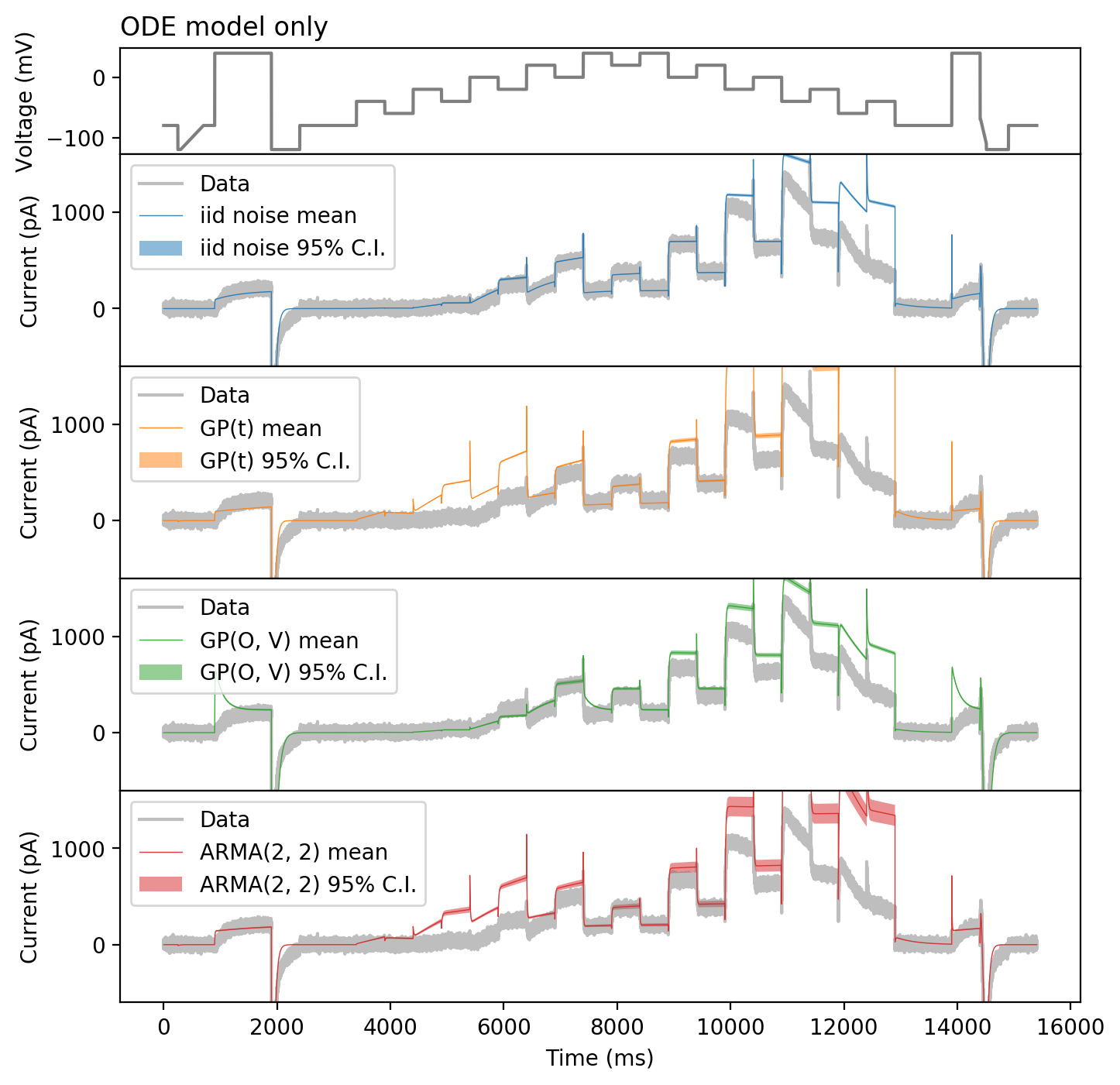}
	\caption{\label{fig:tutorial-ion-channel-prediction2b-model-b}
		Predictions of the ODE model of Model~B, using different discrepancy models: no discrepancy (i.i.d.\ noise), $\GP(t)$, $\GP(O, V)$, and $\ARMA(2, 2)$.
		The voltage clamp protocol for calibration is the staircase protocol \cite{lei_rapid_2019-1}.
	}
\end{figure}

\clearpage
\subsection{GP covariance functions: RBF, OU and Matern3/2}
\label{sub:GPCovFuncs}

\subsubsection*{Full model predictions}

\begin{figure}[htb]
  \centering
  \includegraphics[width=\textwidth]{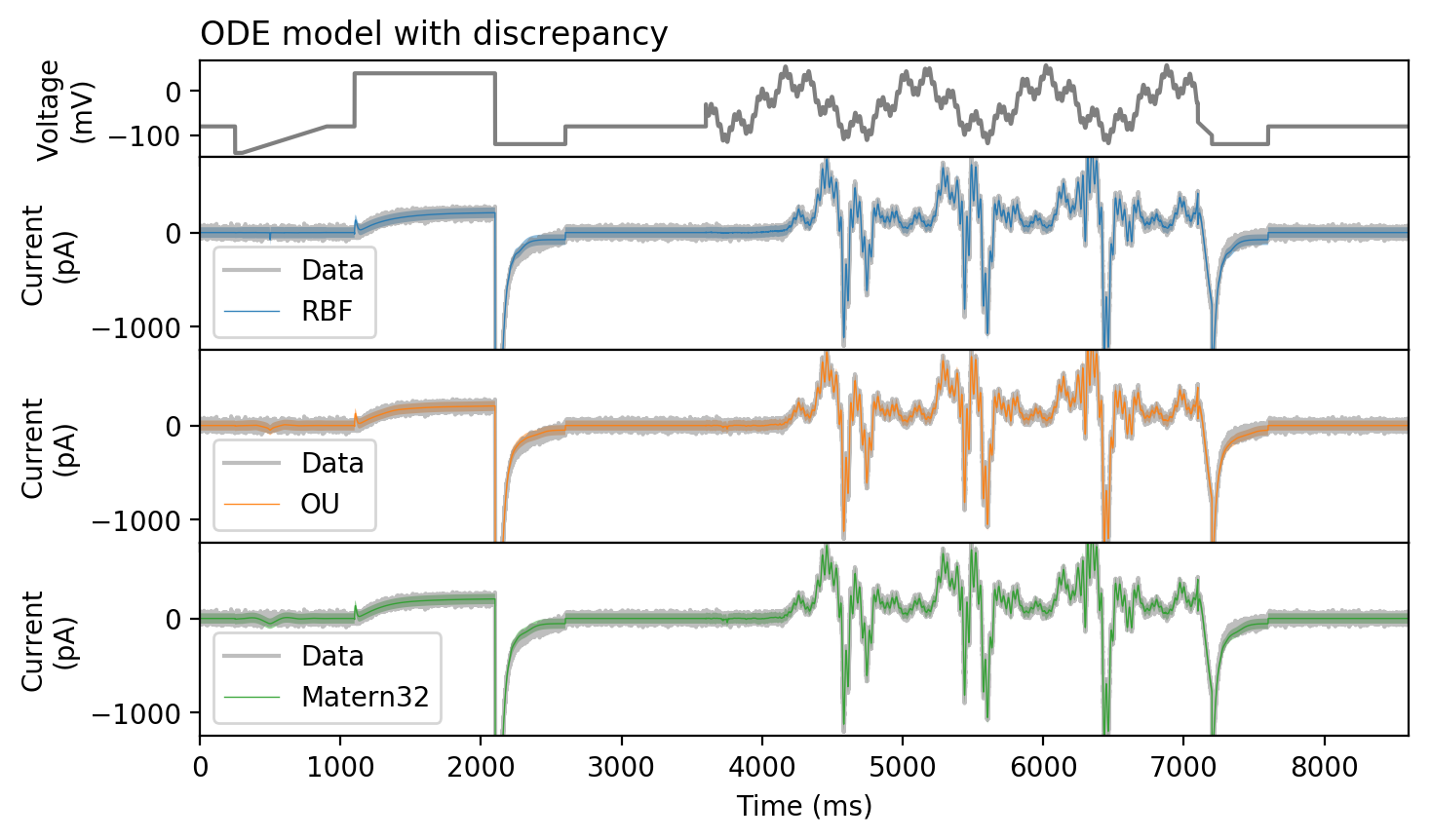}
  \caption{
    Model~A $\GP(O, V)$ discrepancy model fitted with different GP covariance functions: radial basis function (RBF), Ornstein–Uhlenbeck (OU, also known as exponential covariance function), and Mat\'{e}rn 3/2 covariance function.
    The voltage clamp protocol for calibration is the sinusoidal protocol \cite{beattie2018sinusoidal}.
    It shows the posterior predictive with the bounds showing the 95\% credible interval.
  }
\end{figure}

\begin{figure}[tbh]
  \centering
  \includegraphics[width=\textwidth]{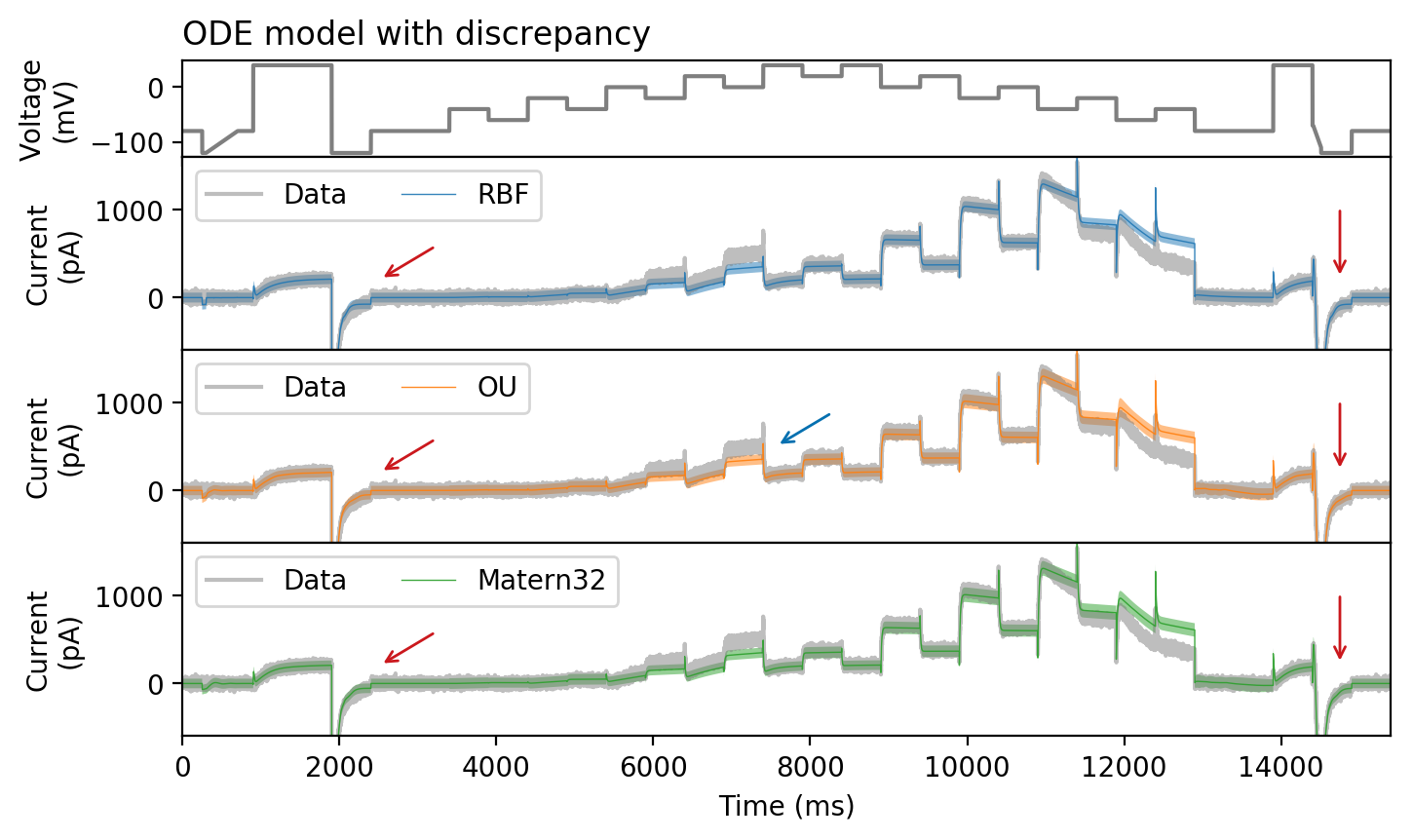}
  \caption{
    Model~A $\GP(O, V)$ discrepancy model fitted with different GP covariance functions: radial basis function (RBF), Ornstein–Uhlenbeck (OU, also known as exponential covariance function), and Mat\'{e}rn 3/2 covariance function.
    The voltage clamp protocol for calibration is the staircase protocol \cite{lei_rapid_2019-1}.
    It shows the posterior predictive with the bounds showing the 95\% credible interval.
      }
\end{figure}

\begin{figure}[htb]
  \centering
  \includegraphics[width=\textwidth]{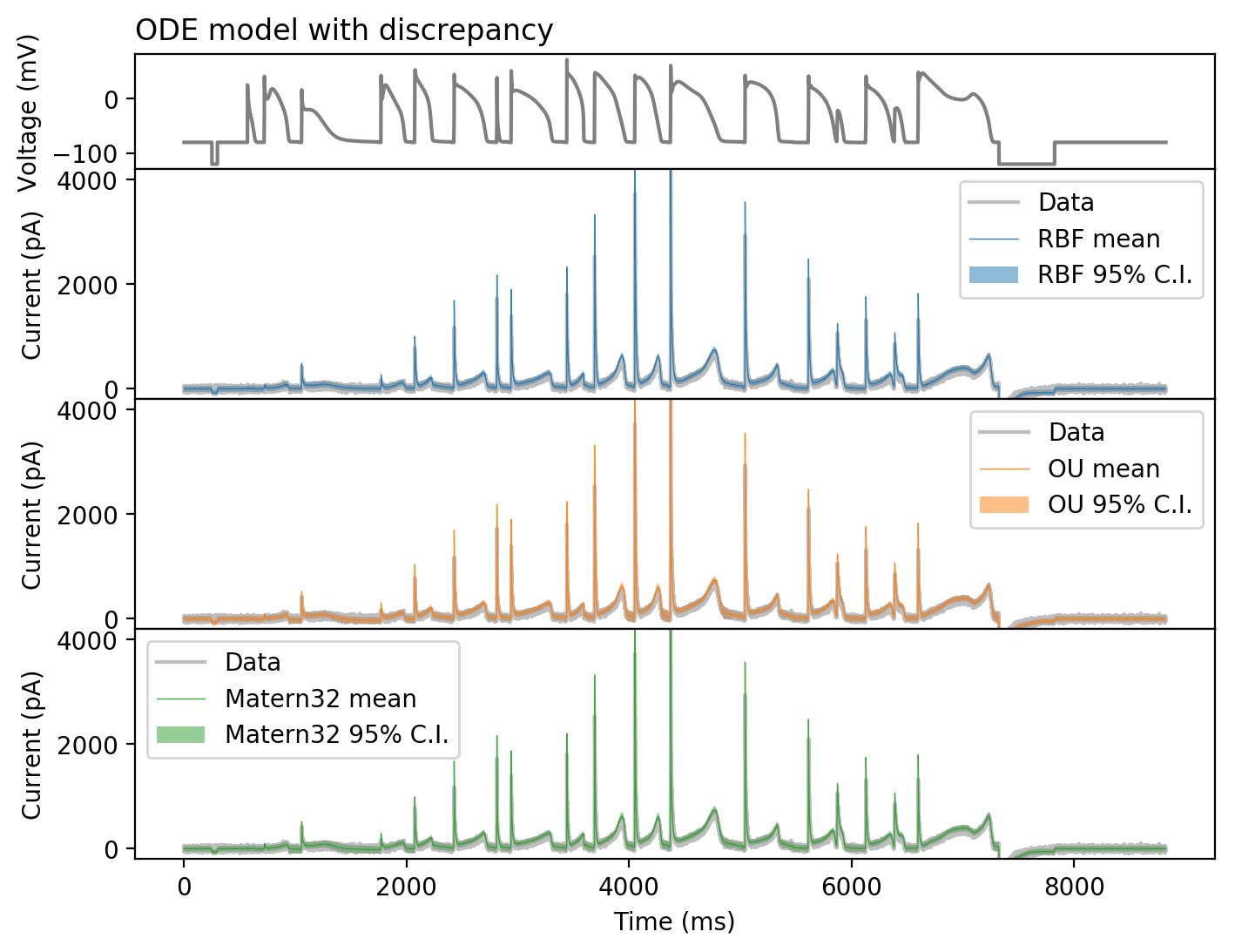}
  \caption{
    Model~A $\GP(O, V)$ discrepancy model fitted with different GP covariance functions: radial basis function (RBF), Ornstein–Uhlenbeck (OU, also known as exponential covariance function), and Mat\'{e}rn 3/2 covariance function.
    The voltage clamp protocol for calibration is the action potential series protocol \cite{beattie2018sinusoidal}.
    It shows the posterior predictive with the bounds showing the 95\% credible interval.
  }
\end{figure}

\clearpage
\subsubsection*{Discrepancy predictions}

\begin{figure}[htb]
  \centering
  \includegraphics[width=\textwidth]{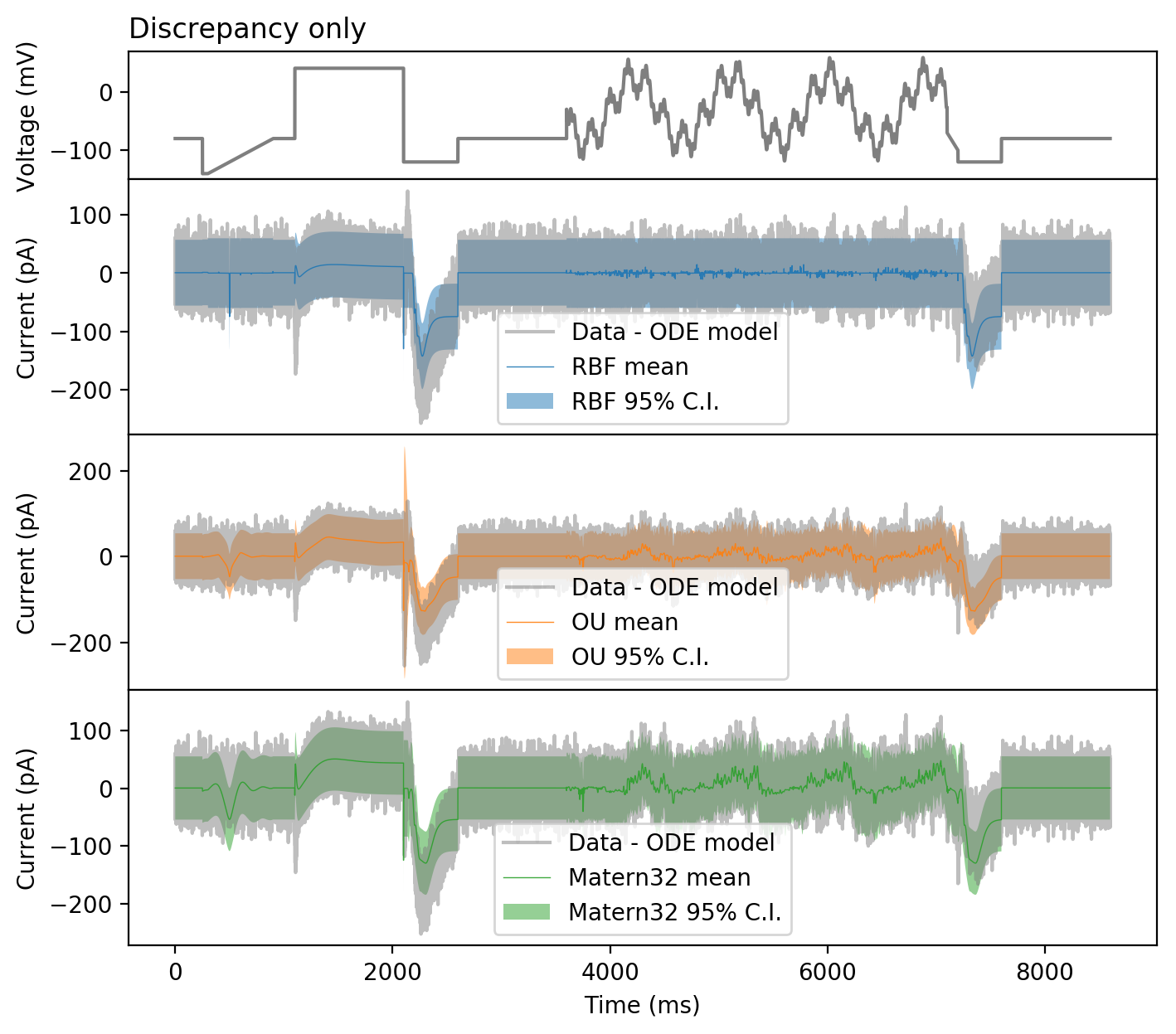}
  \caption{
    Model~A with $\GP(O, V)$ discrepancy model fitting residuals of the MAP estimate accounted by different GP covariance functions: radial basis function (RBF), Ornstein–Uhlenbeck (OU, also known as exponential covariance function), and Mat\'{e}rn 3/2 covariance function.
    The voltage clamp protocol for calibration is the sinusoidal protocol \cite{beattie2018sinusoidal}.
  }
\end{figure}

\begin{figure}[htb]
  \centering
  \includegraphics[width=\textwidth]{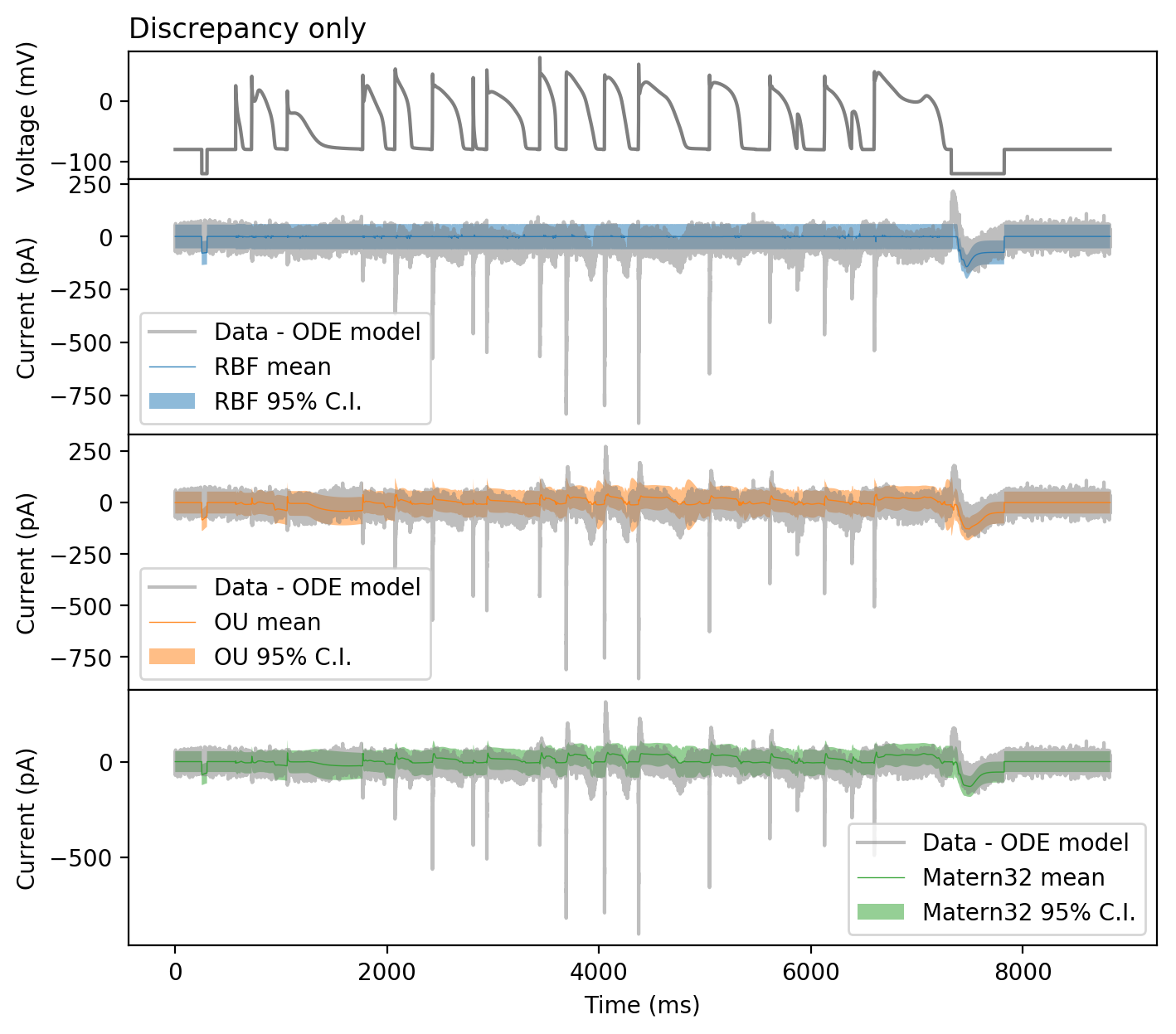}
  \caption{
    Model~A with $\GP(O, V)$ discrepancy model fitting residuals of the MAP estimate accounted by different GP covariance functions: radial basis function (RBF), Ornstein–Uhlenbeck (OU, also known as exponential covariance function), and Mat\'{e}rn 3/2 covariance function.
    The voltage clamp protocol for calibration is the action potential series protocol \cite{beattie2018sinusoidal}.
  }
\end{figure}

\begin{figure}[htb]
  \centering
  \includegraphics[width=\textwidth]{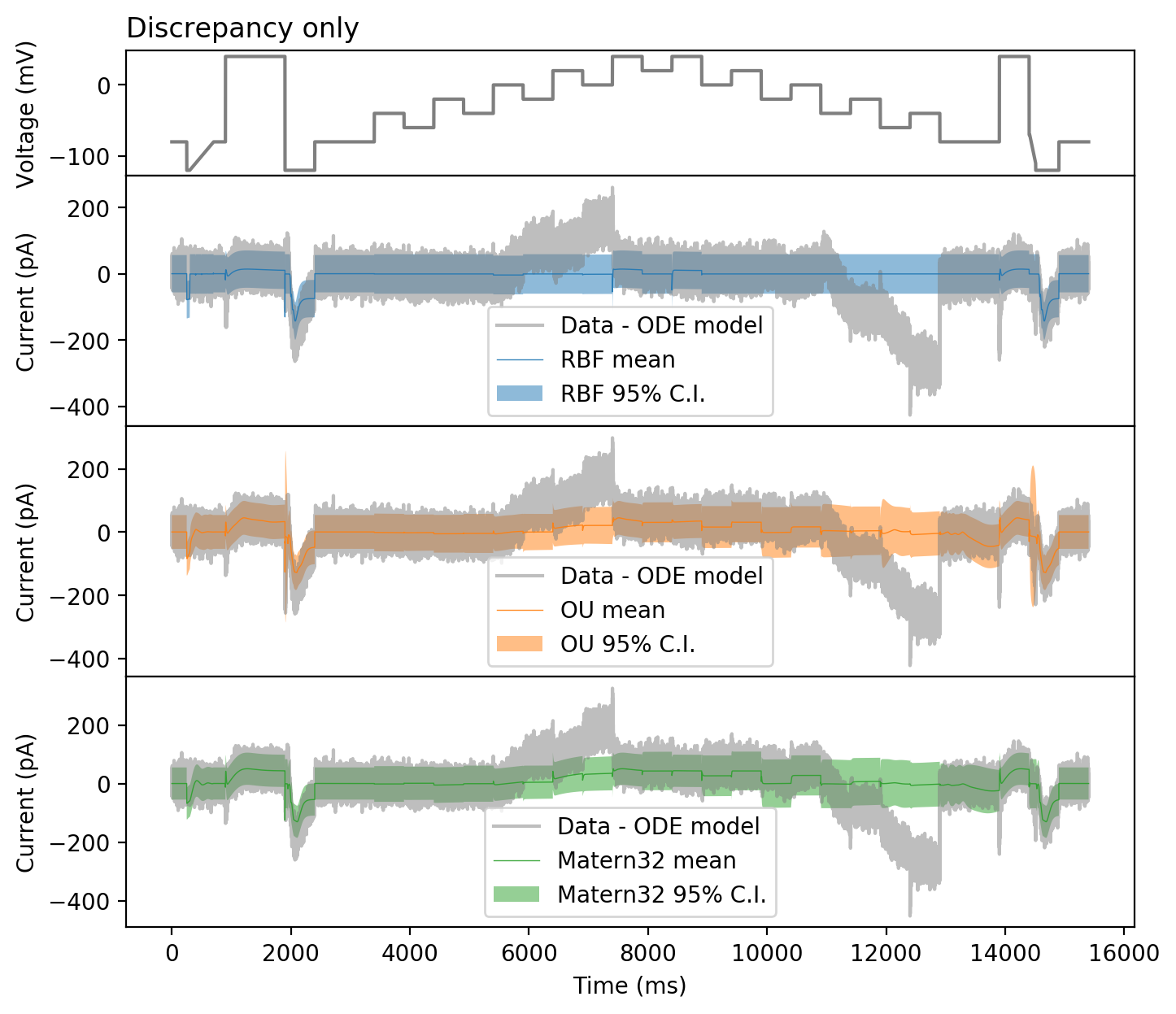}
  \caption{
    Model~A with $\GP(O, V)$ discrepancy model fitting residuals of the MAP estimate accounted by different GP covariance functions: radial basis function (RBF), Ornstein–Uhlenbeck (OU, also known as exponential covariance function), and Mat\'{e}rn 3/2 covariance function.
    The voltage clamp protocol for calibration is the staircase protocol \cite{lei_rapid_2019-1}.
  }
\end{figure}

\clearpage
\subsubsection*{ODE model predictions}

\begin{figure}[htb]
	\centering
	\includegraphics[width=\textwidth]{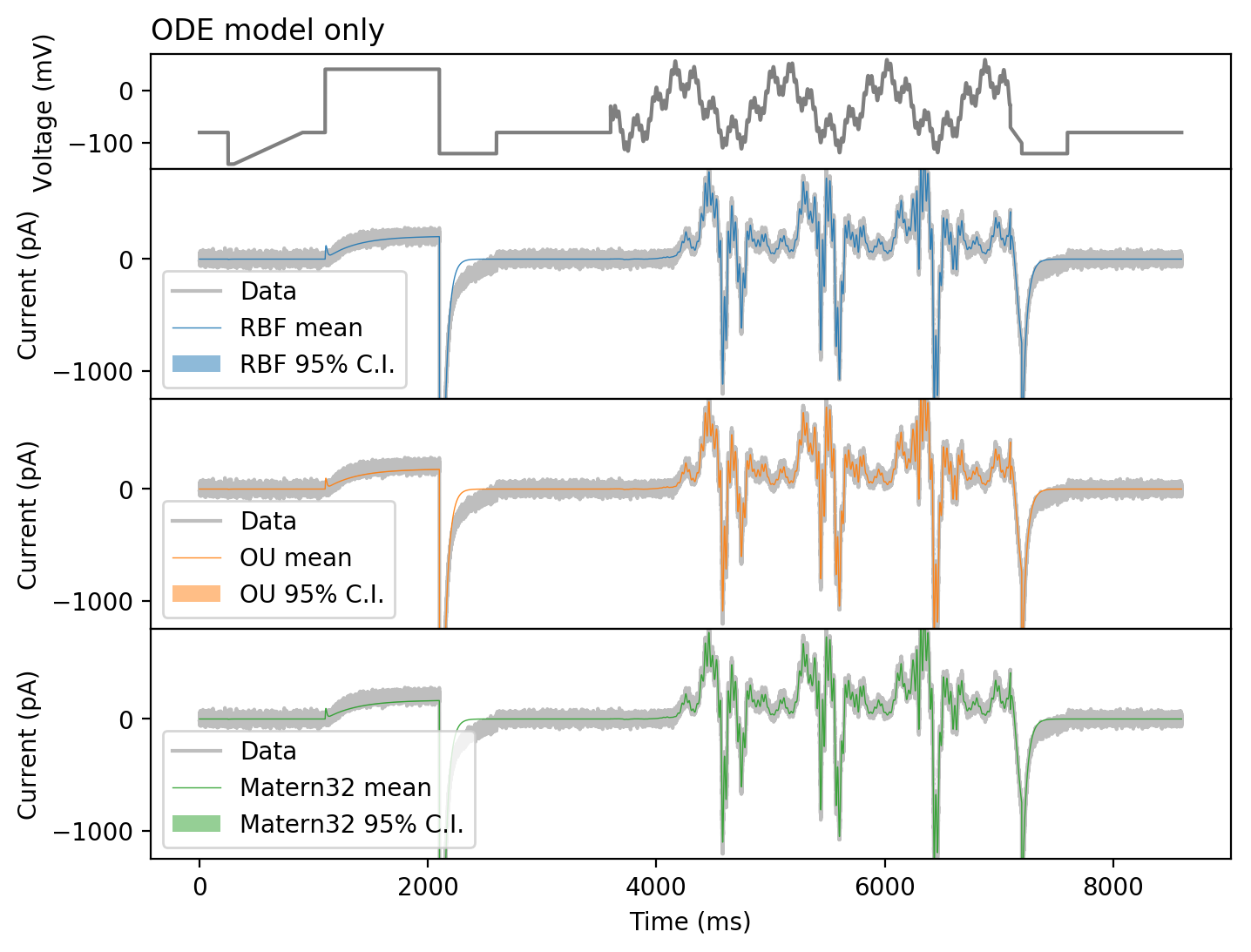}
	\caption{
		Fitting of the ODE model of Model~A with $\GP(O, V)$ discrepancy model, using different GP covariance functions: radial basis function (RBF), Ornstein–Uhlenbeck (OU, also known as exponential covariance function), and Mat\'{e}rn 3/2 covariance function.
		The voltage clamp protocol for calibration is the sinusoidal protocol \cite{beattie2018sinusoidal}.
	}
\end{figure}

\begin{figure}[htb]
	\centering
	\includegraphics[width=\textwidth]{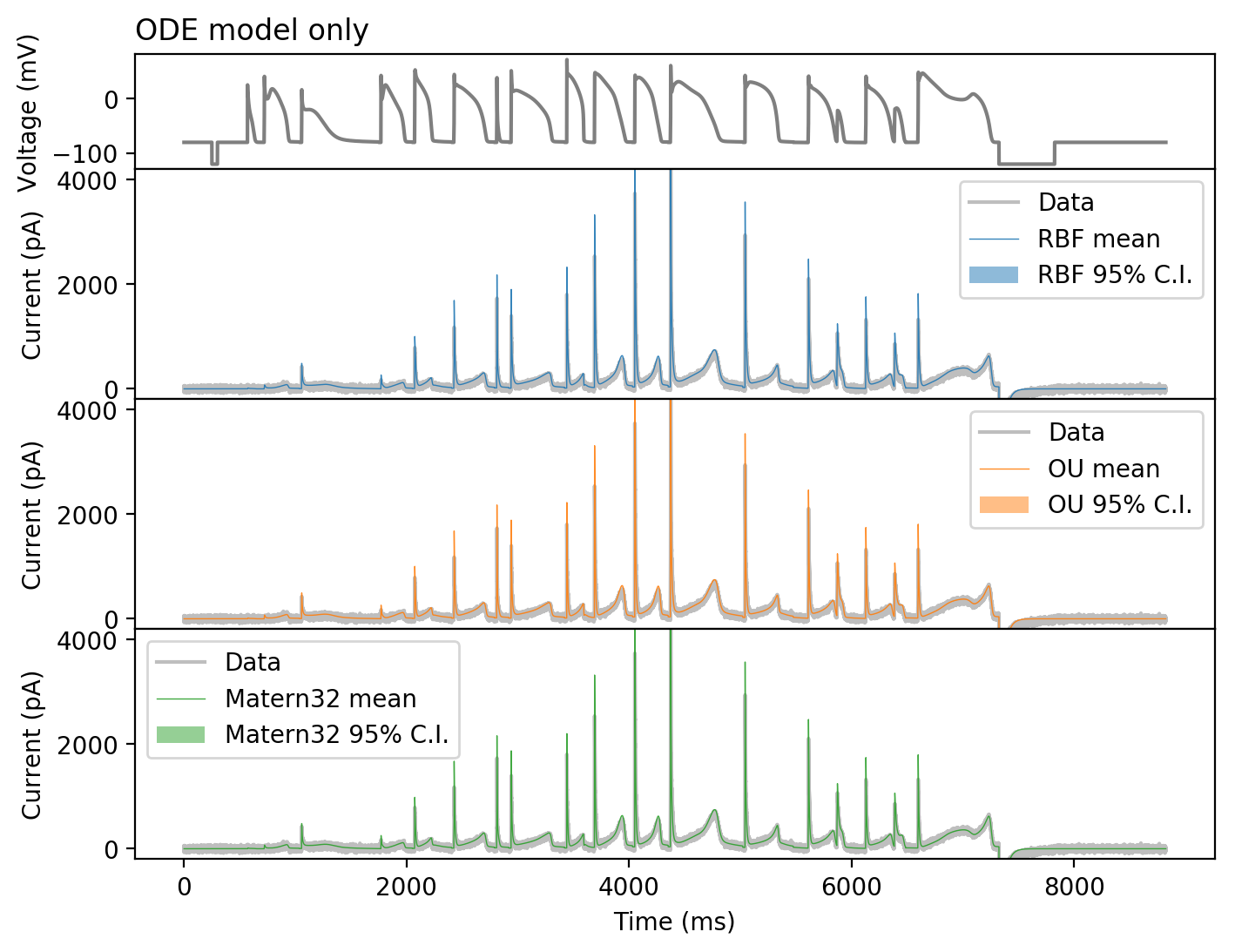}
	\caption{
		Fitting of the ODE model of Model~A with $\GP(O, V)$ discrepancy model, using different GP covariance functions: radial basis function (RBF), Ornstein–Uhlenbeck (OU, also known as exponential covariance function), and Mat\'{e}rn 3/2 covariance function.
		The voltage clamp protocol for calibration is the action potential series protocol \cite{beattie2018sinusoidal}.
	}
\end{figure}

\begin{figure}[htb]
	\centering
	\includegraphics[width=\textwidth]{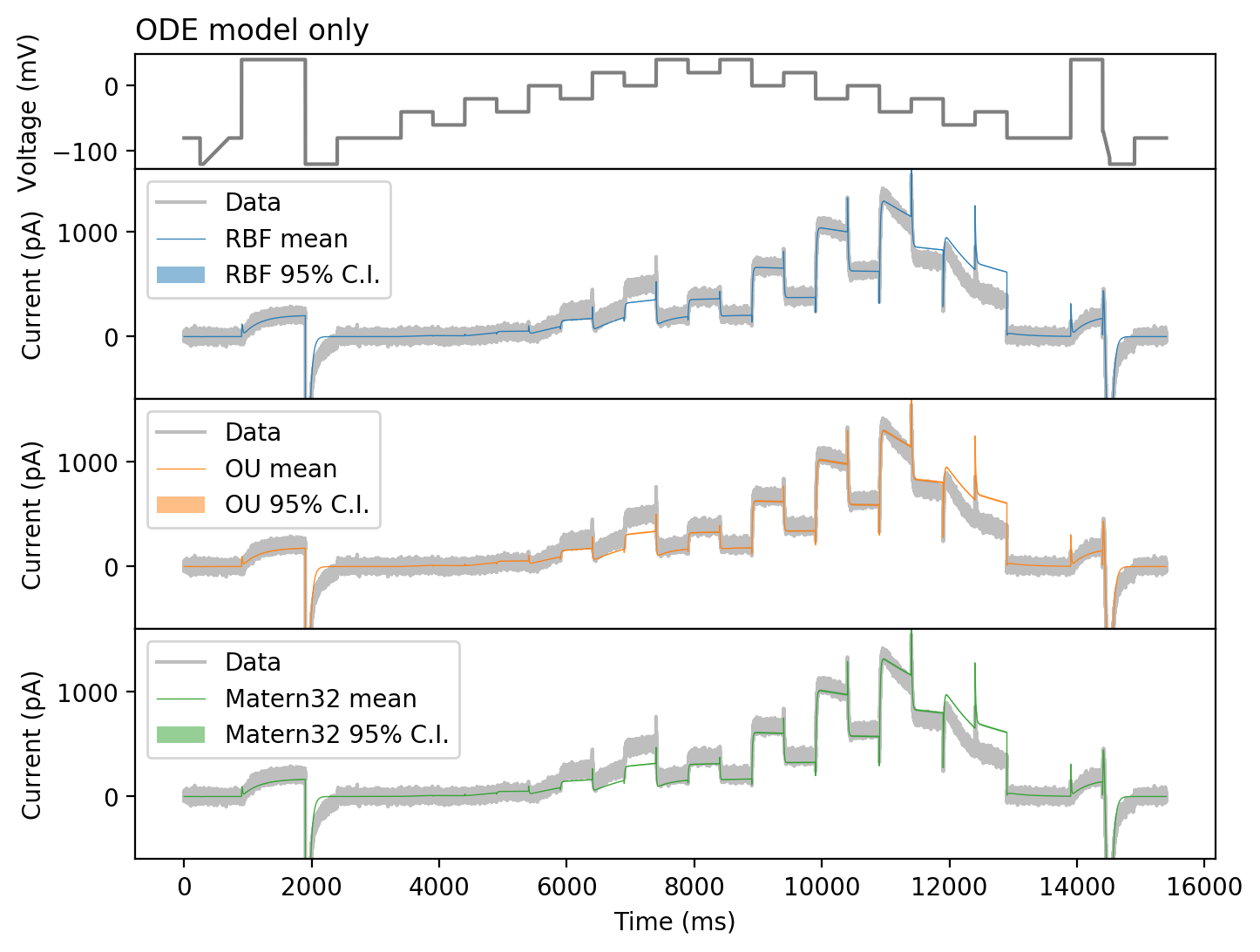}
	\caption{
		Fitting of the ODE model of Model~A with $\GP(O, V)$ discrepancy model, using different GP covariance functions: radial basis function (RBF), Ornstein–Uhlenbeck (OU, also known as exponential covariance function), and Mat\'{e}rn 3/2 covariance function.
		The voltage clamp protocol for calibration is the staircase protocol \cite{lei_rapid_2019-1}.
	}
\end{figure}

\clearpage

\bibliographystyle{ieeetr}
\bibliography{refs}

\makeatletter\@input{yy.tex}\makeatother